\documentclass[aps,prd,preprintnumbers,superscriptaddress,nofootinbib]{revtex4}

\usepackage{hyperref}
\hypersetup{colorlinks=true,linkcolor=purple,anchorcolor=blue,citecolor=blue, filecolor=blue,urlcolor=red,bookmarksnumbered=true,
pdfview=FitB
}
\usepackage{color}
\usepackage{xcolor}
\usepackage{ulem}
\usepackage{csquotes}
\colorlet{purple1}{blue!70!red}
\colorlet{darkred}{red!50!black}

\usepackage{graphicx}
\usepackage[bf,SL,BF]{subfigure}
\usepackage{psfrag}
\usepackage{color}
\usepackage{amssymb}
\usepackage{amsmath}
\usepackage{epstopdf}
\usepackage{natbib}
\newcommand{\be}{\begin{eqnarray}}
\newcommand{\ee}{\end{eqnarray}}

\newcommand{\bfk}{{\bf k}_{\perp}}

\newcommand{\bfP}{{\bf P}_{\perp}} 
 
\newcommand{\bfp}{{\bf p}_{\perp}}

\begin{document}

\title{Sivers and Boer-Mulders TMDs of proton in a light-front quark-diquark model
}

\author{Bheemsehan~Gurjar}
\email{gbheem@iitk.ac.in} 
\affiliation{Indian Institute of Technology Kanpur, Kanpur-208016, India}

\author{Dipankar~Chakrabarti}
\email{dipankar@iitk.ac.in} 
\affiliation{Indian Institute of Technology Kanpur, Kanpur-208016, India}

\author{Chandan~Mondal}
\email{mondal@impcas.ac.cn} 
\affiliation{Institute of Modern Physics, Chinese Academy of Sciences, Lanzhou 730000, China}
\affiliation{School of Nuclear Science and Technology, University of Chinese Academy of Sciences, Beijing 100049, China}

\date{\today}

\begin{abstract}
We obtain the leading twist T-odd quark transverse momentum dependent parton distribution functions (TMDs) of the
proton, namely the Sivers function, $f_{1T}^{\perp q}(x,p_\perp^2)$, and the Boer-Mulders function, $h_1^{\perp q}(x,p_\perp^2)$, in a light-front quark-diquark model constructed with the wave functions predicted by the soft-wall AdS/QCD. The gluon rescattering is crucial to predict a nonzero T-odd TMDs. We study the utility of a nonperturbative $SU(3)$ gluon rescattering kernel going beyond the usual approximation of perturbative $U(1)$ gluons. The spin asymmetries in semi-inclusive deep inelastic scattering (SIDIS) associated with these T-odd TMDs are found to be consistent with HERMES and COMPASS data. We also evaluate the generalized Sivers and Boer-Mulders shifts and compare them with the available lattice QCD simulations.

\end{abstract}

\maketitle

	\section{Introduction}
	The generalized  Parton distributions(GPDs)~\cite{Muller:1994ses,Ji:1996ek,Radyushkin:1997ki} and the transverse momentum dependent parton distribution(TMDs)~\cite{Anselmino:1994gn,Barone:2001sp} provide insights into the three-dimensional structure of proton and the spin and orbital angular momentum distributions at the partonic level and have been studied in different QCD inspired models~\cite{Zhang:2008nu,Gamberg:2007wm,Burkardt:2007xm,Pasquini:2010af,DAlesio:2004eso,Efremov:2004tp,Anselmino:2005ea,Collins:2005rq,Anselmino:2008sga,Anselmino:2013rya,Martin:2017yms,Barone:2009hw}. Various single spin asymmetries (SSAs)~\cite{E581:1991eys,FNAL-E704:1991ovg} measured in the semi-inclusive deep inelastic scattering (SIDIS) and the Drell-Yan (DY) processes can be related with  the T-odd TMDs, namely,  Sivers~\cite{Sivers:1989cc,Collins:2002kn} and Boer-Mulders functions~\cite{Boer:1997nt,Boer:2003cm}. The Sivers and Boer-Mulders asymmetries in the SIDIS require a final state interaction (FSI) in which the struck quark exchanges a gluon with the spectators~\cite{COMPASS:2010hbb,COMPASS:2008isr,HERMES:2009lmz,JeffersonLabHallA:2011ayy,Sbrizzai:2016gro}. In the quark-diquark model of a proton, the FSI involves a gluon exchanged between the struck quark and the diquark. In some works with light-front quark-diquark model~\cite{Hwang:2010dd,Maji:2017wwd,Lyubovitskij:2022vcl}, the effect of FSI has been incorporated in the light-front wave functions to model the Sivers and Boer-Mulders functions. Another way to produce the Sivers and Boer Mulders functions in the overlap of the light-front wave function  formalism is to introduce a kernel that contains the FSI.  In this work, we follow the second path, i.e, to include the FSI effects through a gluon rescattering kernel \cite{Lu:2006kt,PhysRevD.78.074010}.
	
	The FSI, which is a soft gluon exchange between the active quark and the spectators, can be described by a Wilson line included in the quark distribution function \cite{Lu:2006kt,Brodsky:2002cx}. It describes the phase factor of the struck quark as it leaves the proton \cite{Ji:2002aa}. This Wilson line phase factor describes the effect of the transverse component of the force acting on the struck quark and is on average directed towards the center of the proton thus giving rise to the chromodynamics lensing \cite{Burkardt:2003je,Belitsky:2002sm,Boer:2003cm}. In the quark-diquark model of nucleons, the spin-flip GPD ${E}_q(x, t)$  and the Sivers function $f_{1T}^{\perp q}(x,p_\perp^2)$ are expressed as the overlap of same light-front wave functions and enable us to relate the Sivers function with the GPD in the impact parameter space  through the ``lensing function" ${\cal{I}} (x,b_\perp)$ \cite{Maji:2017ill,Gurjar:2021dyv,Meissner:2007rx}.  Again, the first moment of T-odd TMDs ( Sivers and Boer-Mulders functions) can be written as the convolution of the GPD $E$ with the gluon rescattering kernel and thus the lensing function and the gluon rescattering kernel are related to each other \cite{Gurjar:2021dyv,Meissner:2007rx}.
	
	As the Sivers and Boer Mulder's functions are related to the SSAs observed in the SIDIS and the DY processes, they are investigated in several models \cite{Kafer:2008ud,Bressan:2009eu,HERMES:2012kpt,Giordano:2009hi,E581:1991eys,FNAL-E704:1991ovg}. In this work, we consider a light-front quark-diquark model including only scalar diquarks with the wave functions modeled from the prediction of light-front holography \cite{Gutsche:2013zia,Mondal:2015uha}. Rather than modifying the wave functions \cite{Maji:2017wwd,Gurjar:2021dyv}, here  we  consider the gluon rescattering kernel to incorporate the FSI effect \cite{Brodsky:2002cx,Burkardt:2003je}. Both perturbative $U(1)$ and nonperturbative $SU(3)$ gluon rescattering kernels are compared and contrasted in the model. We concentrate only on the SIDIS processes and compare our results for the SSAs with the  HERMES and COMPASS data \cite{Barone:2009hw,Kafer:2008ud,Giordano:2009hi,HERMES:2009lmz}.  The generalized Sivers and Boer-Mulders shifts provide information about the average transverse momentum distributions of unpolarized quarks orthogonal to the transverse spin of the proton and that of the transversely polarized quarks in an unpolarized proton, respectively. We compare our results for these shifts with the available lattice QCD results \cite{Musch:2011er}.
	
	The paper is organized as follows. In Sec.~\ref{model} and Sec.~\ref{tmd}, we give a brief introduction about the light-front quark-diquark model and the TMDs, respectively. The relation between the gluon rescattering kernel and the lensing function is presented in Sec.~\ref{lensing}. Our model results for the  Sivers and Boer-Mulders TMDs are discussed in Sec.~\ref{results}. The Sivers and Boer-Mulders shifts are evaluated and compared with lattice QCD results in Sec.~\ref{shifts} and the results for the SSAs are presented in Sec.~\ref{ssa}. We provide a summary in Sec.~\ref{concl}.
	
\section{Light-front quark-diquark model} \label{model}

Here we consider the generic ansatz for the light-front quark-diquark model for the proton~\cite{Gutsche:2013zia}, where the light-front wave functions (LFWFs)
are constructed from the solution of soft-wall Anti-de Sitter (AdS)/QCD. In this model,  the three valence quarks of the proton are contemplated as an effective system
composed of a  quark (fermion) and a bound state of diquark (boson) having spin zero, i.e., scalar diquark.
Then the two particle Fock state expansion for the proton spin components, $J^z = \pm \frac{1}{2}$ in a frame where the transverse momentum of proton assumed to be zero, i.e., $P \equiv \big(P^+,\textbf{0}_\perp,\frac{M^2}{P^+}\big)$, 
is expressed as
 \be\label{state}
  |P;\uparrow(\downarrow)\rangle
& =& \sum_q \int \frac{{\rm d}x~ {\rm d}^2\textbf{p}_{\perp}}{2(2\pi)^3\sqrt{x(1-x)}}\bigg[ \psi^{\uparrow(\downarrow)}_{+q}(x,\textbf{p}_{\perp})|+\frac{1}{2},0; xP^+,\textbf{p}_{\perp}\rangle + \psi^{\uparrow(\downarrow)}_{-q}(x,\textbf{p}_{\perp})|-\frac{1}{2},0; xP^+,\textbf{p}_{\perp}\rangle\bigg].
  \ee
Note that for nonzero transverse momentum of the proton, i.e., $\bfP\ne0$, the physical transverse momenta of the quark and the diquark are $\bfk^q=x\bfP+\bfp$ and $\bfk^D=(1-x)\bfP-\bfp$, respectively, where $x$ and $\bfp$ correspond to the longitudinal momentum fraction and the relative transverse momentum of the constituents, respectively. $\psi_{\lambda_q}^{\lambda_N}(x,\bfp)$ are the LFWFs with the proton helicities $\lambda_N=\pm$ and for the quark $\lambda_q=\pm$; plus and minus represent $+\frac{1}{2}$ and $-\frac{1}{2}$,  respectively. The LFWFs at the model scale $\mu_0^2=0.32$ GeV$^2$ are given by~\cite{Chakrabarti:2020kdc}
\be\label{WF}
\psi_{+q}^+(x,\bfp) &=&  \varphi_q^{(1)}(x,\bfp) \,,\nonumber\\ 
\quad
\psi_{-q}^+(x,\bfp) &=& -\frac{p^1 + ip^2}{xM}   \, \varphi_q^{(2)}(x,\bfp) \,, \nonumber\\
\psi_{+q}^-(x,\bfp) &=& \frac{p^1 - ip^2}{xM}  \, \varphi_q^{(2)}(x,\bfp)\,. \\
\psi_{-q}^-(x,\bfp) &=& \varphi_q^{(1)}(x,\bfp),\nonumber
\ee
with $\varphi_q^{(i=1,2)}(x,\bfp)$ being the modified form of the soft-wall AdS/QCD wave functions modeled by introducing the parameters $a_q^{(i)}$ and $b_q^{(i)}$ for the quark $q$~\cite{Gutsche:2013zia, Brodsky:2014yha},
\be\label{wf2}
\varphi_q^{(i)}(x,\bfp)&=&N_q^{(i)}\frac{4\pi}{\kappa}\sqrt{\frac{\log(1/x)}{1-x}}x^{a_q^{(i)}}
(1-x)^{b_q^{(i)}}\exp\bigg[-\frac{\bfp^2}{2\kappa^2}\frac{\log(1/x)}{(1-x)^2}\bigg].
\ee
When $a_q^{(i)}=b_q^{(i)}=0$, 
$\varphi_q^{(i)}(x,\bfp)$ reduces to the original AdS/QCD solution~\cite{Brodsky:2014yha}. Note that the modification of the soft-wall AdS/QCD solution in Eq.~(\ref{wf2}) is not unique. A generic reparametrization function $w(x)$ that unifies the description of polarized and unpolarized quark distributions in the proton has been introduced in Refs.~\cite{deTeramond:2018ecg,Liu:2019vsn}.
We take the AdS/QCD scale parameter $\kappa =0.4$ GeV, fixed by fitting the nucleon electromagnetic form factors in the soft-wall model of AdS/QCD \cite{Chakrabarti:2013gra,Chakrabarti:2013dda}. In this model, the quarks are assumed to be massless and the parameters $a^{(i)}_q$ and $b^{(i)}_q$ with the constants $N^{(i)}_q$ are determined by fitting the electromagnetic properties of the nucleons, i.e., $F_1^q(0)=n_q$ and $F_2^q(0)=\kappa_q$, with $n_u=2$ and $n_d=1$ being the number of valence $u$ and $d$ quarks in proton and
the anomalous magnetic moments for the $u$ and $d$ quarks are $\kappa_u=1.673$ and
$\kappa_d=-2.033$~\cite{Chakrabarti:2015ama,Mondal:2017wbf}.  Since no flavor or isospin symmetry is imposed, the parameters for $d$ and $u$ quarks in the model  are different. The parameters are given by  $a^{(1)}_u  = 0.020,~  a^{(1)}_d= 0.10,~
b^{(1)}_u = 0.022,~b^{(1)}_d=0.38,~
a^{(2)}_u=  1.033,~ a^{(2)}_d=  1.087,~
b^{(2)}_u= -0.15, ~b^{(2)}_d= -0.20,
N^{(1)}_u = 2.055,~ N^{(1)}_d = 1.7618,
N^{(2)}_u= 1.322, N^{(2)}_d = -2.4827$.
We estimate a $5\%$ uncertainty
in the model parameters. The model motivated by
soft-wall AdS/QCD has been extensively employed to study and successfully reproduce many interesting properties of the proton
~\cite{Gutsche:2013zia,Chakrabarti:2016yuw,Chakrabarti:2015ama,Chakrabarti:2015lba,Mondal:2015uha,Gutsche:2016gcd,Mondal:2017wbf,Mondal:2016xsm,Maji:2015vsa,Chakrabarti:2020kdc,Choudhary:2022den}.

\section{TMDs}	\label{tmd}
The TMDs of a quark inside the proton are defined through the
quark-quark correlator function defined as~\cite{Goeke:2005hb}
\begin{eqnarray} \label{correlator}
	\Phi^{q[\Gamma]}\left(x, \mathbf{p}_{\perp} ; S\right)=\frac{1}{4} \int \frac{d z^{-}}{(2 \pi)} \frac{d^{2} z_{\perp}}{(2 \pi)^{2}} e^{i p . z}\left\langle P ; S\left|\bar{\psi}^{q}(0) \Gamma \mathcal{W}_{[0, z]} \psi^{q}(z)\right| P ; S\right\rangle \big\vert_{z^+=0}\,,
\end{eqnarray}
where flavor and color indexes and summations are implicit. Here, $\psi$ represents the quark field, and $\Gamma$ denotes the Dirac matrix which in the leading twist, is taken as $ \Gamma=\{\gamma^+\,,\gamma^+\gamma^5\,,i\sigma^{j+}\gamma^5 \}$ corresponding to unpolarized, longitudinally polarized and transversely polarized quarks, respectively. Here, $p$ is the momentum of the active quark inside the proton of momentum $P$, spin $S$ and $x\equiv p^+/P^+$ is the longitudinal momentum fraction carried by the active quark. The gauge link, $\mathcal{W}_{[0, z]}$ that ensures the $SU(3)$ color gauge invariance of the bilocal quark operator is expressed as
\begin{eqnarray}\label{wilsonline}
	\mathcal{W}_{[0,z]}=\mathcal{P} \exp \left(-i g \int_{\mathbf{z}_{\perp}}^{\infty} \mathrm{d} \eta_{\perp} \cdot \mathbf{A}_{\perp}\left(\eta^{-}=n \cdot \infty, \mathbf{z}_{\perp}\right)\right)\,.
\end{eqnarray}
In the current study, we choose the light cone gauge $A^{+}=0$ and only retain the zeroth-order
expansion of the gauge link, i.e., $\mathcal{W}_{[0,z]}\approx 1$.
 The proton with helicity $\lambda$ has spin components $S^{+}=\lambda\frac{P^{+}}{M}$, $S^{-}=\lambda\frac{P^{-}}{M}$ and $S_{T}$. 
The unpolarized TMD, $f_{1}(x,\mathbf{p}_{\perp}^{2})$ is then given by \cite{Bacchetta:2008af}.
\begin{eqnarray}\label{unpolarizedTMD}
f_{1}(x,\mathbf{p}_{\perp}^{2})=\frac{1}{2}Tr\left(\Phi^{[\gamma^{+}]}\right)\,,
\end{eqnarray}
and the T-odd TMDs at the leading twist, the Sivers function $f_{1 T}^{\perp q}(x,\mathbf{p}_{\perp}^{2})$ and the Boer-Mulders function $h_{1}^{\perp q}(x,\mathbf{p}_{\perp}^{2})$ are parameterized as \cite{Bacchetta:2008af} 
\begin{align}\label{phiSivers}
	\Phi^{q\left[\gamma^{+}\right]}\left(x, \mathbf{p}_{\perp} ; S\right)&=\ldots-\frac{\epsilon_{T}^{i j} p_{\perp}^{i} S_{T}^{j}}{M} f_{1 T}^{\perp q}\left(x, \mathbf{p}_{\perp}^{2}\right)\,,\\
\label{phiBM}
	\Phi^q{\left[i \sigma^{j+} \gamma^{5}\right]}\left(x, \mathbf{p}_{\perp} ; S\right)&=\ldots+\frac{\epsilon_{T}^{i j} p_{\perp}^{i}}{M} h_{1}^{\perp q}\left(x, \mathbf{p}_{\perp}^{2}\right)\,.
\end{align}
The T-even TMDs are suppressed in the above equations.
The unpolarized TMD $f_{1}(x,\mathbf{p}_{\perp}^{2})$ describes the momentum distribution of unpolarized quarks within an unpolarized proton, whereas the Sivers function $f_{1T}^{\perp q}(x,\mathbf{p}_{\perp}^{2})$ encodes distribution of unpolarized quarks inside a transversely polarized proton, which comes from a correlation of the proton transverse spin and the quark transverse momentum. The Boer-Mulders function $h_{1}^{\perp q}(x,\mathbf{p}_{\perp}^{2})$ describes the spin-orbit correlations of transversely polarized quarks within an unpolarized proton.

Using LFWFs the unpolarized TMD is expressed as
\begin{eqnarray} \label{unpolarizedTMD1}
f_{1}^{q}\left(x, \mathbf{p}_{\perp}^{2}\right)&=&\frac{1}{16 \pi^{3}} \frac{1}{2} \sum_{\lambda_{N},\lambda_{q}}\left|\psi_{\lambda_{q}}^{\lambda_{N}}\right|^{2}=\frac{1}{16 \pi^{3}}\left(\left|\psi_{+}^{+}\right|^{2}+\left|\psi_{-}^{+}\right|^{2}\right)\,,
\end{eqnarray}
where the proton and the active quark helicities remain unchanged in  the overlap of LFWFs. The Sivers function requires the proton helicity to be flipped from the initial to the final state but the quark helicity remains unchanged. On the other hand, the Boer-Mulders function necessitates the quark helicity to be flipped from the initial to the final state, while keeping the proton helicity same in  the overlap of LFWFs. To calculate these T-odd TMDs, we express them as the overlap integrations of LFWFs differing by one unit orbital angular momentum, i.e., $\Delta L_{z}=\pm 1$ \cite{Brodsky:2002cx}. However, to generate a nonzero T-odd TMDs, one also needs to take into account the gauge link. Physically, this is equivalent to consider initial or final state interactions of the active quark with the target remnant, which we refer to collectively as gluon rescattering. We consider that this physics is encoded in a gluon rescattering kernel $G\left(x,\mathbf{q}_{\perp}\right)$ such that~\cite{Lu:2006kt,PhysRevD.78.074010}
\begin{align}\label{sivers}
	\frac{p^{L}}{ M} f_{1 T}^{\perp q}\left(x, \mathbf{p}_{\perp}^{2}\right)&=i \sum_{\lambda_{q},\lambda_{N},\lambda_{N}^{\prime}} \int \frac{{\rm d}^{2} \mathbf{p}_{\perp}^{\prime}}{16 \pi^{3}} \left[\psi_{\lambda_{q}}^{\lambda_{N} \star}\left(x, \mathbf{p}_{\perp}\right)\, G\left(x,\mathbf{q}_{\perp}\right)\, \psi_{\lambda_{q}}^{\lambda_{N}^{\prime}}\left(x, \mathbf{p}_{\perp}^{\prime} \right)\right]\,,\\
 \label{BMTMD}
	\frac{p^{L}}{ M} h_{1}^{\perp q}\left(x, \mathbf{p}_{\perp}^{2}\right)&=i \sum_{\lambda_{q},\lambda_{q}^{\prime},\lambda_{N}} \int \frac{{\rm d}^{2} \mathbf{p}_{\perp}^{\prime}}{16 \pi^{3}} \left[\psi_{\lambda_{q}}^{\lambda_{N} \star}\left(x, \mathbf{p}_{\perp}\right)\, G\left(x,\mathbf{q}_{\perp}\right)\, \psi_{\lambda_{q}^{\prime}}^{\lambda_{N}}\left(x, \mathbf{p}_{\perp}^{\prime} \right)\right]\,,
\end{align}
where $\mathbf{q}_{\perp}=\mathbf{p}_{\perp} - \mathbf{p}_{\perp}^{\prime}$ and $p^{L}=p^{1}-i p^{2}$. To proceed further we must specify the form of the gluon rescattering kernel $G\left(x,\mathbf{q}_{\perp}\right)$. Alternatively to incorporate the effects of the final-state interaction, the LFWFs can be modified to have a phase factor, which is essential to obtain Sivers or Boer-Mulders functions~\cite{Brodsky:2006ha,Ji:2002xn,Gurjar:2021dyv,Maji:2017wwd}. In this work, we explicitly employ a nonperturbative gluon rescattering kernel $G\left(x,\mathbf{q}_{\perp}\right)$~\cite{Ahmady:2019yvo} to produce nonzero T-odd TMDs.

\section{The gluon rescattering kernel and the lensing function}
\label{lensing}
 The simplest form for the one-gluon exchange approximation
of the gauge-link is to assume that \cite{Brodsky:2002rv,Brodsky:2002cx,Ahmady:2019yvo}
\begin{equation}
 	\Im \mathrm{m} G^{\mathrm{pert.}}(x, q_\perp) \propto \frac{C_F\alpha_s}{q^2_\perp} \;,
 	\label{pert-G}
\end{equation}
with $\alpha_s$ being the coupling constant and $C_F$ is the color factor. Eq.~\eqref{pert-G} is referred to as the perturbative Abelian gluon rescattering kernel, which can be derived by working with perturbative Abelian gluons. By hypothesis, the coupling is weak, i.e. $\alpha_s \ll 1$, though different values of $\alpha_s$ have been used in Eq.~\eqref{pert-G} in the literature. For example, while Ref.~\cite{Lu:2004hu} employes $\alpha_s=0.3$,  other authors prefer to consider much larger values of the coupling constant, e.g., $\alpha_s=0.911$ is used in Ref. \cite{Wang:2017onm}, and $\alpha_s=1.2$ in Ref. \cite{Pasquini:2014ppa}. However, using such large values of $\alpha_s$ disputes the weak coupling hypothesis that leads to Eq. \eqref{pert-G}. Though the perturbative kernel with $\alpha_s \sim 1$ may be considered  in some extent as a \enquote{phenomenological model} which reproduces some data, it indicates the necessity of higher order corrections or nonperturbative kernel.  The main reason for the discrepancy is that
 the perturbative kernel cannot capture precisely the dynamics of soft gluons, which are mainly responsible for producing a nonperturbative quantity like the Sivers and Boer-Mulders TMDs. An exact structure of the nonperturbative gluon rescattering kernel is yet not available
 and, in practice, some approximation procedure is needed.
 Meantime, the gluon rescattering kernel can be expressed in term of the so-called QCD lensing function $I(x, q_\perp)$ as~\cite{Ahmady:2019yvo}
\begin{equation}
i G(x, q_\perp)= -\frac{2}{(2\pi)^2} \frac{(1-x) I(x, q_\perp)}{q_\perp} \;,
\label{Relation-GI}
\end{equation}
which has been derived from the relation between the first moment of the Boer-Mulders function with the chiral-odd GPD.
In Ref.~\cite{Gamberg:2009uk}, the QCD lensing function, $I(x, q_\perp)$, has been obtained from the eikonal amplitude  for final state rescattering via the exchange of soft $U(1)$, $SU(2)$ and $SU(3)$ gluons.

The lensing function in the impact parameter space is given by~\cite{Gamberg:2009uk}
\begin{eqnarray}\label{Ixb}
\mathcal{I}^{i}(x,b_{\perp})=\frac{(1-x)}{2N_{c}}\frac{\mathbf{b_{\perp}^{i}}}{b_{\perp}}\frac{\chi^{\prime}}{4}C\left(\frac{\chi}{4}\right)\,,
\end{eqnarray}
where the colour function $C(\frac{\chi}{4})$ reads
\begin{eqnarray}\label{Cchi}	
C\left(\frac{\chi}{4}\right)=Tr \left\{\Im \mathrm{m} f^{\prime}\left(\frac{\chi}{4}\right)+\frac{1}{2}\left[\Im \mathrm{m} f^{\prime}\left(\frac{\chi}{4}\right) \Re \mathrm{e} f\left(\frac{\chi}{4}\right)-\Im \mathrm{m} f\left(\frac{\chi}{4}\right) \Re \mathrm{e} f^{\prime}\left(\frac{\chi}{4}\right)\right]\right\}\,,
\end{eqnarray}
with
\begin{eqnarray}\label{chi}
\chi\left(\frac{b_{\perp}}{(1-x)}\right)=\frac{g^{2}}{2 \pi} \int \mathrm{d} p_{\perp} p_{\perp} J_{0}\left(\frac{b_{\perp}p_{\perp}}{(1-x)} \right) \mathcal{D}_{1}\left(-p_{\perp}^{2}\right)\,,
\end{eqnarray}
being the eikonal phase and $\mathcal{D}_{1}(-p_{\perp}^{2})$ is the gauge independent part of the gluon propagator. The real and imaginary parts of $f(\chi/4)$ in Eq.~\eqref{Cchi} emerge from the real and imaginary parts of the eikonal amplitude for final state rescattering via the exchange of generalized infinite ladders of gluons. 
For SU(3) gluons,
\begin{equation}
	\Re \mathrm{e} [f^{\mathrm{SU(3)}}_{\alpha \beta} ](a)= \delta_{\alpha \beta} (-c_2 a^2 + c_4 a^4 -c_6a^6-c_8 a^8 + ...)
\label{fSU3-Re}
\end{equation}
and
\begin{equation}
	\Im \mathrm{m} [f^\mathrm{SU(3)}_{\alpha \beta} ](a)= \delta_{\alpha \beta} (c_1 a - c_3 a^3 + c_5a^5-c_7 a^7 + ...)
\label{fSU3-Im}
\end{equation}
where $a \equiv\chi/4$ and $c_i$ are numerical coefficients given in Ref.~\cite{Gamberg:2009uk}. We compute the eikonal phase $\chi$ in Eq.~\eqref{chi} using a nonperturbative Dyson-Schwinger gluon propagator given by 
\begin{equation}
	\mathcal{D}_1 (k_\perp^2, \Lambda_{\mathrm{QCD}}^2)=\frac{1}{k_\perp^2}\left(\frac{\alpha_s(k_\perp^2)}{\alpha_s(\Lambda_{\mathrm{QCD}}^2)}\right)^{1+2\delta}\left( \frac{c(k^2_\perp/\Lambda^2)^\kappa+d(k^2_\perp/\Lambda^2)^{2\kappa}}{1+c(k^2_\perp/\Lambda^2)^\kappa+d(k^2_\perp/\Lambda^2)^{2\kappa}}\right)^2 \,,
\label{DS-gluon}
\end{equation}
with
\begin{equation}
	\alpha_s(\mu^2)=\frac{\alpha_s(0)}{\ln [e + a_1(\mu^2/\Lambda^2)^{a_2} + b_1(\mu^2/\Lambda^2)^{b_2}]}\,,
\end{equation}
where all parameters are taken from Refs.~\cite{Fischer:2003rp,Gamberg:2009uk}.
The lensing function in momentum space is then obtained by the inverse Fourier transform of Eq.~\eqref{Ixb}
\begin{eqnarray}\label{Ixq}
I\left(x, q_{\perp}\right) \frac{\mathbf{q}^{i}}{q_{\perp}}&=&-\frac{i}{(1-x)^{3}} \int \mathrm{d}^{2} \mathbf{b} \exp \left(-i \frac{\mathbf{q} \cdot \mathbf{b}}{(1-x)}\right) \mathcal{I}\left(x, b_{\perp}\right) \frac{\mathbf{b}^{i}}{b_{\perp}}\,.
\end{eqnarray}
This leads to the nonperturbative SU$(3)$ gluon
rescattering kernel followed by Eq.~\eqref{Relation-GI}.  Thus, going beyond the usual approximation of perturbative U$(1)$ gluons, we compute the nonperturbative T-odd TMDs using the nonperturbative SU$(3)$ gluon rescattering kernel.

 The perturbative lensing function for the Abelian gluons is given by~\cite{Ahmady:2019yvo}
\begin{align}
I_{U(1)}^{\rm pert}\left(x,q_{\rm \perp}\right)=-\frac{g^{2}}{2}\frac{1}{(1-x)q_{\perp}}\,,
\end{align}
which leads to 
\begin{eqnarray}
iG_{U(1)}^{\rm pert}(q_\perp)=\frac{g^{2}}{4\pi^{2}}\frac{1}{q_{\perp}^{2}}\,,
\end{eqnarray}
following Eq.~(\ref{Relation-GI}). After replacing $g^{2}\rightarrow4\pi C_{F}\alpha_{s}$, one obtains the perturbative gluon rescattering kernel
\begin{eqnarray}\label{pertG}
iG^{\rm pert}(q_{\perp})=\frac{\alpha_{s}C_{F}}{\pi q_{\perp}^{2}}\,.
\end{eqnarray} 

\section{Results}\label{results}
Using the overlap representation of LFWFs, Eqs.~(\ref{unpolarizedTMD1}), (\ref{sivers}), and (\ref{BMTMD}), we evaluate the TMDs in the light-front quark-diquark model.  With the LFWFs given in Eq.~(\ref{WF}), the explicit expression for the unpolarized TMD reads
\begin{align}
 \label{unpoltmd}
f_{1}^{q}\left(x,\mathbf{p}_{\perp}^{2} \right)=\frac{\log (1 / x)}{\pi \kappa^{2}} \exp \left[-\frac{\mathbf{p}_{\perp}^{2} \log (1 / x)}{\kappa^{2}(1-x)^{2}}\right]\left(F_{1}(x)+\frac{\mathbf{p}_{\perp}^{2}}{M^{2}} F_{2}(x)\right)\,,
\end{align}
with
\begin{align}\label{F12x}
F_{1}(x)&=\left|N_{q}^{(1)}\right|^{2} x^{2 a_{q}^{(1)}}(1-x)^{2 b_{q}^{(1)}-1} \,,\nonumber  \\
F_{2}(x)&=\left|N_{q}^{(2)}\right|^{2} x^{2 a_{q}^{(2)}-2}(1-x)^{2 b_{q}^{(2)}-1}\,.
\end{align}
We obtain the expression for the Sivers and Boer-Mulders functions as
\begin{align}\label{SBM}
f_{1T}^{\perp q}\left(x,\mathbf{p}_{\perp}^2\right)\equiv h_{1}^{\perp q\,(\rm pert)}(x,\mathbf{p}_{\perp}^{2})&=-2F_{3}^{q}(x) \frac{\log(1/x)}{\pi \kappa^{2}} \exp\left[-\frac{\mathbf{p}_{\perp}^{2}\log[1/x]}{\kappa^{2}(1-x)^{2}}\right]\times\int d^{2}\mathbf{q}_{\perp} \left(\frac{q_{\perp} \cos(\theta_{q_\perp}-\theta_{p_\perp})}{p_{\perp}} \right)  \nonumber\\ &\times iG\left(x,\mathbf{q}_{\perp}\right)\, \exp\left[-\frac{\mathbf{q}_{\perp}^{2}\log[1/x]}{2\kappa^{2}(1-x)^{2}}\right]\exp\left(\frac{p_{\perp}q_{\perp}\log[1/x]}{\kappa^{2}(1-x)^{2}}\cos(\theta_{q_\perp}-\theta_{p_\perp})\right)\,,
\end{align}
where 
\begin{eqnarray}\label{F3}
F_{3}^{q}(x)&=&N_{q}^{(1)}N_{q}^{(2)}x^{a_{q}^{(1)}+a_{q}^{(2)}-1}(1-x)^{b_{q}^{(1)}+b_{q}^{(2)}-1}\,.
\end{eqnarray}
Using the perturbative gluon rescattering kernel, Eq.~(\ref{pertG}), we get
\begin{equation}\label{siverstmd}
f_{1 T}^{\perp q\,(\rm pert)}(x,\mathbf{p}_{\perp}^{2})\equiv h_{1}^{\perp q\,(\rm pert)}(x,\mathbf{p}_{\perp}^{2})=\frac{4}{\pi}(C_{F}\alpha_{s})(1-x)^{2} F_{3}^{q}(x)\frac{\exp \left[-\frac{\mathbf{p}_{\perp}^{2}\log[1/x]}{\kappa^{2}(1-x)^{2}}\right]}{p_{\perp}^{2}} \left(1-\exp\left[\frac{\mathbf{p}_{\perp}^{2}\log[1/x]}{2\kappa^{2}(1-x)^{2}}\right]\right)\,.
\end{equation}
Note that in our scalar quark-diquark model, the relations between the Sivers and Boer-Mulders TMDs for both the up and down quarks have the same sign. This is a special property of the quark-scalar diquark model as reported in Refs.~\cite{Lyubovitskij:2022vcl,Bacchetta:2008af,Hwang:2010dd,Boer:2002ju}. Including the axial-vector diquark, one perhaps distinguishes
the relations for the up and the down quarks~\cite{Maji:2017wwd,Bacchetta:2008af,Ellis:2008in}.

\begin{figure}
\centering
		\includegraphics[scale=0.55]{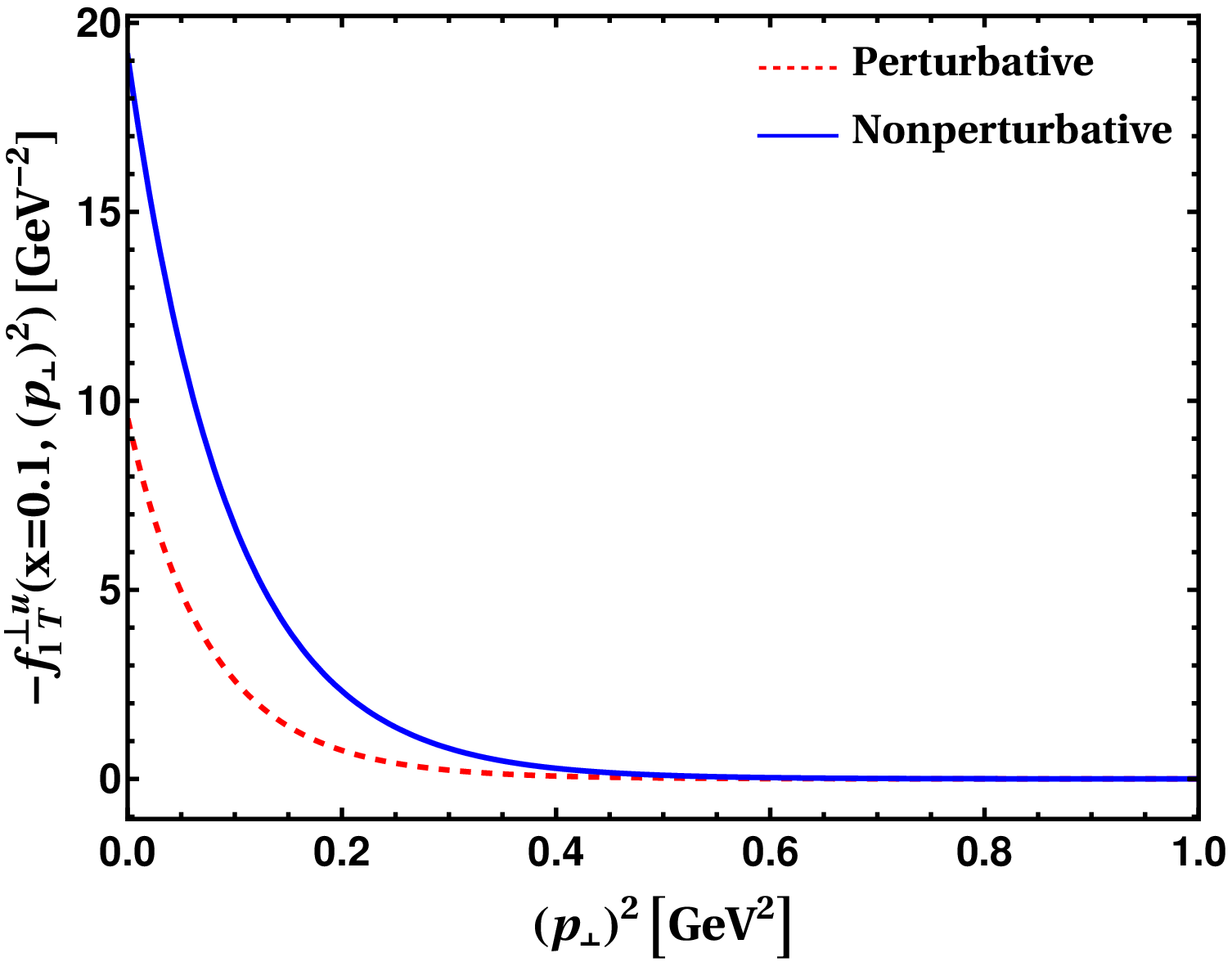}
		\vspace{0.5cm}
		\includegraphics[scale=0.55]{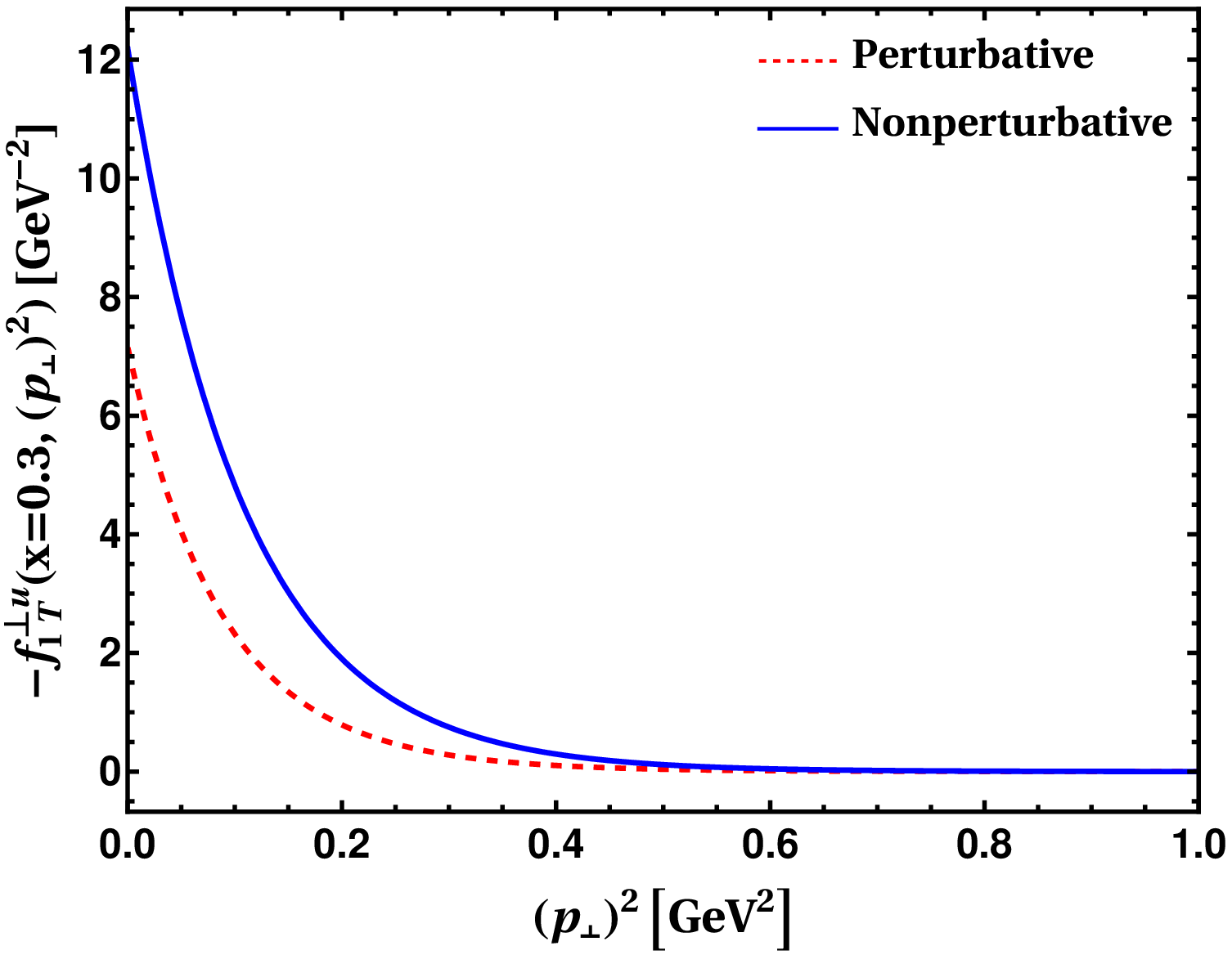}
		\vspace{0.5cm}
		\includegraphics[scale=0.55]{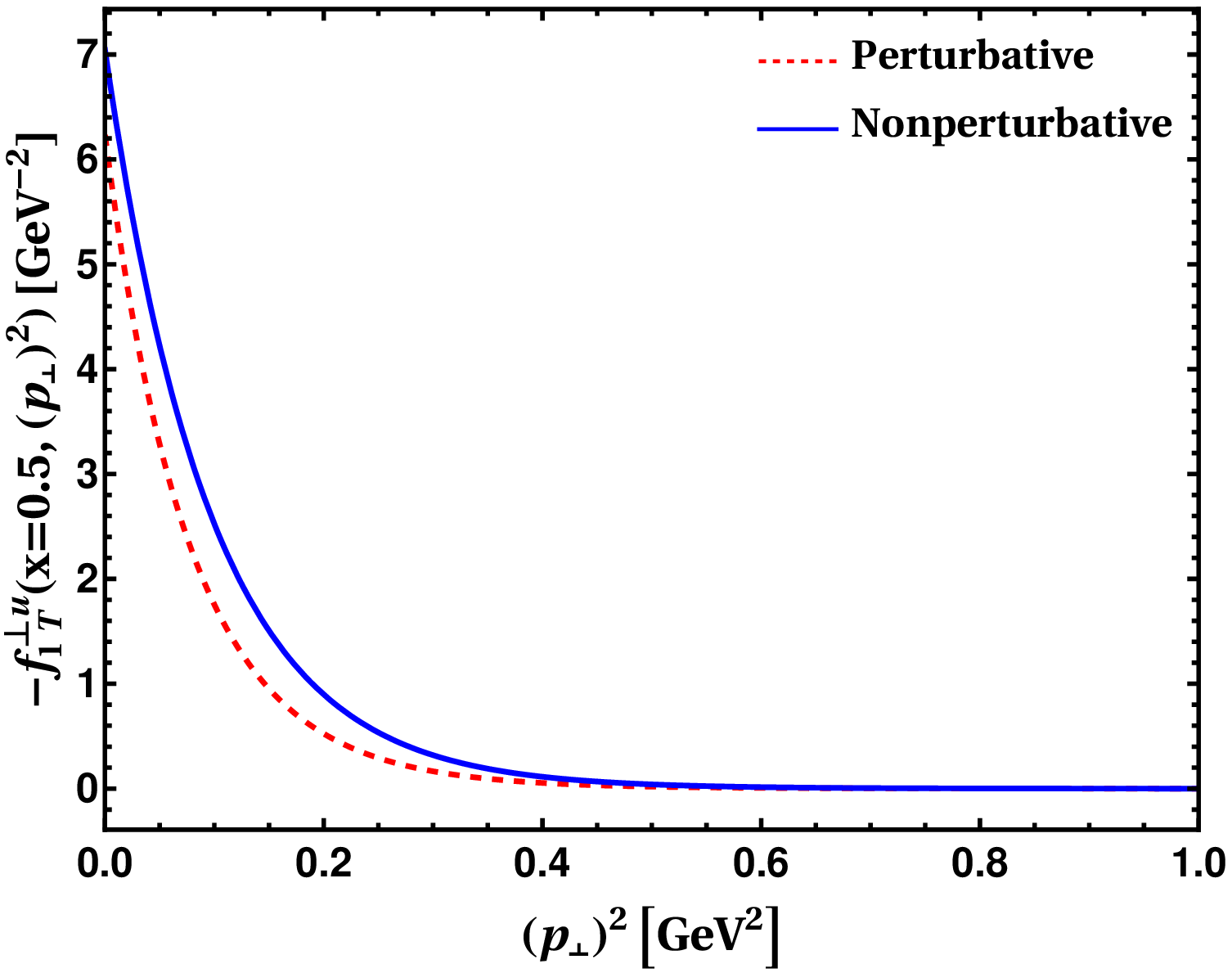}
		\includegraphics[scale=0.55]{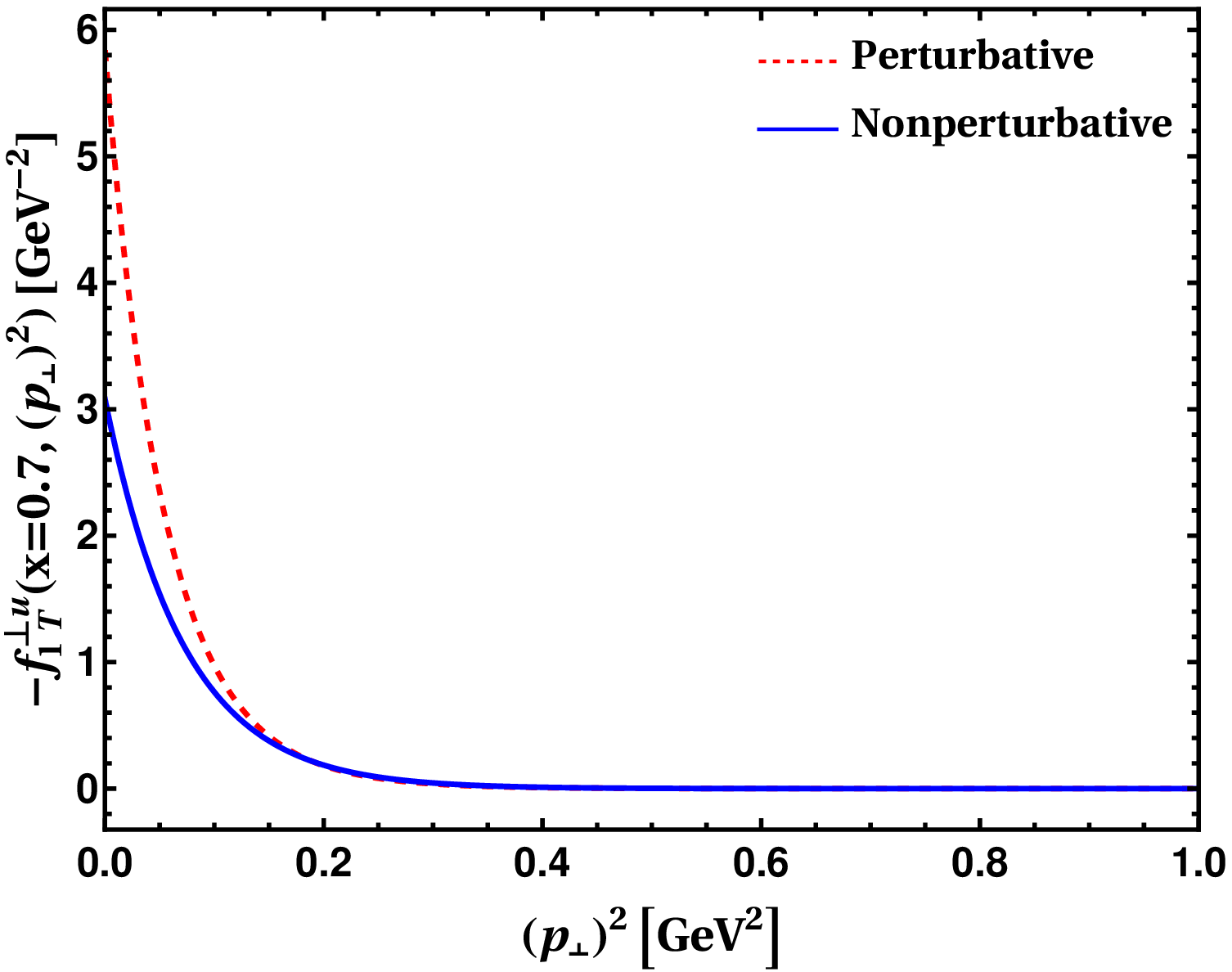}
		\caption{Comparison of the up quark Sivers function generated by the perturbative (dashed red curves) and nonperturbative (solid blue curves) gluon rescattering kernels at different values of $x$. Upper left: $x=0.1$; upper right: $x=0.3$; lower left: $x=0.5$; and lower right: $x=0.7$. The perturbative results are computed with $\alpha_{s}=0.3$.}
		\label{Figure1}
\end{figure}

\begin{figure}
\centering
		\includegraphics[scale=0.55]{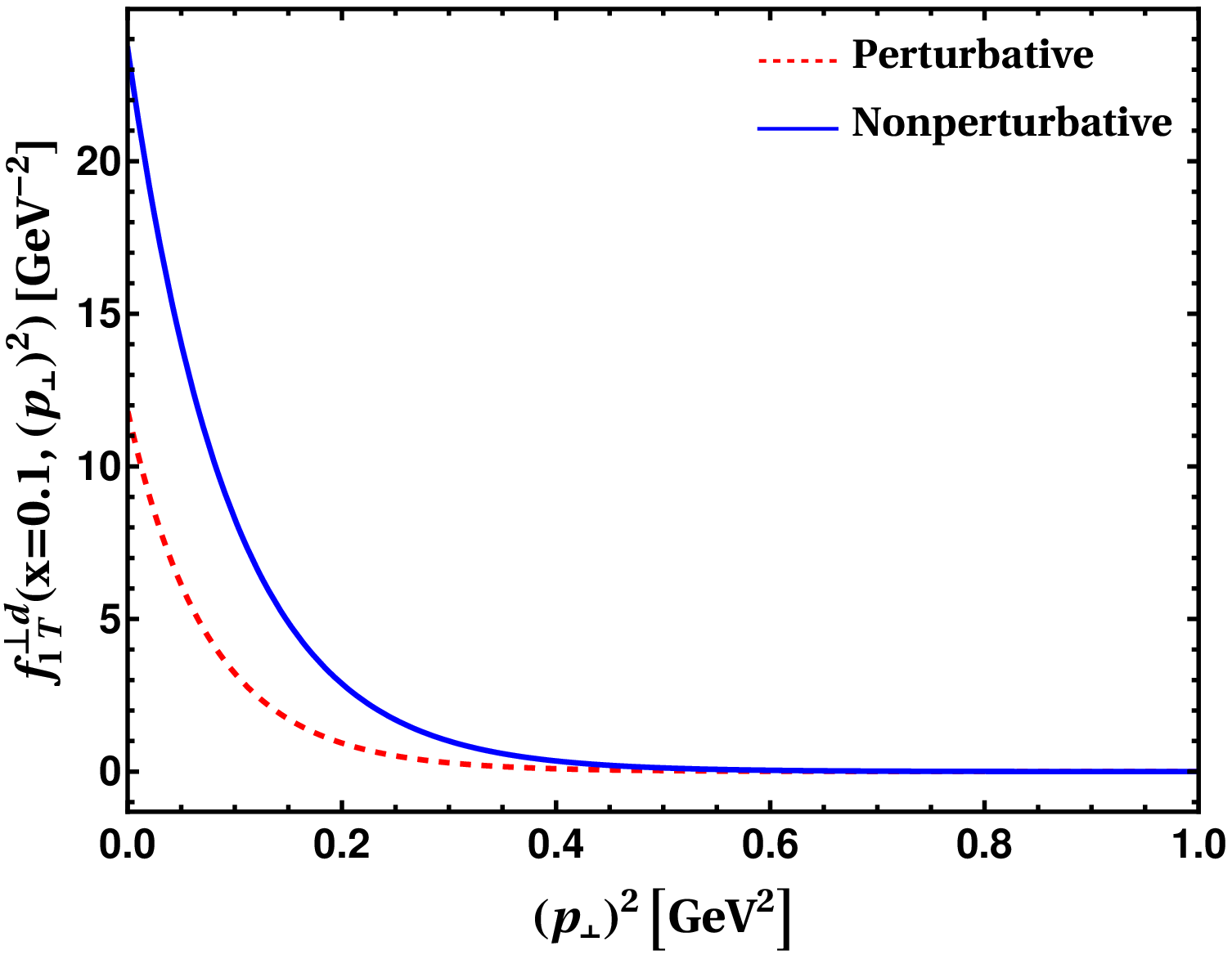}
		\vspace{0.5cm}
		\includegraphics[scale=0.55]{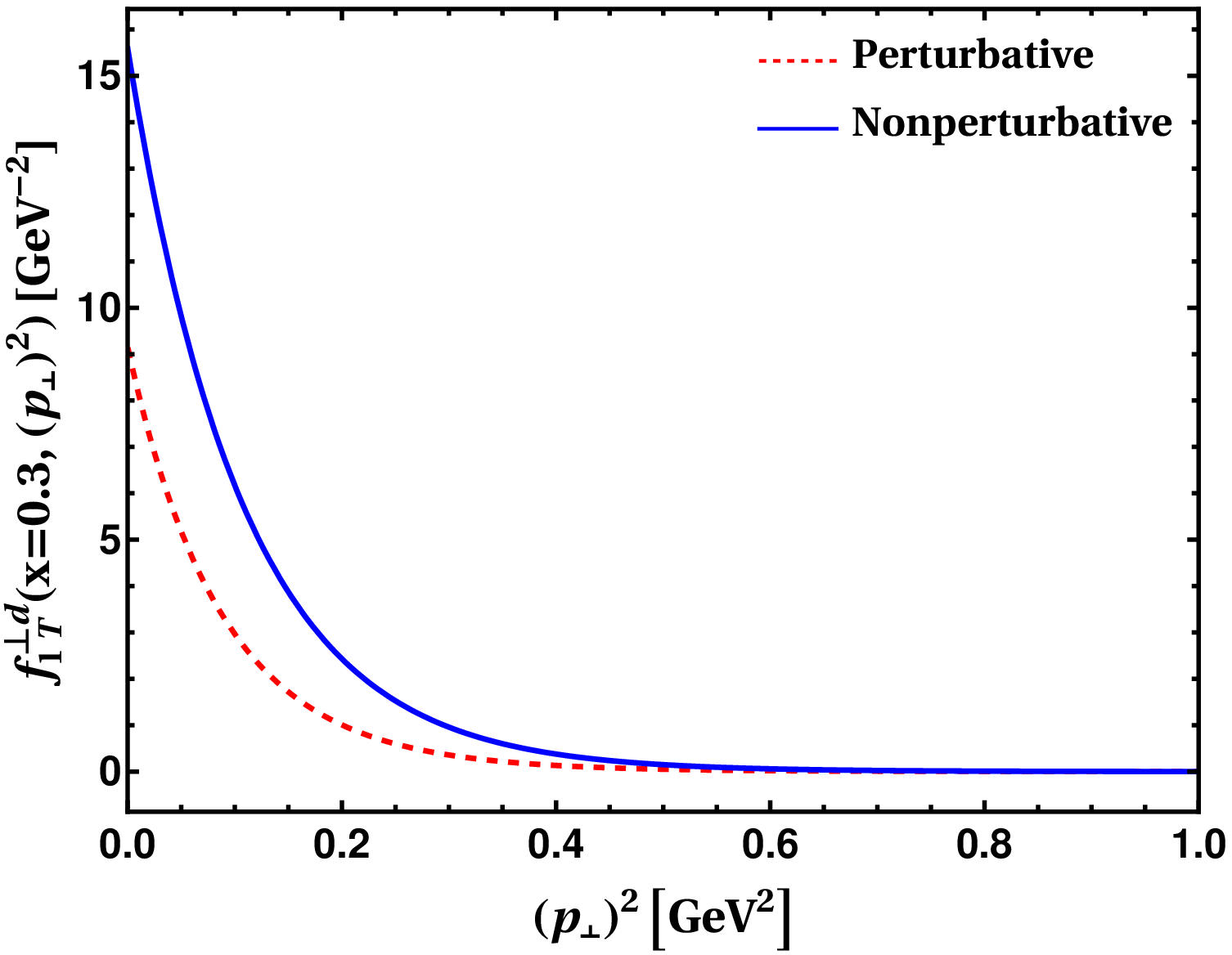}
		\vspace{0.5cm}
		\includegraphics[scale=0.55]{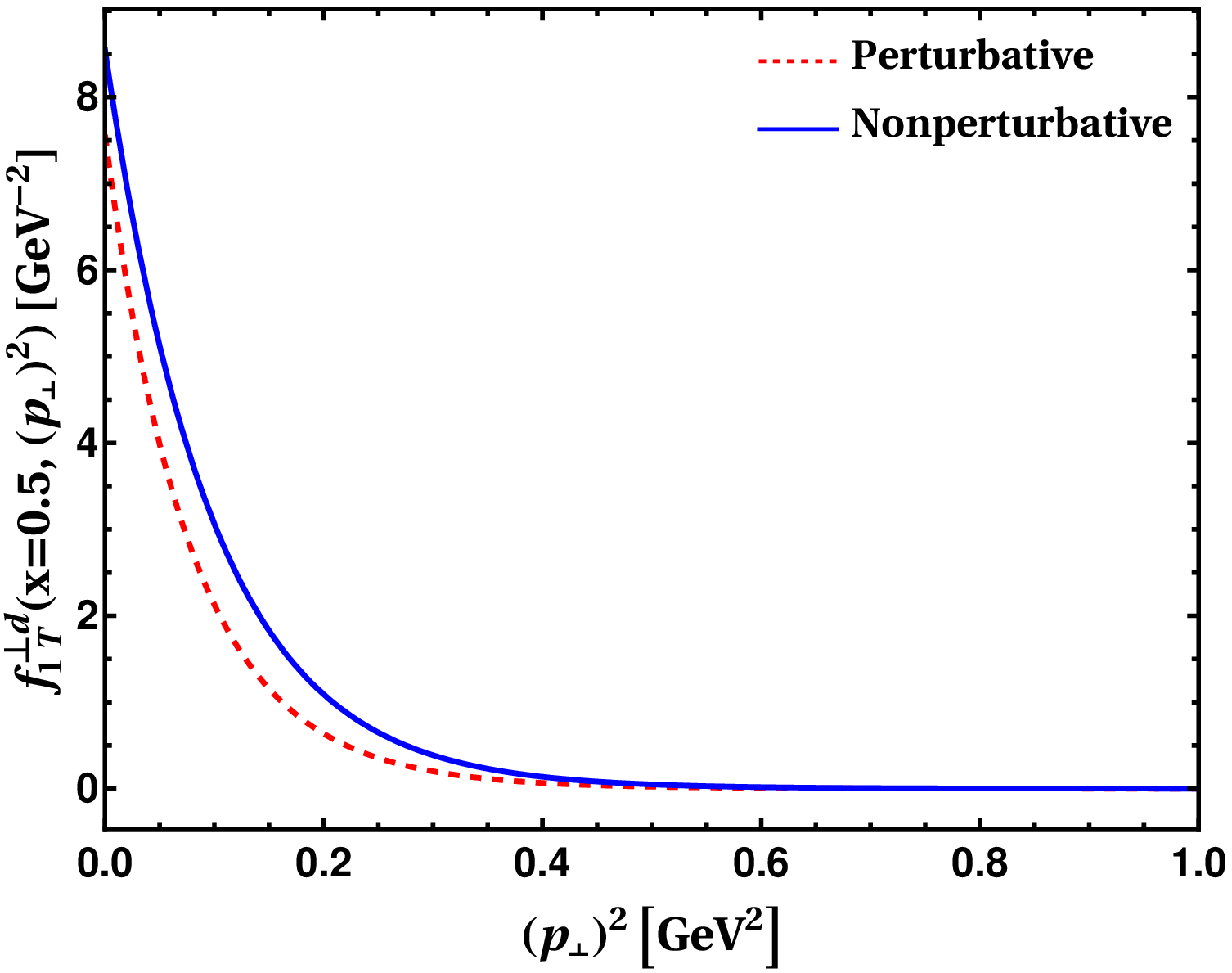}
		\includegraphics[scale=0.55]{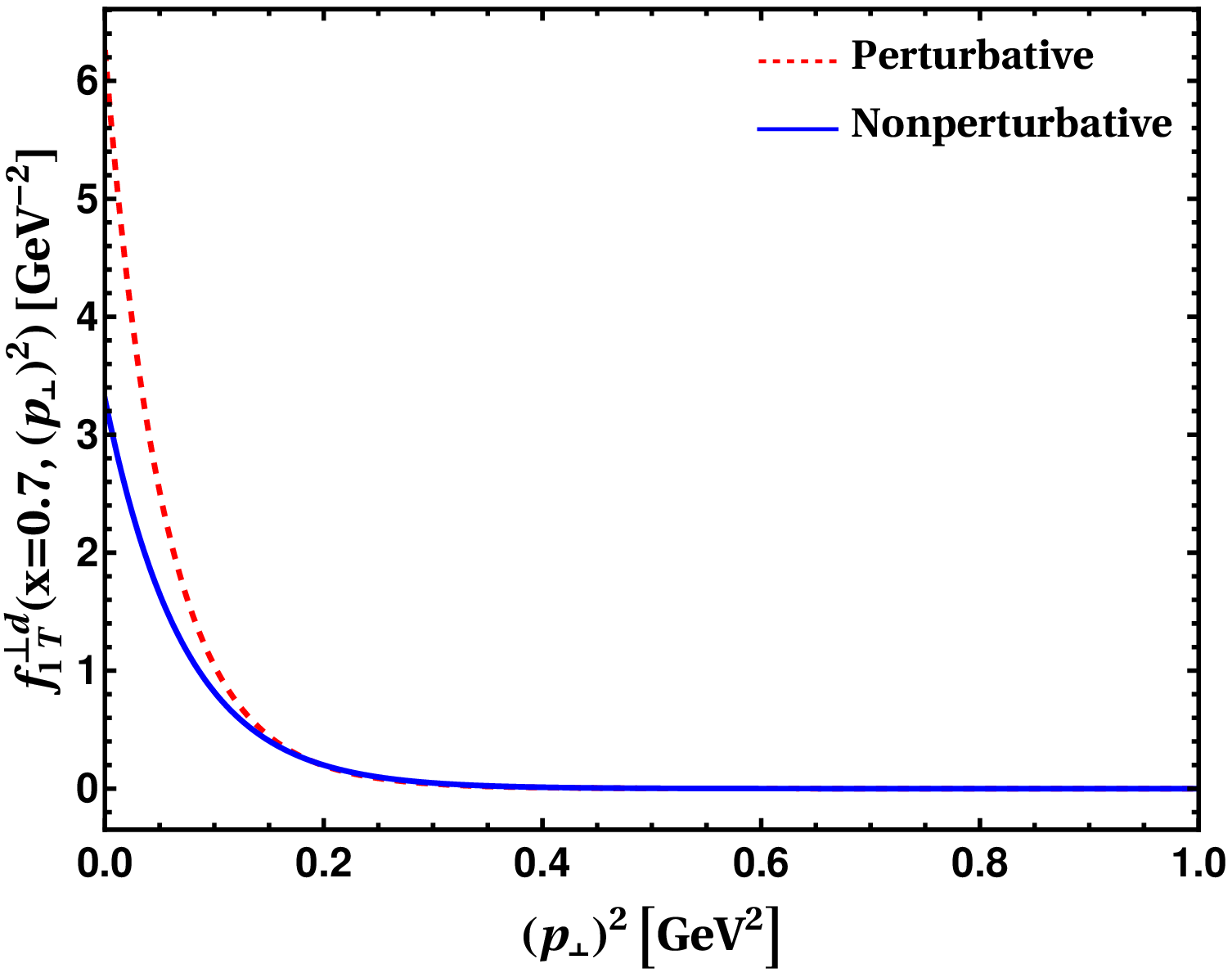}
		\caption{Comparison of the down quark Sivers function generated by the perturbative (dashed red curves) and nonperturbative (solid blue curves) gluon rescattering kernels at different values of $x$. Upper left: $x=0.1$; upper right: $x=0.3$; lower left: $x=0.5$; and lower right: $x=0.7$. The perturbative results are computed with $\alpha_{s}=0.3$.}
		\label{Figure2}
\end{figure}

Figure~\ref{Figure1} illustrates the difference between the Sivers functions generated by the perturbative and nonperturbative gluon rescattering kernels for the up quark, whereas the same for the down quarks are shown in Fig.~\ref{Figure2}.
We notice that a straightforward rescaling of the normalization of the perturbative gluon kernel, say by increasing the coupling $\alpha_s$, cannot fully capture the nonperturbative effects. This is due to the fact that the difference between the two TMDs (perturbative and nonperturbative) is $x$-dependent. At low $x$, the size of the nonperturbatively generated functions are larger than that of the perturbatively generated TMDs while the opposite is true at large $x$.
\begin{figure}
\centering
\includegraphics[scale=0.55]{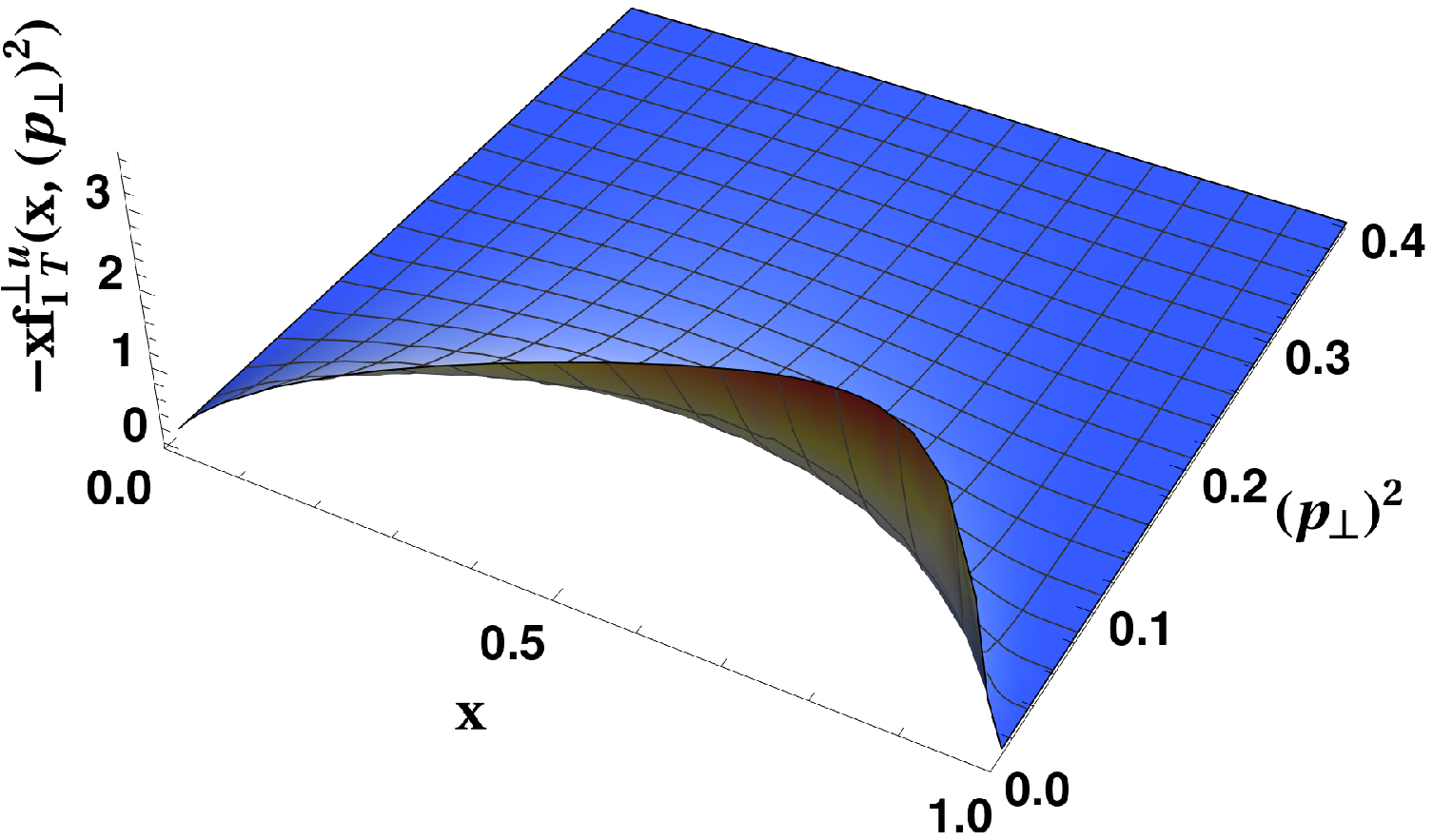}
\hspace{0.5cm}
\includegraphics[scale=0.55]{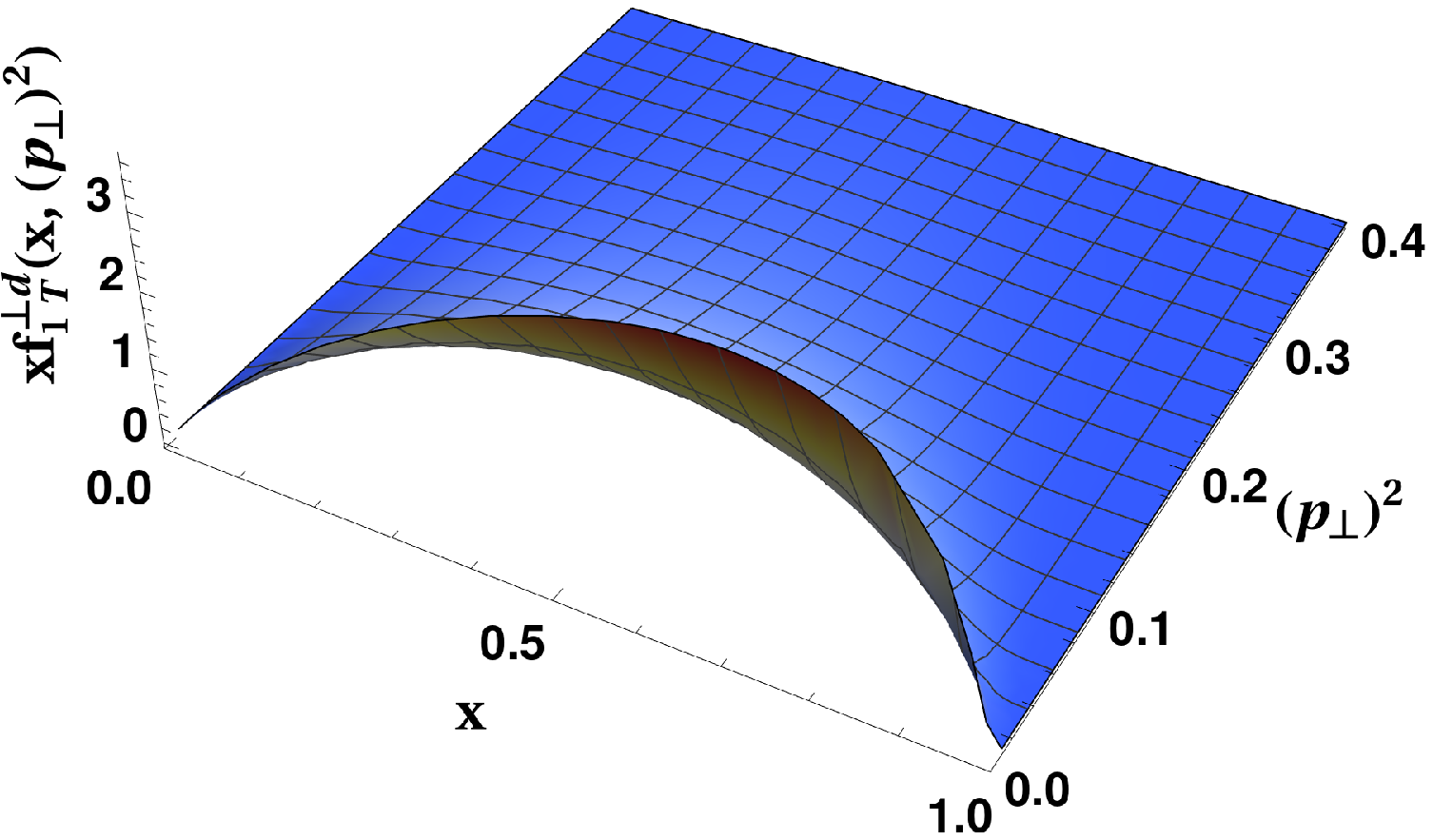}
\caption{Three-dimensional plots of the Sivers TMD $xf_{1T}^{\perp q}(x,\bfp ^{2})$ as a function of $x$ and $\bfp^{2}$ generated by the perturbative kernel with $\alpha=0.3$ Left panel is for the up quark and right panel is for the down quark. The numbers on the vertical axes are in units of $\rm{GeV^{-2}}$ and $\bfp^2$ is in units of GeV$^2$. }
\label{Figure3}
\end{figure}

In Fig.~\ref{Figure3}, we show the three-dimensional structure of the Sivers TMDs computed by the perturbative gluon rescattering kernel with the coupling constant $\alpha=0.3$. 
We note that the overall features of our Sivers functions are similar to those of other theoretical calculations in Refs.~\cite{Lyubovitskij:2022vcl,Bacchetta:2008af,Hwang:2010dd,Boer:2002ju}. 

We further obtain the $f_{1T}^{\perp(1)q}$ and $f_{1T}^{\perp(1/2)q}$ moments of the perturbatively evaluated Sivers function in our model: 
\begin{equation}\label{firstMoment} 
f_{1T}^{\perp(1)q}(x)=\int d^{2}\mathbf{p_{\perp}}\frac{\mathbf{p^2_{\perp}}}{2M^{2}}f_{1T}^{\perp q}(x,\mathbf{p_{\perp}})\\ 
=-2 \alpha_{S}C_{F}\frac{\kappa^{2}}{M^{2}}\frac{(1-x)^{4}F_{3}^{q}(x)}{\log[1/x]}\,,
\end{equation}
\begin{equation}\label{halfMoment} 
f_{1T}^{\perp(1/2)q}(x)=\int d^{2}\mathbf{p_{\perp}}\frac{\mathbf{p_{\perp}}}{M}f_{1T}^{\perp q}(x,\mathbf{p_{\perp}})
\\
=-4(-1+\sqrt{2})\sqrt{\pi} \alpha_{S}C_{F}\frac{\kappa}{M}\frac{(1-x)^{3}F_{3}^{q}(x)}{\sqrt{\log[1/x]}}\,.
\end{equation}

\begin{figure}
\centering
		\includegraphics[scale=0.55]{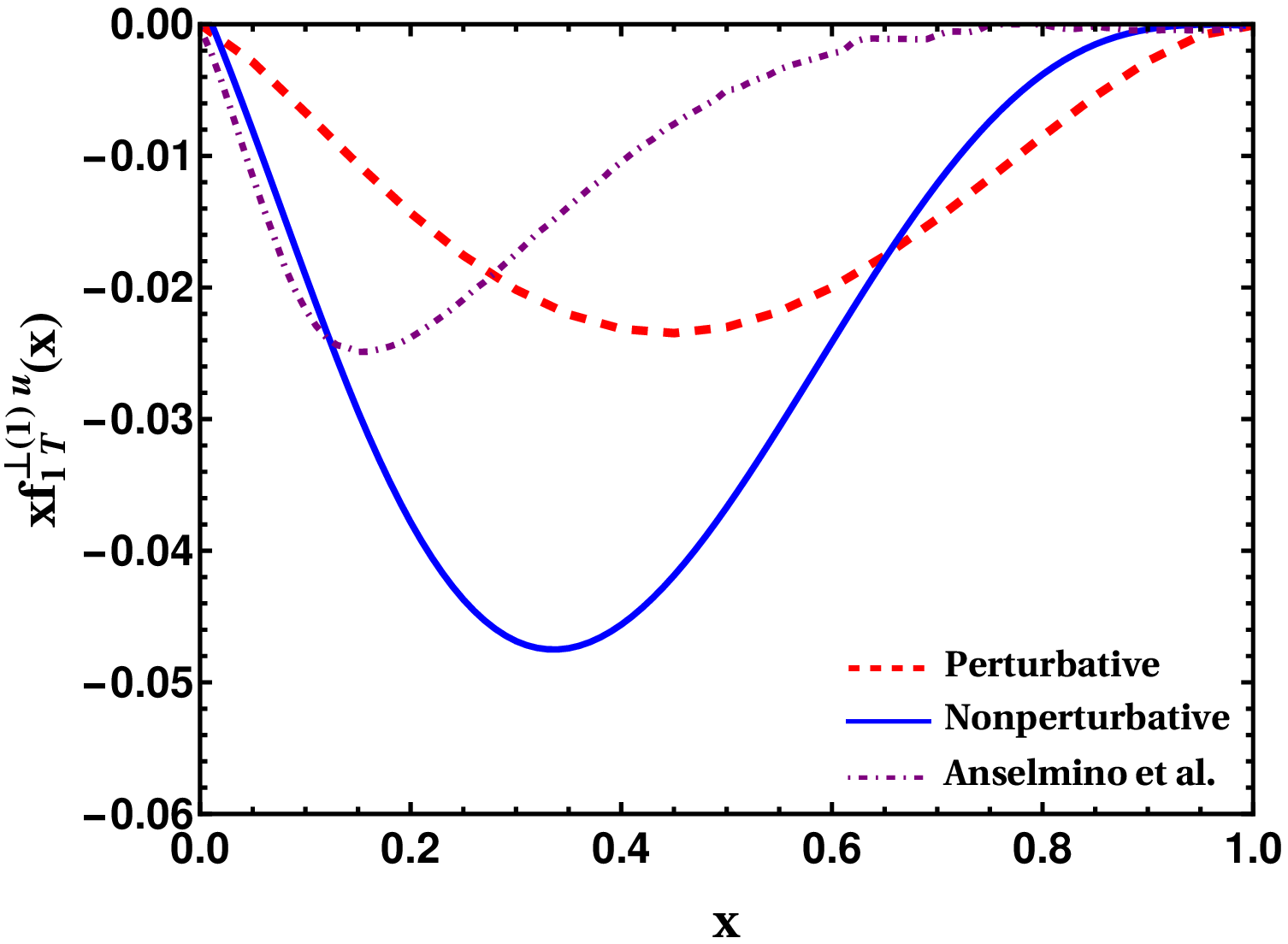}
		\vspace{0.5cm}
		\includegraphics[scale=0.55]{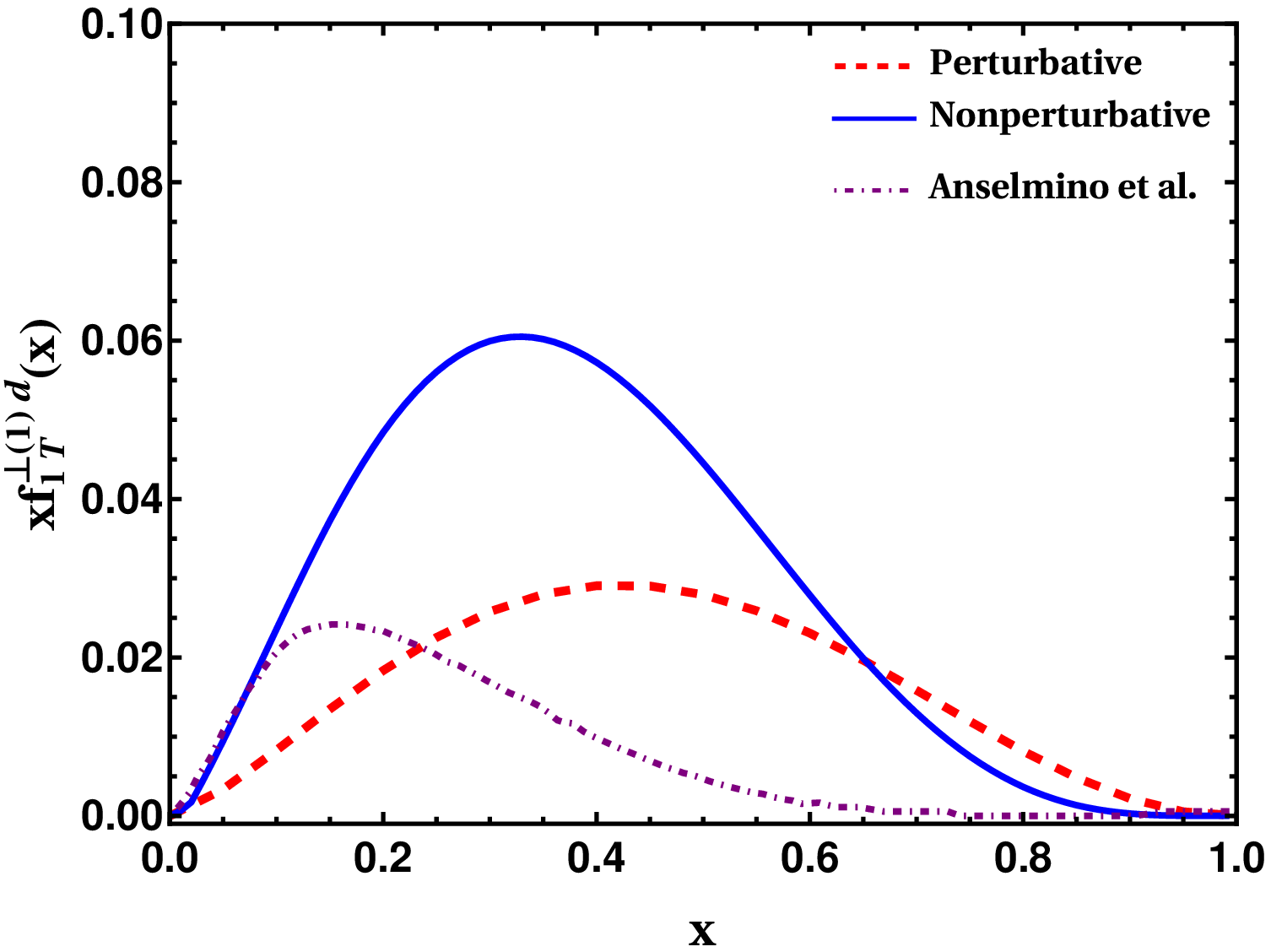}
		\vspace{0.5cm}
		\includegraphics[scale=0.55]{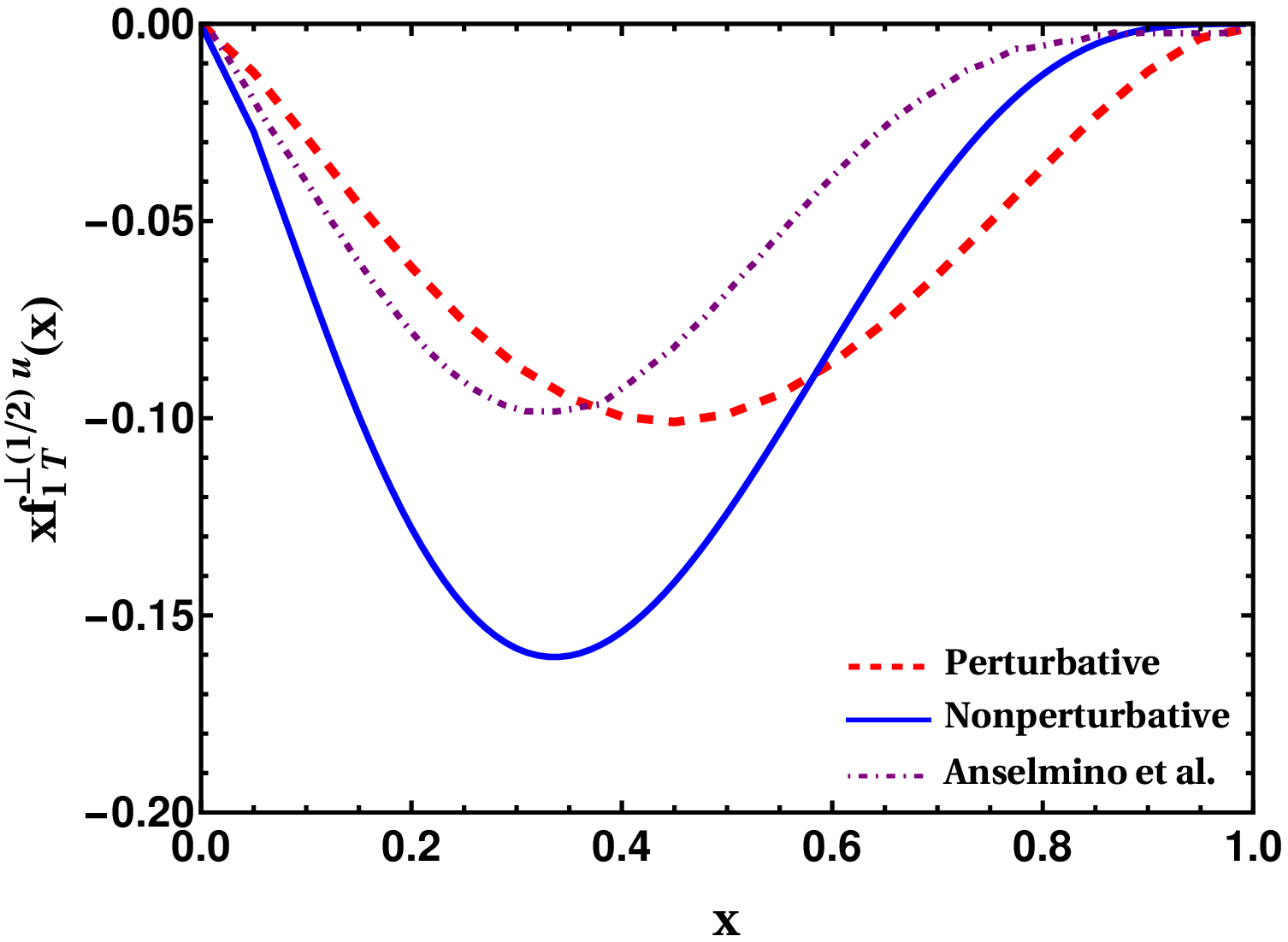}
		\includegraphics[scale=0.55]{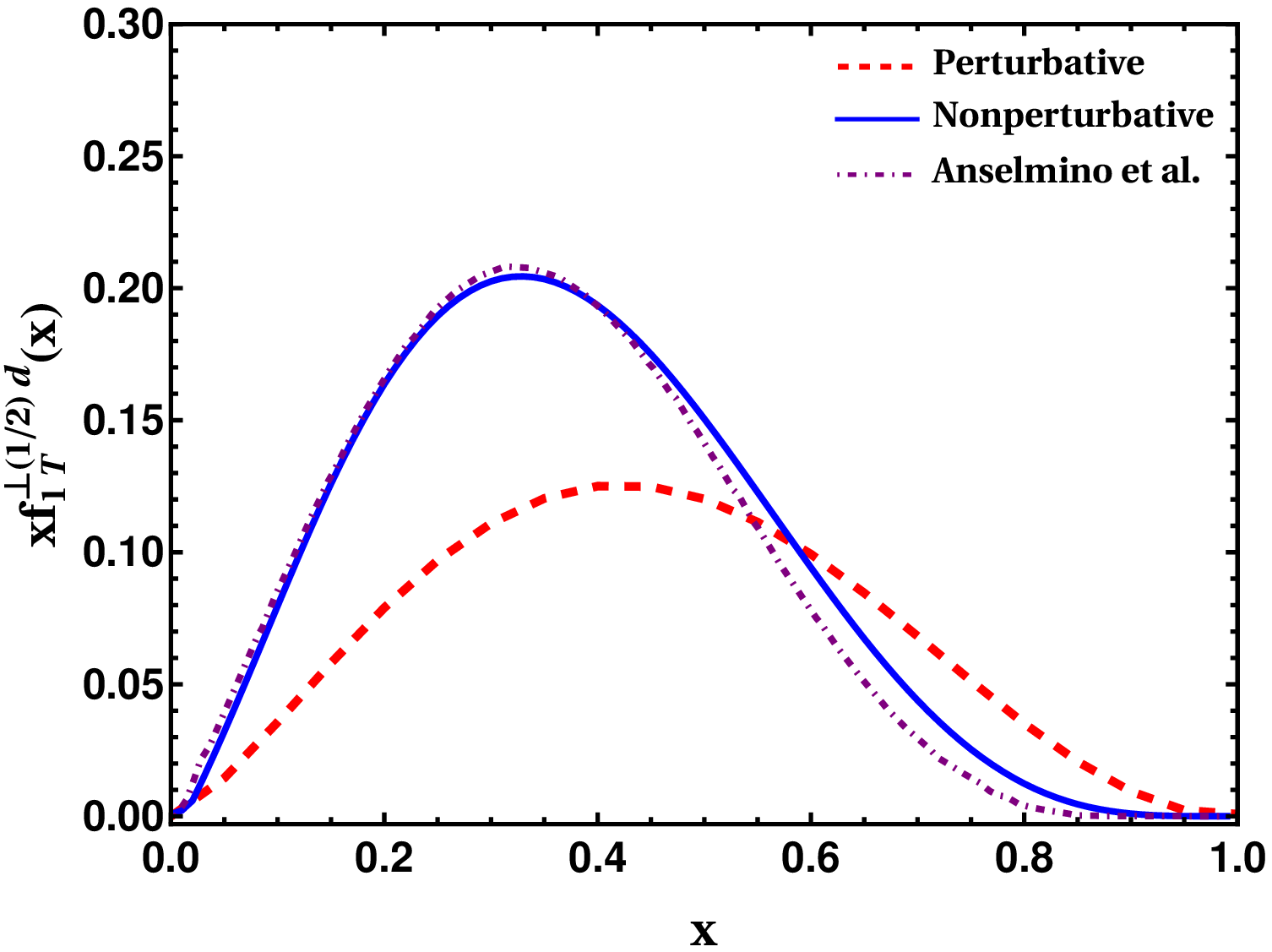}
		\caption{Upper panel: the first moment of the Sivers function  for the up (upper left) and down (upper right) quarks. Lower panel: the $1/2$ moment of the Sivers function for the up (lower left) and down (lower right) quarks after the QCD evolution at $\mu^2=2.5$ GeV$^2$~\cite{Aybat:2011zv,Echevarria:2014xaa,Echevarria:2012pw,Kishore:2019fzb}. The quark Sivers functions are generated by the perturbative ( with $\alpha_{s}=0.3$) and the nonperturbative gluon rescattering kernels. The red dashed and solid blue lines represent the perturbatively and nonperturbatively generated moments, respectively. Our results are compared with the fit to extracted data~\cite{anselmino2006comparing}. }
		\label{Figure4}
\end{figure}
We present our results for the $f_{1T}^{\perp(1)q}$ and $f_{1T}^{\perp(1/2)q}$ transverse moments of the perturbatively and nonperturbatively generated Sivers TMDs for the up and down quarks in Fig.~\ref{Figure4}, where we compare them with the global fit by Anselmino {\it et. al.}~\cite{anselmino2006comparing}. In Table~\ref{FirstmomentTable}, we provide a comparison between our results for the first moments of the Sivers TMDs $|xf_{1T}^{\perp(1)}(x)|$ and the data at various resolution scales $Q^2$ extracted by the COMPASS Collaboration~\cite{COMPASS:2018ofp}. We notice that our moment results for quark Sivers TMDs are quite consistent with the extracted data~\cite{COMPASS:2018ofp} and the global fit~\cite{anselmino2006comparing}.
\begin{table}
	\caption{Our predictions for the first $\mathbf{p}_{\perp}$-moments of the Sivers TMDs at different values of $x$ and different resolution scales $Q^2$. We compare our results with the available COMPASS data~\cite{COMPASS:2018ofp}. Our predictions are computed using the perturbatively and nonperturbatively generated Sivers functions with the $5\%$ uncertainty in the model parameters.}
	\label{FirstmomentTable}
	\centering
\begin{tabular}{c c c c c c c c }
	\hline \hline 
	& & & $|xf_{1T}^{\perp(1)u}(x)|$    &     &     &  $|xf_{1T}^{\perp(1)d}(x)|$  \\
	$x$ &  \text{$Q^{2}$(GeV$^2$)}    & \text{COMPASS~\cite{COMPASS:2018ofp}}   & \text{Our results}  & \text{Our results} & \text{COMPASS~\cite{COMPASS:2018ofp}}   & \text{Our results}  & \text{Our results} \\ 
	&   &   &  \text{(Perturbative)} & \text{(Nonperturbative)} &  & \text{(Perturbative)} & \text{(Nonperturbative)} \\
	\hline
	0.0063 & 1.27 & 0.0022$\pm$0.0051 & 0.0002$\pm$0.0001 & 0.0002$\pm$0.0001 & 0.001$\pm$0.021   & 0.0002$\pm$0.0003 & 0.0004$\pm$0.0002\\
	0.0105 & 1.55 & 0.0029$\pm$0.0040 & 0.0004$\pm$0.0005 & 0.0004$\pm$0.0001 & 0.004$\pm$0.017   & 0.0004$\pm$0.0005 & 0.0007$\pm$0.0003\\
	0.0164  & 1.83  & 0.0058$\pm$0.0037  & 0.0006$\pm$0.0004  & 0.0010$\pm$0.0003  & 0.019$\pm$0.015     & 0.0007$\pm$0.0002  & 0.0013$\pm$0.0001\\
	0.0257 & 2.17 & 0.0097$\pm$0.0033 & 0.0012$\pm$0.0001 & 0.0029$\pm$0.0003      & 0.034$\pm$0.013 &	0.0013$\pm$0.0016  &   0.0030$\pm$0.0004\\
	0.0399 & 2.82 & 0.0179$\pm$0.0036 & 0.0022$\pm$0.0030 & 0.0059$\pm$0.0012       & 0.032$\pm$0.015 & 0.0026$\pm$0.0032    & 0.0068$\pm$0.0023\\
	0.0629 & 4.34 & 0.0224$\pm$0.0046 & 0.0044$\pm$0.0053 &    0.0109$\pm$0.0013    &       0.048$\pm$0.019 & 0.0053$\pm$0.0029 & 0.0130$\pm$0.0029\\
	0.101  & 6.76 & 0.0171$\pm$0.0057 & 0.0087$\pm$0.0028 & 0.019$\pm$0.0011 &   0.025$\pm$0.023 &	0.0107$\pm$0.0013 & 0.0239$\pm$0.0046\\
	0.163  & 10.6 & 0.0295$\pm$0.0070 & 0.0162$\pm$0.0012 & 0.0318$\pm$0.0012 & 0.056$\pm$0.027 & 0.0206$\pm$0.0063 & 0.0404$\pm$0.0038\\
	0.288  & 20.7 & 0.0160$\pm$0.0073 & 0.0303$\pm$0.0073 & 0.0463$\pm$0.0014&       0.017$\pm$0.028  & 0.0388$\pm$0.0018 & 0.0593$\pm$0.0015\\
	\hline \hline
\end{tabular}
\end{table}

A model-independent constraint on our T-odd TMDs is the positivity bound~\cite{Bacchetta:1999kz}. For the Sivers functions the positivity constrain is given by
\begin{eqnarray} \label{PSivers}
P_{\rm S}(x,p_{\perp})&\equiv& f_{1}^{q}(x,p_{\perp})-\frac{p_{\perp}}{M}|f_{1 T}^{\perp q}(x,p_{\perp})|\geq 0\,,
\end{eqnarray}
and the Boer-Mulders TMD follows
\begin{eqnarray}\label{PBM}
P_{\rm BM}(x,p_{\perp})&\equiv& f_{1}^{q}(x,p_{\perp})-\frac{p_{\perp}}{M}|h_{1}^{\perp q}(x,p_{\perp})|\geq 0\,.
\end{eqnarray}

\begin{figure}
\centering
		\includegraphics[scale=0.55]{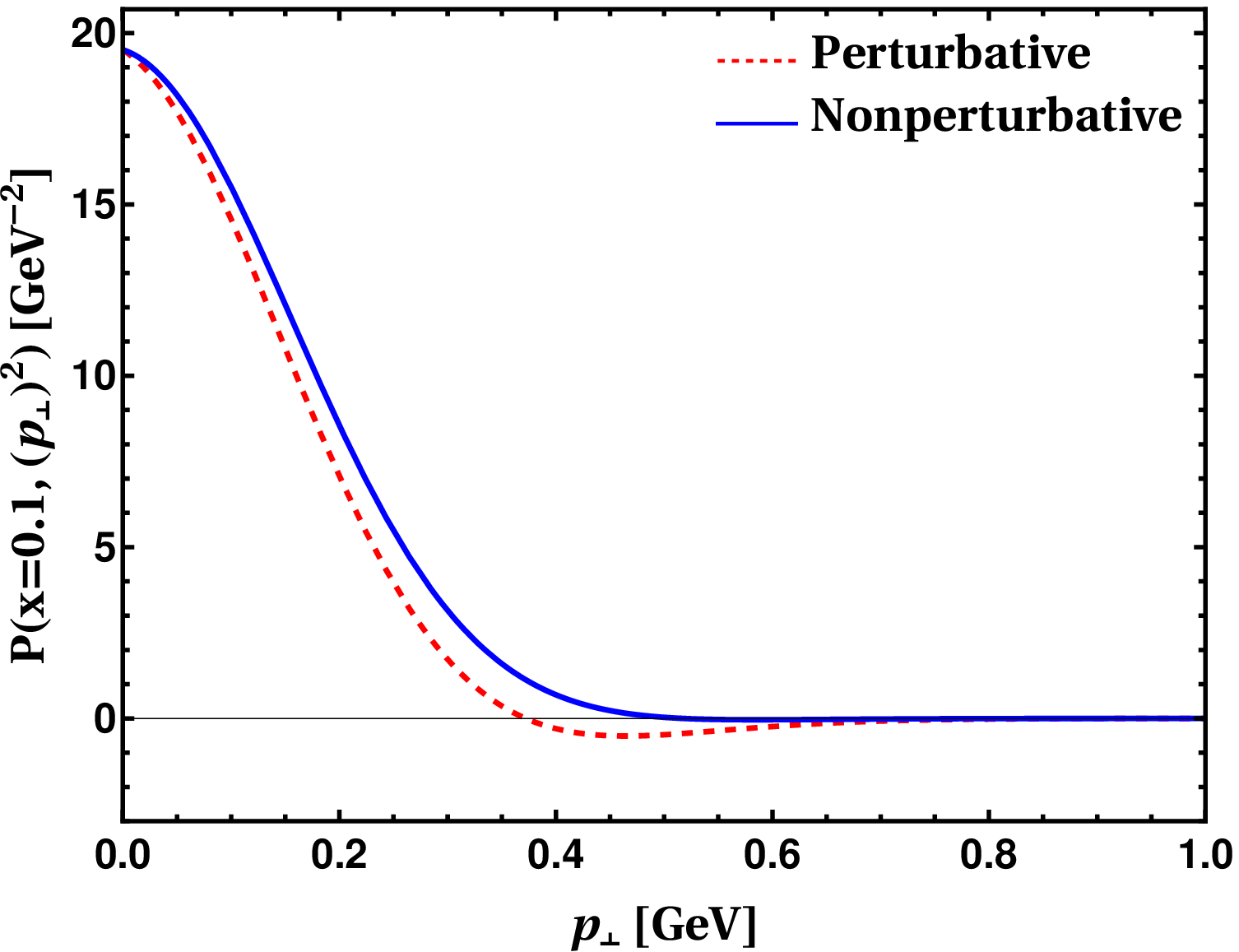}
		\vspace{0.5cm}
		\includegraphics[scale=0.55]{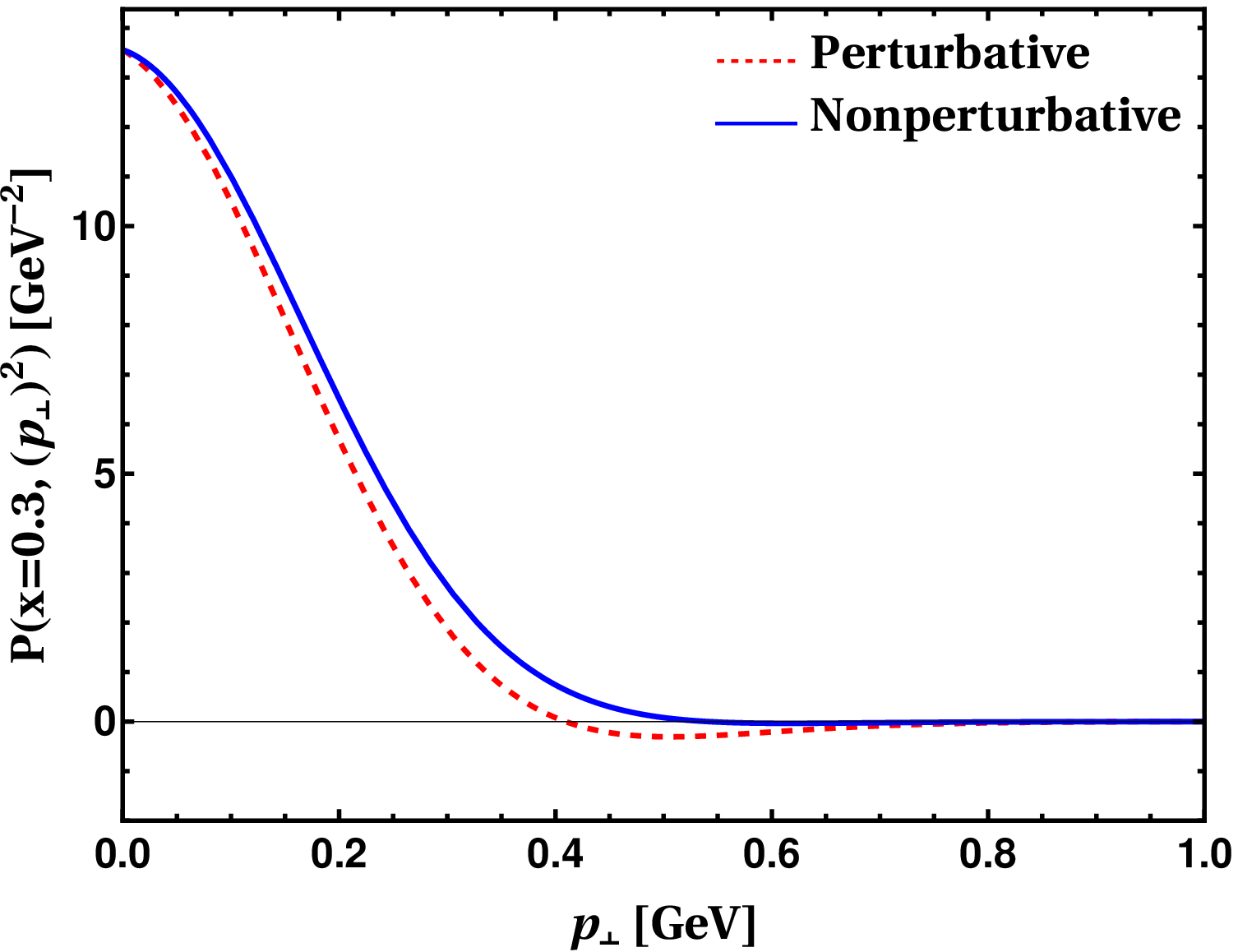}
		\vspace{0.5cm}
		\includegraphics[scale=0.55]{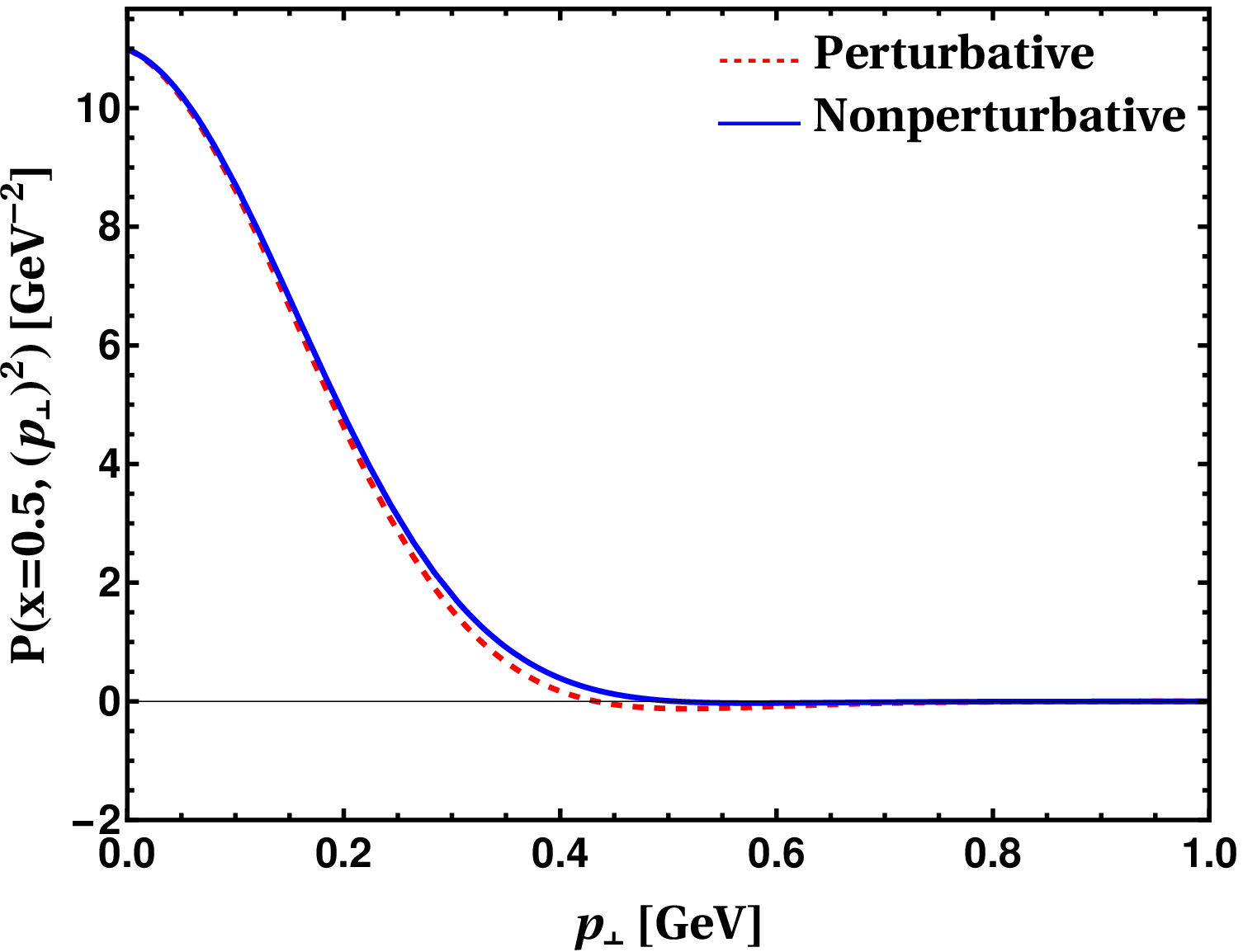}
		\includegraphics[scale=0.55]{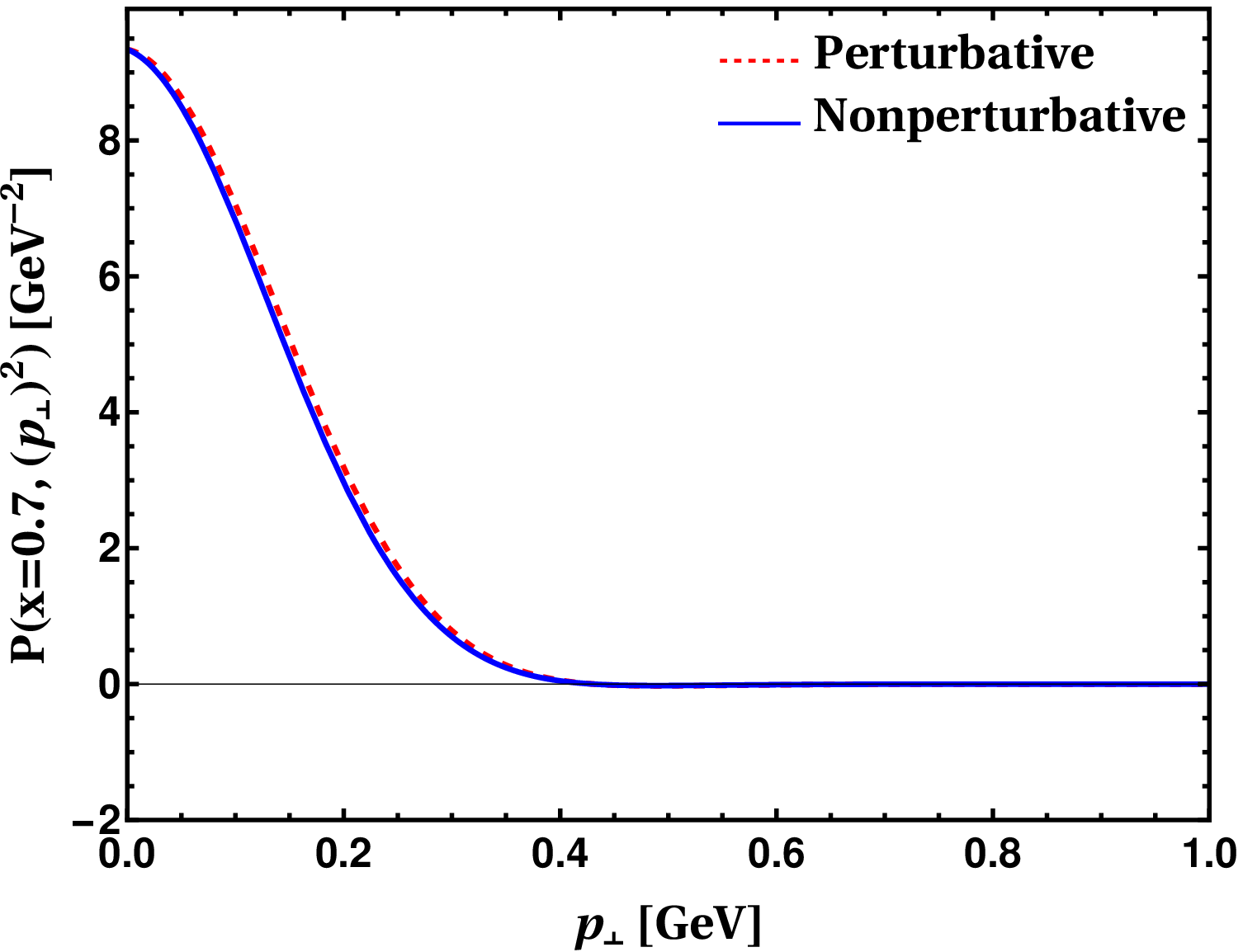}
	\caption{The positivity bound $P_{\rm S}(x,p_{\perp}$) at $x=0.1$ (upper left), $x=0.3$ (upper right), $x=0.5$ (lower left) and $x=0.7$ (lower right) with the
Sivers function is generated by the perturbative kernel (red dashed curves) and the nonperturbative kernel
(solid blue curves).}
		\label{Figure5}
\end{figure}

Figure~\ref{Figure5} confirms that the positivity constraints defined in Eqs.~(\ref{PSivers}) and (\ref{PBM}) are safely satisfied when the  T-odd TMDs are generated by the nonperturbative rescattering kernel. We observe that there is a violation  of the positivity constraints when the TMDs are generated by the perturbative kernel with $\alpha=0.3$, although the violation only occurs for large $p_\perp$. This violation becomes somewhat more pronounced for
small $x$. Similar violation of the positivity constraint for the pion has been reported in the literature~\cite{Pasquini:2014ppa,Wang:2017onm}. It seems to indicate a limitation of the perturbative gluon rescattering kernel to accurately capture the
large $p_\perp$ behavior of the T-odd TMDs.

\subsection{Comparison to lattice QCD}\label{shifts}
\begin{figure}
\centering
\includegraphics[scale=0.58]{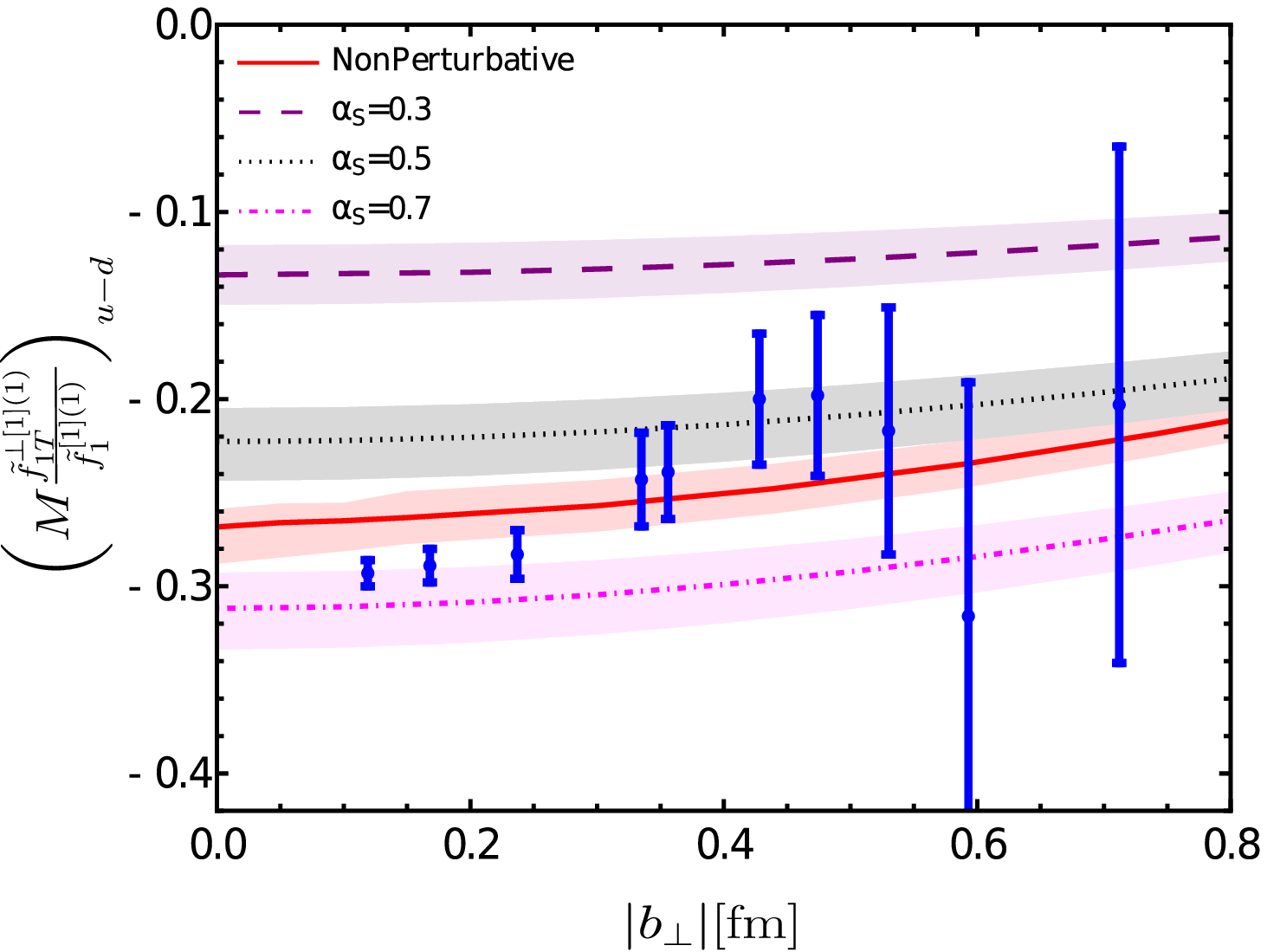} 
\includegraphics[scale=0.58]{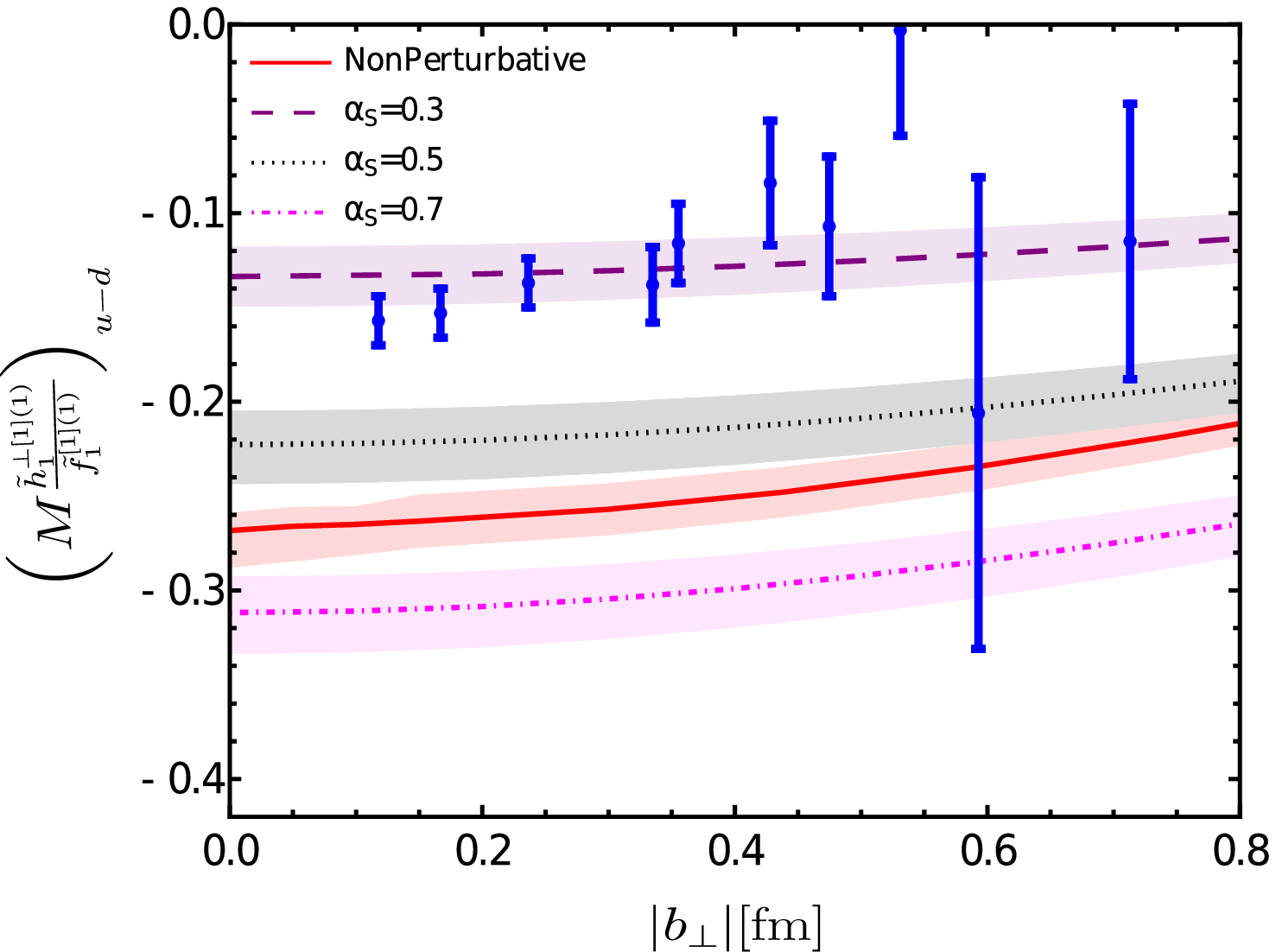}
\caption{The generalized Sivers shift (left panel) and the generalized Boer-Mulders shift (right panel) for $u-d$ as a function of the quark separation $b_{\perp}$ for the SIDIS. We compare the perturbatively (purple-dashed for $\alpha_{s}=0.3$, black-dotted for $\alpha_{s}=0.5$ and magenta-dotdashed for $\alpha_{s}=0.7$) and nonperturbatively (red-solid) generated results with the $5\%$ uncertainty in the model
parameters. Our model results are compared with the lattice QCD simulations (blue points with errorbar)~\cite{Musch:2011er}.}
\label{Fig5}
\end{figure}

To compare with lattice QCD simulations~\cite{Musch:2011er}, we evaluate  the generalized Sivers and Boer-Mulders shifts.
They provide information about the average transverse momentum distributions of unpolarized quarks orthogonal to the transverse spin of the nucleon and that of the transversely polarized quarks in an unpolarized nucleon, respectively.
 The generalized Sivers and  Boer-Mulders shifts are defined as~\cite{Musch:2011er}
\begin{align}
\label{SiversShift}\nonumber
 \langle p_{\perp} \rangle ^{\rm Sivers}& \equiv\langle p_{\perp}\rangle_{TU}\left(\boldsymbol{b}_{\perp}^{2} \right) =  M \frac{\tilde{f}_{1 T}^{\perp[1](1)}\left(\boldsymbol{b}_{\perp}^{2} \right)}{\tilde{f}_{1}^{[1](0)}\left(\boldsymbol{b}_{\perp}^{2} \right)} \,,\\
  \langle p_{\perp} \rangle ^{\rm BM}& \equiv 	\left\langle p_{\perp}\right\rangle_{UT}\left(\boldsymbol{b}_{\perp}^{2} \right) = M \frac{\tilde{h}_{1}^{\perp[1](1)}\left(\boldsymbol{b}_{\perp}^{2} \right)}{\tilde{f}_{1}^{[1](0)}\left(\boldsymbol{b}_{\perp}^{2} \right)} \,,
\end{align}
respectively, where the generalized moments of the TMDs read 
\begin{equation} \label{generalized TMD Moments}
		\tilde{f}^{[m](n)}\left(b_{\perp}^{2}\right)= \frac{2 \pi n !}{M^{2 n}} \int \mathrm{d} x x^{m-1} \int \mathrm{d} p_{\perp} p_{\perp}\left(\frac{p_{\perp}}{b_{\perp}}\right)^{n}  \times J_{n}\left(b_{\perp}.p_{\perp}\right) f\left(x, p_{\perp}^{2}\right)\,.
\end{equation}
Table~\ref{Table2} presents our results for the generalized Sivers shift for $(u-d)$, i.e., $\langle p_{\perp}\rangle^{\rm Sivers}_{u}-\langle p_{\perp}\rangle^{\rm Sivers}_{d}$. As can be seen from Table~\ref{Table2}, it is possible to fit the lattice QCD data by employing a large $\alpha_{s}$ with the perturbative kernel. Since $\alpha_{s}=0.7$ or beyond is not consistent with the weak coupling hypothesis, we prefer to deem the predictions with $\alpha_{s}=0.3$ as a more realistic result with the perturbative kernel. Then it becomes apparent that the nonperturbative kernel does a better job, bringing our predictions closer to the lattice QCD results. In Table~\ref{Table3}, we present the generalized Boer-Mulders shift in our model and observe that the perturbative kernel with $\alpha_s=0.3$ provides a better description of the lattice QCD results compared to the nonperturbative kernel.   In Fig.~\ref{Fig5}, we compare our results for the generalized Sivers and  Boer-Mulders shifts evaluated using both the perturbative and the nonperturbative gluon rescattering kernel with lattice QCD simulations.

\begin{table}
	\caption{Our predictions, in GeV, for the generalized Sivers shifts given by Eq.(\ref{SiversShift}) at different values of $|b_{\perp}|$ in fm.
		The lattice data are taken from Ref.~\cite{Musch:2011er}. Our predictions are
		computed using the nonperturbatively and perturbatively generated Sivers functions with the $5\%$ uncertainty in the model parameters. The “perturbative”
		predictions are given at three different values of $\alpha_{s}$.}
	\label{Table2}
	\centering
\begin{tabular}{c c c c c c}
	\hline \hline 
	~~&~~ ~~&~~ \text{ Non} ~~&~~ \text{perturbative}  ~~&~~  \text{Perturbative}  ~~&~~  \text{Perturbative}  \\
	$|b_{\perp}|$ ~~&~~  \text{lattice QCD}  ~~&~~ \text{Perturbative} ~~&~~ [$\alpha_{s}=0.3$] ~~&~~ [$\alpha_{s}=0.5$] ~~&~~ [$\alpha_{s}=0.7$] \\
	\hline
	$0.15$ & $-0.310 \pm 0.009$ & $-0.2633\pm0.0132$ & $-0.1328\pm0.0149$ & $-0.2214\pm0.0183$ & $-0.3099\pm0.0191$\\
	$0.30$ & $-0.307\pm 0.020$ & $-0.2570 \pm 0.0128$ & $-0.1305\pm0.0145$ & $-0.2175^\pm0.0178$ & $-0.3046\pm0.0188$\\
	$0.44$ & $-0.253 \pm 0.040$ & $-0.2478 \pm 0.0124$ & $-0.1270\pm0.0139$ & 
	$-0.2117\pm0.0171$ & $-0.2965\pm0.0180$\\
	$0.59$ & $-0.210 \pm 0.125$ & $-0.2347 \pm 0.0118$ & $-0.1221\pm0.0132$ & 
	$-0.2035\pm0.0162$ & $-0.2850\pm0.0169$\\
	$0.74$ & $-0.342 \pm 0.158$ & $-0.2187\pm 0.0109$ & $-0.1161\pm0.0124$ & 
	$-0.1935\pm0.0151$ & $-0.2709\pm0.0157$\\
	\hline \hline
\end{tabular}
\end{table}
\begin{table}
	\caption{Our predictions, in GeV, for the generalized Boer Mulders shifts given by Eq.(\ref{SiversShift}) at different values of $|b_{\perp}|$ in fm.
		The lattice data are taken from Ref.~\cite{Musch:2011er}. Our predictions are
		computed using the nonperturbatively and perturbatively generated Sivers functions with the $5\%$ uncertainty in the model parameters. The “perturbative”
		predictions are given at three different values of $\alpha_{s}$.}
	\label{Table3}
	\centering
\begin{tabular}{c c c c c c}
	\hline \hline 
	~~&~~ ~~&~~ \text{Non} ~~&~~ \text{Perturbative}  ~~&~~  \text{Perturbative}  ~~&~~  \text{Perturbative}  \\
	$|b_{\perp}|$ ~~&~~  \text{Lattice QCD}  ~~&~~ \text{ Perturbative} ~~&~~ [$\alpha_{s}=0.3$] ~~&~~ [$\alpha_{s}=0.5$] ~~&~~ [$\alpha_{s}=0.7$] \\
	\hline
$0.15$ & $-0.14 \pm 0.013$ & $-0.2633\pm0.0132$ & $-0.1328\pm0.0149$ &      $-0.2214\pm0.0183$ & $-0.3099\pm0.0191$\\
$0.30$ & $-0.13 \pm 0.020$ & $-0.2570 \pm 0.0128$ & $-0.1305\pm0.0145$ & $-0.2175^\pm0.0178$ & $-0.3046\pm0.0188$\\
$0.44$ & $-0.11\pm 0.035$ & $-0.2478 \pm 0.0124$ & $-0.1270\pm0.0139$ & 
	$-0.2117\pm0.0171$ & $-0.2965\pm0.0180$\\
	$0.59$ & $-0.10 \pm 0.125$ &$-0.2347 \pm 0.0118$ & $-0.1221\pm0.0132$ & 
	$-0.2035\pm0.0162$ & $-0.2850\pm0.0169$\\
	$0.74$ & $-0.19 \pm 0.075$ & $-0.2187\pm 0.0109$ & $-0.1161\pm0.0124$ & 
	$-0.1935\pm0.0151$ & $-0.2709\pm0.0157$\\
	\hline \hline
\end{tabular}
\end{table}

\subsection{Sivers and Boer-Mulders asymmetries}\label{ssa}
The correlation between the transverse momentum of the parton and the transverse spin of the proton is described by the Sivers asymmetry. In the SIDIS procedure, the Sivers asymmetry can be determined by incorporating the weight factor of $\sin(\phi_{h}-\phi_{s})$ as \cite{Boffi:2009sh,Anselmino:2005ea,Anselmino:2005nn,HERMES:2009lmz}
\begin{eqnarray}\label{Eq27}
A_{UT}^{\sin(\phi_{h}-\phi_{s})}=\frac{\int d\phi_{h}d\phi_{s}\left[d\sigma^{\ell P^{\uparrow}\rightarrow\ell' h X}-d\sigma^{\ell P^{\downarrow}\rightarrow\ell' h X}\right]\sin(\phi_{h}-\phi_{s})}{\int d\phi_{h}d\phi_{s}\left[d\sigma^{\ell P^{\uparrow}\rightarrow\ell' h X}+d\sigma^{\ell P^{\downarrow}\rightarrow\ell' h X}\right]}\,,
\end{eqnarray}
where $\uparrow,\downarrow$ at the superscript of $P$ correspond to the up and down transverse spins of the target proton. Using the QCD factorization theorem, the SIDIS cross-section for the one-photon exchange process can be expressed as \cite{Boffi:2009sh,Anselmino:2005ea,Anselmino:2005nn}
\begin{eqnarray}\label{Eq29}
d\sigma^{\ell N\rightarrow\ell' hX}= \sum_{\nu} \hat{f}_{\nu/P}(x,\mathbf{p}_{\perp};Q^2)\otimes d\hat{\sigma}^{\ell q\rightarrow\ell q}\otimes\hat{D}_{h/\nu}(z,\mathbf{k}_{\perp};Q^2)\,.
\end{eqnarray}
In the above expression the hard scattering part, $d\hat{\sigma}^{\ell q \rightarrow \ell q}$, is calculable in perturbative QCD. The soft part is factorized into TMDs designated by $\hat{f}_{ \nu/P}(x,\bfp;Q^2)$ and fragmentation functions (FFs) denoted by $\hat{D}_{ h/\nu}(z,\bfk;Q^2)$. This procedure is valid in the region where $\mathbf{P}^{2}_{h\perp}\simeq \Lambda^{2}_{\rm QCD}\ll Q^{2}$~\cite{Ji:2006br,Anselmino:2006rv}. The
TMD factorization is presented for the SIDIS and the DY processes in Refs.~\cite{Ji:2004wu,Ji:2004xq,Echevarria:2012js} and extensively used in literature. The kinematic variables relevant to the $\gamma*-N$ process are defined in the center-of-mass reference frame as

\begin{eqnarray}
x=\frac{Q^2}{2P.q}=x_{B}\,,\hspace*{1cm}
 z=\frac{P.P_{h}}{P.q}=z_{h}\,, \hspace*{1cm}
  y=\frac{P.q}{P.l}=\frac{Q^2}{sx}\,, 
\end{eqnarray}
with $x_B$ being the Bjorken variable and $Q^2=-q^2$.
The fraction of energy transferred by the photon in the laboratory frame is denoted by $y$, whereas the fraction of energy carried by the produced hadron is given by $z=\mathbf{P}^{-}_{h}/k^{-}$. The transverse momenta of the fragmented quark and the produced hadron are denoted by $\mathbf{k}_{\perp}$ and $\mathbf{P}_{h\perp}$, respectively. The relation between $\mathbf{p}_{\perp}$, $\mathbf{k}_{\perp}$, and $\mathbf{P}_{h\perp}$ is given by $\mathbf{k}_{\perp}=\mathbf{P}_{h\perp}-z\mathbf{p}_{\perp}$. The transverse momentum of the produced hadron makes an azimuthal angle $\phi_{h}$ and the transverse spin ($S_{p}$) of the proton has an azimuthal angle $\phi_{s}$ with respect to the lepton plane. The SIDIS cross-section difference in the numerator of Eq.~(\ref{Eq27}) can be written as~\cite{PhysRevD.83.114019}
\begin{equation} 
\begin{aligned}\label{cross-section numerator}
\frac{d\sigma^{\ell P^{\uparrow}\rightarrow\ell^{'}hX}-d\sigma^{\ell P^{\downarrow}\rightarrow\ell^{'}hX}}{dx_{B}dydzd^{2}\mathbf{P}^{2}_{h\perp}d\phi_{s}}=&\frac{2\alpha^{2}}{s x y^{2}}2\Bigg[\frac{1+(1-y)^{2}}{2}\sin(\phi_{h}-\phi_{s})F_{UT}^{\sin(\phi_{h}-\phi_{s})}\\
&+(1-y)\left(\sin(\phi_{h}+\phi_{s})F_{UT}^{\sin(\phi_{h}+\phi_{s})}+\sin(3\phi_{h}-\phi_{s})F_{UT}^{\sin(3\phi_{h}-\phi_{s})}\right) \\
& +(2-y)\sqrt{(1-y)}\left(\sin(\phi_{s})F_{UT}^{\sin(\phi_{s})}+2\sin(2\phi_{h}-\phi_{s})F_{UT}^{\sin(2\phi_{h}-\phi_{s})}\right)  \Bigg] \,,
\end{aligned}  
\end{equation}
where the weighted structure functions $F_{S_{\ell}S}^{\mathcal{W}(\phi_{h},\phi_{s})}$ are defined as
\begin{equation}\label{structure function}
F_{S_{\ell}S}^{\mathcal{W}(\phi_{h},\phi_{s})}= \sum_{\nu} e_{\nu}^{2}\int d^{2}\mathbf{p}_{\perp} d^{2}\mathbf{k}_{\perp}\delta^{(2)}\, (\mathbf{P}_{\perp}-z\mathbf{p}_{\perp}-k_{\perp})\mathcal{W}(\mathbf{p}_{\perp},\mathbf{P}_{h\perp}) 
\, \hat{f}^{\nu}(x,\mathbf{p}_{\perp})\,\hat{D}^{\nu}(z,\mathbf{k}_{\perp})\,,
\end{equation}
with $\hat{f}^{\nu}(x,\mathbf{p}_{\perp})$ and $\hat{D}^{\nu}(z,\mathbf{k}_{\perp})$ being the leading twist TMDs and the FFs, respectively. By integrating the numerator over $\phi_{h}$ and $\phi_{s}$ with a particular weight factor $\mathcal{W}(\phi_{h},\phi_{s})$, one obtains the corresponding structure function $F_{S_{\ell}S}^{\mathcal{W}(\phi_{h},\phi_{s})}$ and hence, the particular asymmetry can be calculated. For example, the $\phi_{h}$ and $\phi_{s}$ integration with the weight factors $\sin(\phi_{h}-\phi_{s})$ and $\cos(2\phi_{h})$ ends up with the
Sivers and Boer-Mulders asymmetries, respectively.

\begin{figure}
\centering
\includegraphics[scale=1]{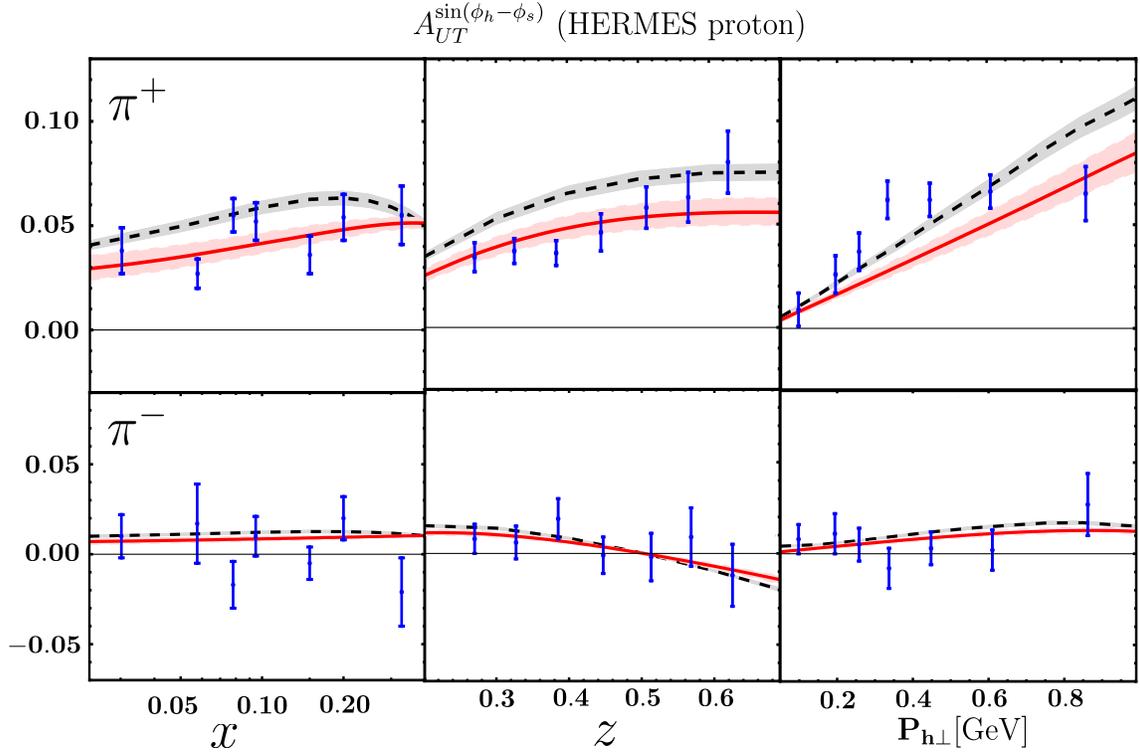}
\caption{Model results of the Sivers asymmetries, $A_{UT}^{\sin(\phi_{h}-\phi_{s})}$, in the $\pi^{+}$ (upper row) and $\pi^{-}$(lower row) channels are compared with the HERMES data~\cite{HERMES:2009lmz}. The solid-red lines represent the results calculated using perturbatively generated Sivers TMDs, whereas the black-dashed lines correspond to the results evaluated employing nonperturbatively generated Sivers TMDs. Both $f_{1T}^{\perp \nu}(x,\mathbf{p}_{\perp}^{2})$ and $f_{1}^{\nu}(x,\mathbf{p}_{\perp}^{2})$ are evolved from the model scale to $\mu^{2}=2.5$ GeV$^{2}$ relevant to the experimental scale following the QCD evolution~\cite{Ji:2020jeb,Aybat:2011zv,Echevarria:2014xaa,Echevarria:2012pw,Kishore:2019fzb}. The fragmentation function $D_{1}^{h/\nu}(z,\mathbf{k}_{\perp})$ are taken as a phenomenological~\cite{Kretzer:2001pz} input at $\mu^{2}=2.5$ GeV$^{2}$.}
\label{SiversSSA}
\end{figure}

Similarly, the denominator of Eq.~(\ref{Eq27}) can be written as~\cite{PhysRevD.83.114019} 
\begin{equation} 
\begin{aligned}\label{cross-section denominator}
\frac{d\sigma^{\ell P^{\uparrow}\rightarrow\ell^{'}hX}+d\sigma^{\ell P^{\downarrow}\rightarrow\ell^{'}hX}}{dx_{B}dydzd^{2}\mathbf{P}^{2}_{h\perp}d\phi_{s}}=& \frac{2\alpha^{2}}{s x y^{2}}2 \Bigg[\frac{1+(1-y)^{2}}{2}F_{UU}+(2-y)\sqrt{1-y}\cos(\phi_{h})F_{UU}^{\cos(\phi_{h})} \\
&+(1-y)\cos(2\phi_{h})F_{UU}^{\cos(2\phi_{h})}\Bigg]\,. 
\end{aligned}  
\end{equation} 
The Sivers asymmetry in terms of structure functions is expressed as~\cite{PhysRevD.83.114019}
\begin{equation}
	\begin{aligned}\label{SSA}
A_{UT}^{\sin(\phi_{h}-\phi_{s})}(x,z,\mathbf{P}_{h\perp},y)=&\frac{2\pi^{2}\alpha^{2}\frac{1+(1-y)^{2}}{sxy^{2}}F^{\sin(\phi_{h}-\phi_{s})}_{UT}(x,z,\mathbf{P}_{h\perp})}{2\pi^{2}\alpha^{2}\frac{1+(1-y)^{2}}{sxy^{2}}F_{UU}(x,z,\mathbf{P}_{h\perp})} \\
=& \frac{{2\pi^{2}\alpha^{2}\frac{1+(1-y)^{2}}{sxy^{2}}\sum_{\nu}e_{\nu}^{2}\int d^{2}\mathbf{p}_{\perp}}\{\frac{-\hat{\mathbf{P}}_{h\perp}.\mathbf{p}_{\perp}}{M}\}f_{1T}^{\perp \nu}(x,\mathbf{p}_{\perp}^{2})D_{1}^{h/\nu}(z,\mathbf{P}_{h\perp}-z\mathbf{p}_{\perp})}{2\pi^{2}\alpha^{2}\frac{1+(1-y)^{2}}{sxy^{2}}\sum_{\nu}e_{\nu}^{2}\int d^{2}\mathbf{p}_{\perp}f_{1}^{ \nu}(x,\mathbf{p}_{\perp}^{2})D_{1}^{h/\nu}(z,\mathbf{P}_{h\perp}-z\mathbf{p}_{\perp})}\,.
	\end{aligned}
\end{equation}
We evaluate the Sivers asymmetries by employing our model TMDs and the unpolarized FF $D_{1}^{h / \nu}\left(z,\left|\mathbf{P}_{h}-z \mathbf{p}_{\perp}\right|\right)$ as phenomenological input taken from Refs.~\cite{Kretzer:2001pz,Anselmino:2013vqa}. The model results for the Sivers asymmetries in the $\pi^{+}$ and $\pi^{-}$ channels are presented in Fig.~\ref{SiversSSA}, where we compare our predictions with the HERMES data \cite{HERMES:2009lmz} in the kinematical region 
\begin{eqnarray}
0.023<x<0.4\,, \hspace{1cm} 0.2<z<0.7,\hspace{1cm} 
 0.31<y<0.95\,, \hspace*{0.5cm} \rm{and} \hspace{0.5cm} \mathbf{P}_{h\perp}>0.05\,.
\end{eqnarray}
Note that we evolve our $f_{1T}^{\perp \nu}(x,\mathbf{p}_{\perp}^{2})$ and  $f_{1}^{ \nu}(x,\mathbf{p}_{\perp}^{2})$ from the model scale to the scale $\mu^{2}=2.5$ GeV$^{2}$ relevant to the experimental data for the asymmetries following QCD evolutions reported in Refs.~\cite{Ji:2020jeb,Aybat:2011zv,Echevarria:2014xaa,Echevarria:2012pw,Kishore:2019fzb}. We illustrate the differences between the asymmetries generated by using the perturbative and nonperturbative gluon rescattering kernels for the Sivers TMDs. We find that both the perturbatively and nonperturbatively generated asymmetries are reasonably consistent with the experimental data.

\begin{figure}
	\centering
	\includegraphics[scale=1]{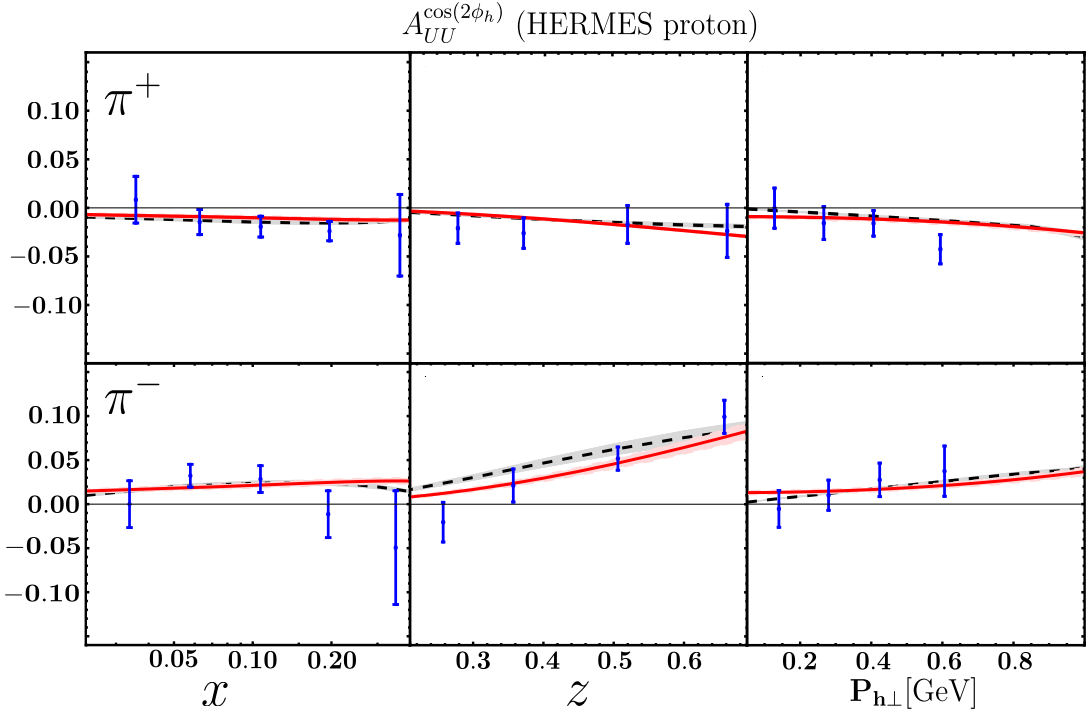}
	\caption{Model results of the Boer-Mulders, $A_{UU}^{\cos(2\phi_{h})}$, in the $\pi^{+}$ (upper row) and $\pi^{-}$(lower row) channels are compared with the HERMES data~\cite{Barone:2009hw,Giordano:2009hi}. The solid-red lines represent the results calculated using perturbatively generated Boer Mulders TMDs, whereas the black-dashed lines correspond to the results evaluated employing nonperturbatively generated Boer Mulders. Both $h_{1}^{\perp \nu}(x,\mathbf{p}_{\perp}^{2})$ and $f_{1}^{\nu}(x,\mathbf{p}_{\perp}^{2})$ are evolved from the model scale to $\mu^{2}=2.5$ GeV$^{2}$ relevant to the experimental scale following the QCD evolution~\cite{Wang:2018naw,Echevarria:2014xaa}. The fragmentation function $H_{1}^{\perp \nu}(z,\mathbf{k}_{\perp})$ are taken as a phenomenological \cite{Anselmino:2013vqa,Anselmino:2007fs} input at $\mu^{2}=2.5$ GeV$^{2}$.}
	\label{BMasymmetry}
\end{figure}

The Boer-Mulders asymmetry can be calculated by using the weight factor of $\cos(2\phi_{h})$ and expressed in terms of structure functions as~\cite{Anselmino:2011ch}
\begin{equation} \label{BMSSA}
	\begin{aligned}
		A_{U U}^{\cos \left(2 \phi_{h}\right)} &=\frac{4 \pi^{2} \alpha^{2} \frac{(1-y)}{s x y^{2}} F_{U U}^{\cos 2 \phi_{h}}\left(x, z, \mathbf{P}_{h \perp}\right)}{2 \pi^{2} \alpha^{2} \frac{1+(1-y)^{2}}{s x y^{2}} F_{U U}\left(x, z, \mathbf{P}_{h \perp}\right)} \\
		&=\frac{4 \pi^{2} \alpha^{2} \frac{(1-y)}{s x y^{2}} \sum_{\nu} e_{\nu}^{2} \int d^{2} p{ }_{\perp}\left\{\frac{\left(\mathbf{P}_{h \perp} \cdot \mathbf{p}_{\perp}\right)-2 z\left(\hat{\mathbf{P}}_{h \perp} \cdot \mathbf{p}_{\perp}\right)^{2}+z p_{\perp}^{2}}{z M_{h} M}\right\} h_{1}^{\perp \nu}\left(x, \mathbf{p}_{\perp}^{2}\right) H_{1}^{\perp \nu}\left(z,\left|\mathbf{P}_{h}-z \mathbf{p}_{\perp}\right|\right)}{2 \pi^{2} \alpha^{2} \frac{1+(1-y)^{2}}{s x y^{2}} \sum_{\nu} e_{\nu}^{2} \int d^{2} p_{\perp} f_{1}^{\nu}\left(x, \mathbf{p}_{\perp}^{2}\right) D_{1}^{h / \nu}\left(z,\left|\mathbf{P}_{h}-z \mathbf{p}_{\perp}\right|\right)}\,.
	\end{aligned}
\end{equation}
The Boer-Mulders TMDs $h_{1}^{\perp \nu}\left(x, \mathbf{p}_{\perp}^{2}\right)$ are obtained in our model and given in Eq.~(\ref{SBM}), whereas the unpolarized FF $D_{1}^{h / \nu}\left(z,\left|\mathbf{P}_{h}-z \mathbf{p}_{\perp}\right|\right)$ and the Collins function $H_{1}^{\perp \nu}\left(z,\left|\mathbf{P}_{h}-z \mathbf{p}_{\perp}\right|\right)$ are taken as phenomenological inputs \cite{Kretzer:2001pz,Anselmino:2013vqa},
\begin{equation}
	\begin{array}{c}
		D_{1}^{h / \nu}\left(z, \mathbf{k}_{\perp}\right)=D_{1}^{h / \nu}(z) \frac{e^{-\mathbf{k}_{\perp}^{2} /\left\langle k_ {\perp}^{2}\right\rangle}}{\pi\left\langle k_ {\perp}^{2}\right\rangle}\,, \\
		H_{1}^{\perp \nu}\left(z, \mathbf{k}_{\perp}\right)=\left(\frac{z M_{h}}{2 k_{\perp}}\right) 2 \mathcal{N}_{\nu}^{C}(z) D_{1}^{h / \nu}(z) h\left(k_{\perp}\right) \frac{e^{-\mathbf{k}_{\perp}^{2} /\left\langle k_ {\perp}^{2}\right\rangle}}{\pi\left\langle k_{\perp}^{2}\right\rangle}\,,
	\end{array}
\end{equation}
with 
	\begin{eqnarray}
		\mathcal{N}_{\nu}^{C}(z)&=&N_{\nu}^{C} z^{\rho_{1}}(1-z)^{\rho_{2}} \frac{\left(\rho_{1}+\rho_{2}\right)^{\left(\rho_{1}+\rho_{2}\right)}}{\rho_{1}^{\rho_{1}} \rho_{2}^{\rho_{2}}}\,, \nonumber \\
		h\left(k_{\perp}\right)&=&\sqrt{2 e} \frac{k_{\perp}}{M_{h}} e^{-\mathbf{k}_{\perp}^{2} / M_{h}^{2}}\,.
	\end{eqnarray}
Here, $z=P_{h}^{-}/k^{-}$ is the energy fraction carried by the fragmenting quark having transverse momentum $\mathbf{k_{\perp}}$. The numerical values of the parameters can be found in Ref.~\cite{Anselmino:2013vqa}.
 The Boer-Mulders asymmetries in the $\pi^{+}$ and $\pi^{-}$ channels are shown in Fig.~\ref{BMasymmetry}. We compare our model results with the HERMES data~\cite{Barone:2009hw,Giordano:2009hi} in the kinematical region
\begin{eqnarray}
	0.023<x<1.0\,, \hspace{1cm} 0.2<z<1.0\,, \hspace{1cm}
	 0.3<y<0.85\,, \hspace{0.5cm}\rm{and} \hspace{0.5cm} \mathbf{P}_{h\perp}>0.05\,.
\end{eqnarray}
We observe that our predictions for the Boer-Mulders asymmetries with both the perturbatively and nonperturbatively generated Boer-Mulders TMDs are fairly consistent with the HERMES data within the uncertainties.
\section{Conclusion}\label{concl}

We have presented a detailed study of the T-odd TMDs in a quark-diquark model of proton written as overlaps of light-front wave functions with a soft gluon rescattering kernel which incorporates the effect of the FSI. The generalized Sivers and Boer Mulders shifts and the SSAs are also studied in the model. In the generalized Sivers and Boer-Mulders shifts, the disagreements between perturbative and nonperturbative kernels become prominent. The generalized shifts with the nonperturbative kernel are found to be consistent with the lattice QCD results, whereas the results with the perturbative kernel varies widely with $\alpha_s$. The perturbative kernel with some intermediate value $\alpha_s\approx 0.5$ produces the shifts close to the nonperturbative kernel, but it is too sensitive to the variation of $\alpha_s$ and requires higher-order corrections to make any reliable predictions.
The Sivers asymmetry $A_{UT}^{\sin(\phi_h-\phi_s)}$ and the Boer-Mulders asymmetry $A_{UU}^{\cos(2\phi_h)}$ for $\pi^+$ and $\pi^-$ channels are found to be in good agreement with the HERMES data. Our study shows that the nonperturbative kernel does a much better job than the perturbative kernel if in accordance with the weak coupling hypothesis, $\alpha_s$ is considered to be small ($\alpha_s \ll 1$). 

\begin{acknowledgments}
The work of DC is supported by Science and Engineering Research Board under the Grant No. CRG/2019/000895. CM is supported by new faculty start up funding by the Institute of Modern Physics, Chinese Academy of Sciences, Grant No. E129952YR0.  CM also thanks the Chinese Academy of Sciences Presidents International Fellowship Initiative for the support via Grants No. 2021PM0023.
\end{acknowledgments}

\bibliography{REF.bib}

\begin{thebibliography}{95}
\expandafter\ifx\csname natexlab\endcsname\relax\def\natexlab#1{#1}\fi
\expandafter\ifx\csname bibnamefont\endcsname\relax
  \def\bibnamefont#1{#1}\fi
\expandafter\ifx\csname bibfnamefont\endcsname\relax
  \def\bibfnamefont#1{#1}\fi
\expandafter\ifx\csname citenamefont\endcsname\relax
  \def\citenamefont#1{#1}\fi
\expandafter\ifx\csname url\endcsname\relax
  \def\url#1{\texttt{#1}}\fi
\expandafter\ifx\csname urlprefix\endcsname\relax\def\urlprefix{URL }\fi
\providecommand{\bibinfo}[2]{#2}
\providecommand{\eprint}[2][]{\url{#2}}

\bibitem[{\citenamefont{M\"uller et~al.}(1994)\citenamefont{M\"uller,
  Robaschik, Geyer, Dittes, and Ho\v{r}ej\v{s}i}}]{Muller:1994ses}
\bibinfo{author}{\bibfnamefont{D.}~\bibnamefont{M\"uller}},
  \bibinfo{author}{\bibfnamefont{D.}~\bibnamefont{Robaschik}},
  \bibinfo{author}{\bibfnamefont{B.}~\bibnamefont{Geyer}},
  \bibinfo{author}{\bibfnamefont{F.~M.} \bibnamefont{Dittes}},
  \bibnamefont{and}
  \bibinfo{author}{\bibfnamefont{J.}~\bibnamefont{Ho\v{r}ej\v{s}i}},
  \bibinfo{journal}{Fortsch. Phys.} \textbf{\bibinfo{volume}{42}},
  \bibinfo{pages}{101} (\bibinfo{year}{1994}), \eprint{hep-ph/9812448}.

\bibitem[{\citenamefont{Ji}(1997)}]{Ji:1996ek}
\bibinfo{author}{\bibfnamefont{X.-D.} \bibnamefont{Ji}},
  \bibinfo{journal}{Phys. Rev. Lett.} \textbf{\bibinfo{volume}{78}},
  \bibinfo{pages}{610} (\bibinfo{year}{1997}), \eprint{hep-ph/9603249}.

\bibitem[{\citenamefont{Radyushkin}(1997)}]{Radyushkin:1997ki}
\bibinfo{author}{\bibfnamefont{A.~V.} \bibnamefont{Radyushkin}},
  \bibinfo{journal}{Phys. Rev. D} \textbf{\bibinfo{volume}{56}},
  \bibinfo{pages}{5524} (\bibinfo{year}{1997}), \eprint{hep-ph/9704207}.

\bibitem[{\citenamefont{Anselmino et~al.}(1995)\citenamefont{Anselmino,
  Efremov, and Leader}}]{Anselmino:1994gn}
\bibinfo{author}{\bibfnamefont{M.}~\bibnamefont{Anselmino}},
  \bibinfo{author}{\bibfnamefont{A.}~\bibnamefont{Efremov}}, \bibnamefont{and}
  \bibinfo{author}{\bibfnamefont{E.}~\bibnamefont{Leader}},
  \bibinfo{journal}{Phys. Rept.} \textbf{\bibinfo{volume}{261}},
  \bibinfo{pages}{1} (\bibinfo{year}{1995}), \bibinfo{note}{[Erratum:
  Phys.Rept. 281, 399--400 (1997)]}, \eprint{hep-ph/9501369}.

\bibitem[{\citenamefont{Barone et~al.}(2002)\citenamefont{Barone, Drago, and
  Ratcliffe}}]{Barone:2001sp}
\bibinfo{author}{\bibfnamefont{V.}~\bibnamefont{Barone}},
  \bibinfo{author}{\bibfnamefont{A.}~\bibnamefont{Drago}}, \bibnamefont{and}
  \bibinfo{author}{\bibfnamefont{P.~G.} \bibnamefont{Ratcliffe}},
  \bibinfo{journal}{Phys. Rept.} \textbf{\bibinfo{volume}{359}},
  \bibinfo{pages}{1} (\bibinfo{year}{2002}), \eprint{hep-ph/0104283}.

\bibitem[{\citenamefont{Zhang et~al.}(2008)\citenamefont{Zhang, Lu, Ma, and
  Schmidt}}]{Zhang:2008nu}
\bibinfo{author}{\bibfnamefont{B.}~\bibnamefont{Zhang}},
  \bibinfo{author}{\bibfnamefont{Z.}~\bibnamefont{Lu}},
  \bibinfo{author}{\bibfnamefont{B.-Q.} \bibnamefont{Ma}}, \bibnamefont{and}
  \bibinfo{author}{\bibfnamefont{I.}~\bibnamefont{Schmidt}},
  \bibinfo{journal}{Phys. Rev. D} \textbf{\bibinfo{volume}{77}},
  \bibinfo{pages}{054011} (\bibinfo{year}{2008}), \eprint{0803.1692}.

\bibitem[{\citenamefont{Gamberg et~al.}(2008)\citenamefont{Gamberg, Goldstein,
  and Schlegel}}]{Gamberg:2007wm}
\bibinfo{author}{\bibfnamefont{L.~P.} \bibnamefont{Gamberg}},
  \bibinfo{author}{\bibfnamefont{G.~R.} \bibnamefont{Goldstein}},
  \bibnamefont{and} \bibinfo{author}{\bibfnamefont{M.}~\bibnamefont{Schlegel}},
  \bibinfo{journal}{Phys. Rev. D} \textbf{\bibinfo{volume}{77}},
  \bibinfo{pages}{094016} (\bibinfo{year}{2008}), \eprint{0708.0324}.

\bibitem[{\citenamefont{Burkardt and Hannafious}(2008)}]{Burkardt:2007xm}
\bibinfo{author}{\bibfnamefont{M.}~\bibnamefont{Burkardt}} \bibnamefont{and}
  \bibinfo{author}{\bibfnamefont{B.}~\bibnamefont{Hannafious}},
  \bibinfo{journal}{Phys. Lett. B} \textbf{\bibinfo{volume}{658}},
  \bibinfo{pages}{130} (\bibinfo{year}{2008}), \eprint{0705.1573}.

\bibitem[{\citenamefont{Pasquini and Yuan}(2010)}]{Pasquini:2010af}
\bibinfo{author}{\bibfnamefont{B.}~\bibnamefont{Pasquini}} \bibnamefont{and}
  \bibinfo{author}{\bibfnamefont{F.}~\bibnamefont{Yuan}},
  \bibinfo{journal}{Phys. Rev. D} \textbf{\bibinfo{volume}{81}},
  \bibinfo{pages}{114013} (\bibinfo{year}{2010}), \eprint{1001.5398}.

\bibitem[{\citenamefont{D'Alesio and Murgia}(2004)}]{DAlesio:2004eso}
\bibinfo{author}{\bibfnamefont{U.}~\bibnamefont{D'Alesio}} \bibnamefont{and}
  \bibinfo{author}{\bibfnamefont{F.}~\bibnamefont{Murgia}},
  \bibinfo{journal}{Phys. Rev. D} \textbf{\bibinfo{volume}{70}},
  \bibinfo{pages}{074009} (\bibinfo{year}{2004}), \eprint{hep-ph/0408092}.

\bibitem[{\citenamefont{Efremov et~al.}(2005)\citenamefont{Efremov, Goeke,
  Menzel, Metz, and Schweitzer}}]{Efremov:2004tp}
\bibinfo{author}{\bibfnamefont{A.~V.} \bibnamefont{Efremov}},
  \bibinfo{author}{\bibfnamefont{K.}~\bibnamefont{Goeke}},
  \bibinfo{author}{\bibfnamefont{S.}~\bibnamefont{Menzel}},
  \bibinfo{author}{\bibfnamefont{A.}~\bibnamefont{Metz}}, \bibnamefont{and}
  \bibinfo{author}{\bibfnamefont{P.}~\bibnamefont{Schweitzer}},
  \bibinfo{journal}{Phys. Lett. B} \textbf{\bibinfo{volume}{612}},
  \bibinfo{pages}{233} (\bibinfo{year}{2005}), \eprint{hep-ph/0412353}.

\bibitem[{\citenamefont{Anselmino
  et~al.}(2005{\natexlab{a}})\citenamefont{Anselmino, Boglione, D'Alesio,
  Kotzinian, Murgia, and Prokudin}}]{Anselmino:2005ea}
\bibinfo{author}{\bibfnamefont{M.}~\bibnamefont{Anselmino}},
  \bibinfo{author}{\bibfnamefont{M.}~\bibnamefont{Boglione}},
  \bibinfo{author}{\bibfnamefont{U.}~\bibnamefont{D'Alesio}},
  \bibinfo{author}{\bibfnamefont{A.}~\bibnamefont{Kotzinian}},
  \bibinfo{author}{\bibfnamefont{F.}~\bibnamefont{Murgia}}, \bibnamefont{and}
  \bibinfo{author}{\bibfnamefont{A.}~\bibnamefont{Prokudin}},
  \bibinfo{journal}{Phys. Rev. D} \textbf{\bibinfo{volume}{72}},
  \bibinfo{pages}{094007} (\bibinfo{year}{2005}{\natexlab{a}}),
  \bibinfo{note}{[Erratum: Phys.Rev.D 72, 099903 (2005)]},
  \eprint{hep-ph/0507181}.

\bibitem[{\citenamefont{Collins et~al.}(2006)\citenamefont{Collins, Efremov,
  Goeke, Grosse~Perdekamp, Menzel, Meredith, Metz, and
  Schweitzer}}]{Collins:2005rq}
\bibinfo{author}{\bibfnamefont{J.~C.} \bibnamefont{Collins}},
  \bibinfo{author}{\bibfnamefont{A.~V.} \bibnamefont{Efremov}},
  \bibinfo{author}{\bibfnamefont{K.}~\bibnamefont{Goeke}},
  \bibinfo{author}{\bibfnamefont{M.}~\bibnamefont{Grosse~Perdekamp}},
  \bibinfo{author}{\bibfnamefont{S.}~\bibnamefont{Menzel}},
  \bibinfo{author}{\bibfnamefont{B.}~\bibnamefont{Meredith}},
  \bibinfo{author}{\bibfnamefont{A.}~\bibnamefont{Metz}}, \bibnamefont{and}
  \bibinfo{author}{\bibfnamefont{P.}~\bibnamefont{Schweitzer}},
  \bibinfo{journal}{Phys. Rev. D} \textbf{\bibinfo{volume}{73}},
  \bibinfo{pages}{094023} (\bibinfo{year}{2006}), \eprint{hep-ph/0511272}.

\bibitem[{\citenamefont{Anselmino et~al.}(2009)\citenamefont{Anselmino,
  Boglione, D'Alesio, Kotzinian, Melis, Murgia, Prokudin, and
  Turk}}]{Anselmino:2008sga}
\bibinfo{author}{\bibfnamefont{M.}~\bibnamefont{Anselmino}},
  \bibinfo{author}{\bibfnamefont{M.}~\bibnamefont{Boglione}},
  \bibinfo{author}{\bibfnamefont{U.}~\bibnamefont{D'Alesio}},
  \bibinfo{author}{\bibfnamefont{A.}~\bibnamefont{Kotzinian}},
  \bibinfo{author}{\bibfnamefont{S.}~\bibnamefont{Melis}},
  \bibinfo{author}{\bibfnamefont{F.}~\bibnamefont{Murgia}},
  \bibinfo{author}{\bibfnamefont{A.}~\bibnamefont{Prokudin}}, \bibnamefont{and}
  \bibinfo{author}{\bibfnamefont{C.}~\bibnamefont{Turk}},
  \bibinfo{journal}{Eur. Phys. J. A} \textbf{\bibinfo{volume}{39}},
  \bibinfo{pages}{89} (\bibinfo{year}{2009}), \eprint{0805.2677}.

\bibitem[{\citenamefont{Anselmino
  et~al.}(2013{\natexlab{a}})\citenamefont{Anselmino, Boglione, D'Alesio,
  Melis, Murgia, and Prokudin}}]{Anselmino:2013rya}
\bibinfo{author}{\bibfnamefont{M.}~\bibnamefont{Anselmino}},
  \bibinfo{author}{\bibfnamefont{M.}~\bibnamefont{Boglione}},
  \bibinfo{author}{\bibfnamefont{U.}~\bibnamefont{D'Alesio}},
  \bibinfo{author}{\bibfnamefont{S.}~\bibnamefont{Melis}},
  \bibinfo{author}{\bibfnamefont{F.}~\bibnamefont{Murgia}}, \bibnamefont{and}
  \bibinfo{author}{\bibfnamefont{A.}~\bibnamefont{Prokudin}},
  \bibinfo{journal}{Phys. Rev. D} \textbf{\bibinfo{volume}{88}},
  \bibinfo{pages}{054023} (\bibinfo{year}{2013}{\natexlab{a}}),
  \eprint{1304.7691}.

\bibitem[{\citenamefont{Martin et~al.}(2017)\citenamefont{Martin, Bradamante,
  and Barone}}]{Martin:2017yms}
\bibinfo{author}{\bibfnamefont{A.}~\bibnamefont{Martin}},
  \bibinfo{author}{\bibfnamefont{F.}~\bibnamefont{Bradamante}},
  \bibnamefont{and} \bibinfo{author}{\bibfnamefont{V.}~\bibnamefont{Barone}},
  \bibinfo{journal}{Phys. Rev. D} \textbf{\bibinfo{volume}{95}},
  \bibinfo{pages}{094024} (\bibinfo{year}{2017}), \eprint{1701.08283}.

\bibitem[{\citenamefont{Barone et~al.}(2010)\citenamefont{Barone, Melis, and
  Prokudin}}]{Barone:2009hw}
\bibinfo{author}{\bibfnamefont{V.}~\bibnamefont{Barone}},
  \bibinfo{author}{\bibfnamefont{S.}~\bibnamefont{Melis}}, \bibnamefont{and}
  \bibinfo{author}{\bibfnamefont{A.}~\bibnamefont{Prokudin}},
  \bibinfo{journal}{Phys. Rev. D} \textbf{\bibinfo{volume}{81}},
  \bibinfo{pages}{114026} (\bibinfo{year}{2010}), \eprint{0912.5194}.

\bibitem[{\citenamefont{Adams et~al.}(1991{\natexlab{a}})}]{E581:1991eys}
\bibinfo{author}{\bibfnamefont{D.~L.} \bibnamefont{Adams}} \bibnamefont{et~al.}
  (\bibinfo{collaboration}{E581, E704}), \bibinfo{journal}{Phys. Lett. B}
  \textbf{\bibinfo{volume}{261}}, \bibinfo{pages}{201}
  (\bibinfo{year}{1991}{\natexlab{a}}).

\bibitem[{\citenamefont{Adams et~al.}(1991{\natexlab{b}})}]{FNAL-E704:1991ovg}
\bibinfo{author}{\bibfnamefont{D.~L.} \bibnamefont{Adams}} \bibnamefont{et~al.}
  (\bibinfo{collaboration}{FNAL-E704}), \bibinfo{journal}{Phys. Lett. B}
  \textbf{\bibinfo{volume}{264}}, \bibinfo{pages}{462}
  (\bibinfo{year}{1991}{\natexlab{b}}).

\bibitem[{\citenamefont{Sivers}(1990)}]{Sivers:1989cc}
\bibinfo{author}{\bibfnamefont{D.~W.} \bibnamefont{Sivers}},
  \bibinfo{journal}{Phys. Rev. D} \textbf{\bibinfo{volume}{41}},
  \bibinfo{pages}{83} (\bibinfo{year}{1990}).

\bibitem[{\citenamefont{Collins}(2002)}]{Collins:2002kn}
\bibinfo{author}{\bibfnamefont{J.~C.} \bibnamefont{Collins}},
  \bibinfo{journal}{Phys. Lett. B} \textbf{\bibinfo{volume}{536}},
  \bibinfo{pages}{43} (\bibinfo{year}{2002}), \eprint{hep-ph/0204004}.

\bibitem[{\citenamefont{Boer and Mulders}(1998)}]{Boer:1997nt}
\bibinfo{author}{\bibfnamefont{D.}~\bibnamefont{Boer}} \bibnamefont{and}
  \bibinfo{author}{\bibfnamefont{P.~J.} \bibnamefont{Mulders}},
  \bibinfo{journal}{Phys. Rev. D} \textbf{\bibinfo{volume}{57}},
  \bibinfo{pages}{5780} (\bibinfo{year}{1998}), \eprint{hep-ph/9711485}.

\bibitem[{\citenamefont{Boer et~al.}(2003{\natexlab{a}})\citenamefont{Boer,
  Mulders, and Pijlman}}]{Boer:2003cm}
\bibinfo{author}{\bibfnamefont{D.}~\bibnamefont{Boer}},
  \bibinfo{author}{\bibfnamefont{P.~J.} \bibnamefont{Mulders}},
  \bibnamefont{and} \bibinfo{author}{\bibfnamefont{F.}~\bibnamefont{Pijlman}},
  \bibinfo{journal}{Nucl. Phys. B} \textbf{\bibinfo{volume}{667}},
  \bibinfo{pages}{201} (\bibinfo{year}{2003}{\natexlab{a}}),
  \eprint{hep-ph/0303034}.

\bibitem[{\citenamefont{Alekseev et~al.}(2010)}]{COMPASS:2010hbb}
\bibinfo{author}{\bibfnamefont{M.~G.} \bibnamefont{Alekseev}}
  \bibnamefont{et~al.} (\bibinfo{collaboration}{COMPASS}),
  \bibinfo{journal}{Phys. Lett. B} \textbf{\bibinfo{volume}{692}},
  \bibinfo{pages}{240} (\bibinfo{year}{2010}), \eprint{1005.5609}.

\bibitem[{\citenamefont{Alekseev et~al.}(2009)}]{COMPASS:2008isr}
\bibinfo{author}{\bibfnamefont{M.}~\bibnamefont{Alekseev}} \bibnamefont{et~al.}
  (\bibinfo{collaboration}{COMPASS}), \bibinfo{journal}{Phys. Lett. B}
  \textbf{\bibinfo{volume}{673}}, \bibinfo{pages}{127} (\bibinfo{year}{2009}),
  \eprint{0802.2160}.

\bibitem[{\citenamefont{Airapetian et~al.}(2009)}]{HERMES:2009lmz}
\bibinfo{author}{\bibfnamefont{A.}~\bibnamefont{Airapetian}}
  \bibnamefont{et~al.} (\bibinfo{collaboration}{HERMES}),
  \bibinfo{journal}{Phys. Rev. Lett.} \textbf{\bibinfo{volume}{103}},
  \bibinfo{pages}{152002} (\bibinfo{year}{2009}), \eprint{0906.3918}.

\bibitem[{\citenamefont{Qian et~al.}(2011)}]{JeffersonLabHallA:2011ayy}
\bibinfo{author}{\bibfnamefont{X.}~\bibnamefont{Qian}} \bibnamefont{et~al.}
  (\bibinfo{collaboration}{Jefferson Lab Hall A}), \bibinfo{journal}{Phys. Rev.
  Lett.} \textbf{\bibinfo{volume}{107}}, \bibinfo{pages}{072003}
  (\bibinfo{year}{2011}), \eprint{1106.0363}.

\bibitem[{\citenamefont{Sbrizzai}(2016)}]{Sbrizzai:2016gro}
\bibinfo{author}{\bibfnamefont{G.}~\bibnamefont{Sbrizzai}}
  (\bibinfo{collaboration}{COMPASS}), \bibinfo{journal}{Int. J. Mod. Phys.
  Conf. Ser.} \textbf{\bibinfo{volume}{40}}, \bibinfo{pages}{1660032}
  (\bibinfo{year}{2016}).

\bibitem[{\citenamefont{Hwang}(2013)}]{Hwang:2010dd}
\bibinfo{author}{\bibfnamefont{D.~S.} \bibnamefont{Hwang}},
  \bibinfo{journal}{J. Korean Phys. Soc.} \textbf{\bibinfo{volume}{62}},
  \bibinfo{pages}{581} (\bibinfo{year}{2013}), \eprint{1003.0867}.

\bibitem[{\citenamefont{Maji et~al.}(2018)\citenamefont{Maji, Chakrabarti, and
  Mukherjee}}]{Maji:2017wwd}
\bibinfo{author}{\bibfnamefont{T.}~\bibnamefont{Maji}},
  \bibinfo{author}{\bibfnamefont{D.}~\bibnamefont{Chakrabarti}},
  \bibnamefont{and}
  \bibinfo{author}{\bibfnamefont{A.}~\bibnamefont{Mukherjee}},
  \bibinfo{journal}{Phys. Rev. D} \textbf{\bibinfo{volume}{97}},
  \bibinfo{pages}{014016} (\bibinfo{year}{2018}), \eprint{1711.02930}.

\bibitem[{\citenamefont{Lyubovitskij et~al.}(2022)\citenamefont{Lyubovitskij,
  Schmidt, and Brodsky}}]{Lyubovitskij:2022vcl}
\bibinfo{author}{\bibfnamefont{V.~E.} \bibnamefont{Lyubovitskij}},
  \bibinfo{author}{\bibfnamefont{I.}~\bibnamefont{Schmidt}}, \bibnamefont{and}
  \bibinfo{author}{\bibfnamefont{S.~J.} \bibnamefont{Brodsky}},
  \bibinfo{journal}{Phys. Rev. D} \textbf{\bibinfo{volume}{105}},
  \bibinfo{pages}{114032} (\bibinfo{year}{2022}), \eprint{2205.08986}.

\bibitem[{\citenamefont{Lu and Schmidt}(2007)}]{Lu:2006kt}
\bibinfo{author}{\bibfnamefont{Z.}~\bibnamefont{Lu}} \bibnamefont{and}
  \bibinfo{author}{\bibfnamefont{I.}~\bibnamefont{Schmidt}},
  \bibinfo{journal}{Phys. Rev. D} \textbf{\bibinfo{volume}{75}},
  \bibinfo{pages}{073008} (\bibinfo{year}{2007}), \eprint{hep-ph/0611158}.

\bibitem[{\citenamefont{Bacchetta
  et~al.}(2008{\natexlab{a}})\citenamefont{Bacchetta, Conti, and
  Radici}}]{PhysRevD.78.074010}
\bibinfo{author}{\bibfnamefont{A.}~\bibnamefont{Bacchetta}},
  \bibinfo{author}{\bibfnamefont{F.}~\bibnamefont{Conti}}, \bibnamefont{and}
  \bibinfo{author}{\bibfnamefont{M.}~\bibnamefont{Radici}},
  \bibinfo{journal}{Phys. Rev. D} \textbf{\bibinfo{volume}{78}},
  \bibinfo{pages}{074010} (\bibinfo{year}{2008}{\natexlab{a}}).

\bibitem[{\citenamefont{Brodsky
  et~al.}(2002{\natexlab{a}})\citenamefont{Brodsky, Hwang, and
  Schmidt}}]{Brodsky:2002cx}
\bibinfo{author}{\bibfnamefont{S.~J.} \bibnamefont{Brodsky}},
  \bibinfo{author}{\bibfnamefont{D.~S.} \bibnamefont{Hwang}}, \bibnamefont{and}
  \bibinfo{author}{\bibfnamefont{I.}~\bibnamefont{Schmidt}},
  \bibinfo{journal}{Phys. Lett. B} \textbf{\bibinfo{volume}{530}},
  \bibinfo{pages}{99} (\bibinfo{year}{2002}{\natexlab{a}}),
  \eprint{hep-ph/0201296}.

\bibitem[{\citenamefont{Ji and Yuan}(2002)}]{Ji:2002aa}
\bibinfo{author}{\bibfnamefont{X.-d.} \bibnamefont{Ji}} \bibnamefont{and}
  \bibinfo{author}{\bibfnamefont{F.}~\bibnamefont{Yuan}},
  \bibinfo{journal}{Phys. Lett. B} \textbf{\bibinfo{volume}{543}},
  \bibinfo{pages}{66} (\bibinfo{year}{2002}), \eprint{hep-ph/0206057}.

\bibitem[{\citenamefont{Burkardt and Hwang}(2004)}]{Burkardt:2003je}
\bibinfo{author}{\bibfnamefont{M.}~\bibnamefont{Burkardt}} \bibnamefont{and}
  \bibinfo{author}{\bibfnamefont{D.~S.} \bibnamefont{Hwang}},
  \bibinfo{journal}{Phys. Rev. D} \textbf{\bibinfo{volume}{69}},
  \bibinfo{pages}{074032} (\bibinfo{year}{2004}), \eprint{hep-ph/0309072}.

\bibitem[{\citenamefont{Belitsky et~al.}(2003)\citenamefont{Belitsky, Ji, and
  Yuan}}]{Belitsky:2002sm}
\bibinfo{author}{\bibfnamefont{A.~V.} \bibnamefont{Belitsky}},
  \bibinfo{author}{\bibfnamefont{X.}~\bibnamefont{Ji}}, \bibnamefont{and}
  \bibinfo{author}{\bibfnamefont{F.}~\bibnamefont{Yuan}},
  \bibinfo{journal}{Nucl. Phys. B} \textbf{\bibinfo{volume}{656}},
  \bibinfo{pages}{165} (\bibinfo{year}{2003}), \eprint{hep-ph/0208038}.

\bibitem[{\citenamefont{Maji et~al.}(2017)\citenamefont{Maji, Mondal, and
  Chakrabarti}}]{Maji:2017ill}
\bibinfo{author}{\bibfnamefont{T.}~\bibnamefont{Maji}},
  \bibinfo{author}{\bibfnamefont{C.}~\bibnamefont{Mondal}}, \bibnamefont{and}
  \bibinfo{author}{\bibfnamefont{D.}~\bibnamefont{Chakrabarti}},
  \bibinfo{journal}{Phys. Rev. D} \textbf{\bibinfo{volume}{96}},
  \bibinfo{pages}{013006} (\bibinfo{year}{2017}), \eprint{1702.02493}.

\bibitem[{\citenamefont{Gurjar et~al.}(2021)\citenamefont{Gurjar, Chakrabarti,
  Choudhary, Mukherjee, and Talukdar}}]{Gurjar:2021dyv}
\bibinfo{author}{\bibfnamefont{B.}~\bibnamefont{Gurjar}},
  \bibinfo{author}{\bibfnamefont{D.}~\bibnamefont{Chakrabarti}},
  \bibinfo{author}{\bibfnamefont{P.}~\bibnamefont{Choudhary}},
  \bibinfo{author}{\bibfnamefont{A.}~\bibnamefont{Mukherjee}},
  \bibnamefont{and} \bibinfo{author}{\bibfnamefont{P.}~\bibnamefont{Talukdar}},
  \bibinfo{journal}{Phys. Rev. D} \textbf{\bibinfo{volume}{104}},
  \bibinfo{pages}{076028} (\bibinfo{year}{2021}), \eprint{2107.02216}.

\bibitem[{\citenamefont{Meissner et~al.}(2007)\citenamefont{Meissner, Metz, and
  Goeke}}]{Meissner:2007rx}
\bibinfo{author}{\bibfnamefont{S.}~\bibnamefont{Meissner}},
  \bibinfo{author}{\bibfnamefont{A.}~\bibnamefont{Metz}}, \bibnamefont{and}
  \bibinfo{author}{\bibfnamefont{K.}~\bibnamefont{Goeke}},
  \bibinfo{journal}{Phys. Rev. D} \textbf{\bibinfo{volume}{76}},
  \bibinfo{pages}{034002} (\bibinfo{year}{2007}), \eprint{hep-ph/0703176}.

\bibitem[{\citenamefont{Kafer}(2008)}]{Kafer:2008ud}
\bibinfo{author}{\bibfnamefont{W.}~\bibnamefont{Kafer}}
  (\bibinfo{collaboration}{COMPASS}) (\bibinfo{year}{2008}),
  \eprint{0808.0114}.

\bibitem[{\citenamefont{Bressan}(2009)}]{Bressan:2009eu}
\bibinfo{author}{\bibfnamefont{A.}~\bibnamefont{Bressan}}
  (\bibinfo{collaboration}{COMPASS}) (\bibinfo{year}{2009}), p.
  \bibinfo{pages}{211}, \eprint{0907.5511}.

\bibitem[{\citenamefont{Airapetian et~al.}(2013)}]{HERMES:2012kpt}
\bibinfo{author}{\bibfnamefont{A.}~\bibnamefont{Airapetian}}
  \bibnamefont{et~al.} (\bibinfo{collaboration}{HERMES}),
  \bibinfo{journal}{Phys. Rev. D} \textbf{\bibinfo{volume}{87}},
  \bibinfo{pages}{012010} (\bibinfo{year}{2013}), \eprint{1204.4161}.

\bibitem[{\citenamefont{Giordano and Lamb}(2009)}]{Giordano:2009hi}
\bibinfo{author}{\bibfnamefont{F.}~\bibnamefont{Giordano}} \bibnamefont{and}
  \bibinfo{author}{\bibfnamefont{R.}~\bibnamefont{Lamb}}
  (\bibinfo{collaboration}{HERMES}), \bibinfo{journal}{AIP Conf. Proc.}
  \textbf{\bibinfo{volume}{1149}}, \bibinfo{pages}{423} (\bibinfo{year}{2009}),
  \eprint{0901.2438}.

\bibitem[{\citenamefont{Gutsche et~al.}(2014)\citenamefont{Gutsche,
  Lyubovitskij, Schmidt, and Vega}}]{Gutsche:2013zia}
\bibinfo{author}{\bibfnamefont{T.}~\bibnamefont{Gutsche}},
  \bibinfo{author}{\bibfnamefont{V.~E.} \bibnamefont{Lyubovitskij}},
  \bibinfo{author}{\bibfnamefont{I.}~\bibnamefont{Schmidt}}, \bibnamefont{and}
  \bibinfo{author}{\bibfnamefont{A.}~\bibnamefont{Vega}},
  \bibinfo{journal}{Phys. Rev. D} \textbf{\bibinfo{volume}{89}},
  \bibinfo{pages}{054033} (\bibinfo{year}{2014}), \bibinfo{note}{[Erratum:
  Phys.Rev.D 92, 019902 (2015)]}, \eprint{1306.0366}.

\bibitem[{\citenamefont{Mondal and Chakrabarti}(2015)}]{Mondal:2015uha}
\bibinfo{author}{\bibfnamefont{C.}~\bibnamefont{Mondal}} \bibnamefont{and}
  \bibinfo{author}{\bibfnamefont{D.}~\bibnamefont{Chakrabarti}},
  \bibinfo{journal}{Eur. Phys. J. C} \textbf{\bibinfo{volume}{75}},
  \bibinfo{pages}{261} (\bibinfo{year}{2015}), \eprint{1501.05489}.

\bibitem[{\citenamefont{Musch et~al.}(2012)\citenamefont{Musch, Hagler,
  Engelhardt, Negele, and Schafer}}]{Musch:2011er}
\bibinfo{author}{\bibfnamefont{B.~U.} \bibnamefont{Musch}},
  \bibinfo{author}{\bibfnamefont{P.}~\bibnamefont{Hagler}},
  \bibinfo{author}{\bibfnamefont{M.}~\bibnamefont{Engelhardt}},
  \bibinfo{author}{\bibfnamefont{J.~W.} \bibnamefont{Negele}},
  \bibnamefont{and} \bibinfo{author}{\bibfnamefont{A.}~\bibnamefont{Schafer}},
  \bibinfo{journal}{Phys. Rev. D} \textbf{\bibinfo{volume}{85}},
  \bibinfo{pages}{094510} (\bibinfo{year}{2012}), \eprint{1111.4249}.

\bibitem[{\citenamefont{Chakrabarti et~al.}(2020)\citenamefont{Chakrabarti,
  Mondal, Mukherjee, Nair, and Zhao}}]{Chakrabarti:2020kdc}
\bibinfo{author}{\bibfnamefont{D.}~\bibnamefont{Chakrabarti}},
  \bibinfo{author}{\bibfnamefont{C.}~\bibnamefont{Mondal}},
  \bibinfo{author}{\bibfnamefont{A.}~\bibnamefont{Mukherjee}},
  \bibinfo{author}{\bibfnamefont{S.}~\bibnamefont{Nair}}, \bibnamefont{and}
  \bibinfo{author}{\bibfnamefont{X.}~\bibnamefont{Zhao}},
  \bibinfo{journal}{Phys. Rev. D} \textbf{\bibinfo{volume}{102}},
  \bibinfo{pages}{113011} (\bibinfo{year}{2020}), \eprint{2010.04215}.

\bibitem[{\citenamefont{Brodsky et~al.}(2015)\citenamefont{Brodsky,
  de~Teramond, Dosch, and Erlich}}]{Brodsky:2014yha}
\bibinfo{author}{\bibfnamefont{S.~J.} \bibnamefont{Brodsky}},
  \bibinfo{author}{\bibfnamefont{G.~F.} \bibnamefont{de~Teramond}},
  \bibinfo{author}{\bibfnamefont{H.~G.} \bibnamefont{Dosch}}, \bibnamefont{and}
  \bibinfo{author}{\bibfnamefont{J.}~\bibnamefont{Erlich}},
  \bibinfo{journal}{Phys. Rept.} \textbf{\bibinfo{volume}{584}},
  \bibinfo{pages}{1} (\bibinfo{year}{2015}), \eprint{1407.8131}.

\bibitem[{\citenamefont{de~Teramond et~al.}(2018)\citenamefont{de~Teramond,
  Liu, Sufian, Dosch, Brodsky, and Deur}}]{deTeramond:2018ecg}
\bibinfo{author}{\bibfnamefont{G.~F.} \bibnamefont{de~Teramond}},
  \bibinfo{author}{\bibfnamefont{T.}~\bibnamefont{Liu}},
  \bibinfo{author}{\bibfnamefont{R.~S.} \bibnamefont{Sufian}},
  \bibinfo{author}{\bibfnamefont{H.~G.} \bibnamefont{Dosch}},
  \bibinfo{author}{\bibfnamefont{S.~J.} \bibnamefont{Brodsky}},
  \bibnamefont{and} \bibinfo{author}{\bibfnamefont{A.}~\bibnamefont{Deur}}
  (\bibinfo{collaboration}{HLFHS}), \bibinfo{journal}{Phys. Rev. Lett.}
  \textbf{\bibinfo{volume}{120}}, \bibinfo{pages}{182001}
  (\bibinfo{year}{2018}), \eprint{1801.09154}.

\bibitem[{\citenamefont{Liu et~al.}(2020)\citenamefont{Liu, Sufian,
  de~T\'eramond, Dosch, Brodsky, and Deur}}]{Liu:2019vsn}
\bibinfo{author}{\bibfnamefont{T.}~\bibnamefont{Liu}},
  \bibinfo{author}{\bibfnamefont{R.~S.} \bibnamefont{Sufian}},
  \bibinfo{author}{\bibfnamefont{G.~F.} \bibnamefont{de~T\'eramond}},
  \bibinfo{author}{\bibfnamefont{H.~G.} \bibnamefont{Dosch}},
  \bibinfo{author}{\bibfnamefont{S.~J.} \bibnamefont{Brodsky}},
  \bibnamefont{and} \bibinfo{author}{\bibfnamefont{A.}~\bibnamefont{Deur}},
  \bibinfo{journal}{Phys. Rev. Lett.} \textbf{\bibinfo{volume}{124}},
  \bibinfo{pages}{082003} (\bibinfo{year}{2020}), \eprint{1909.13818}.

\bibitem[{\citenamefont{Chakrabarti and
  Mondal}(2013{\natexlab{a}})}]{Chakrabarti:2013gra}
\bibinfo{author}{\bibfnamefont{D.}~\bibnamefont{Chakrabarti}} \bibnamefont{and}
  \bibinfo{author}{\bibfnamefont{C.}~\bibnamefont{Mondal}},
  \bibinfo{journal}{Phys. Rev. D} \textbf{\bibinfo{volume}{88}},
  \bibinfo{pages}{073006} (\bibinfo{year}{2013}{\natexlab{a}}),
  \eprint{1307.5128}.

\bibitem[{\citenamefont{Chakrabarti and
  Mondal}(2013{\natexlab{b}})}]{Chakrabarti:2013dda}
\bibinfo{author}{\bibfnamefont{D.}~\bibnamefont{Chakrabarti}} \bibnamefont{and}
  \bibinfo{author}{\bibfnamefont{C.}~\bibnamefont{Mondal}},
  \bibinfo{journal}{Eur. Phys. J. C} \textbf{\bibinfo{volume}{73}},
  \bibinfo{pages}{2671} (\bibinfo{year}{2013}{\natexlab{b}}),
  \eprint{1307.7995}.

\bibitem[{\citenamefont{Chakrabarti and Mondal}(2015)}]{Chakrabarti:2015ama}
\bibinfo{author}{\bibfnamefont{D.}~\bibnamefont{Chakrabarti}} \bibnamefont{and}
  \bibinfo{author}{\bibfnamefont{C.}~\bibnamefont{Mondal}},
  \bibinfo{journal}{Phys. Rev. D} \textbf{\bibinfo{volume}{92}},
  \bibinfo{pages}{074012} (\bibinfo{year}{2015}), \eprint{1509.00598}.

\bibitem[{\citenamefont{Mondal}(2017)}]{Mondal:2017wbf}
\bibinfo{author}{\bibfnamefont{C.}~\bibnamefont{Mondal}},
  \bibinfo{journal}{Eur. Phys. J. C} \textbf{\bibinfo{volume}{77}},
  \bibinfo{pages}{640} (\bibinfo{year}{2017}), \eprint{1709.06877}.

\bibitem[{\citenamefont{Chakrabarti et~al.}(2016)\citenamefont{Chakrabarti,
  Maji, Mondal, and Mukherjee}}]{Chakrabarti:2016yuw}
\bibinfo{author}{\bibfnamefont{D.}~\bibnamefont{Chakrabarti}},
  \bibinfo{author}{\bibfnamefont{T.}~\bibnamefont{Maji}},
  \bibinfo{author}{\bibfnamefont{C.}~\bibnamefont{Mondal}}, \bibnamefont{and}
  \bibinfo{author}{\bibfnamefont{A.}~\bibnamefont{Mukherjee}},
  \bibinfo{journal}{Eur. Phys. J. C} \textbf{\bibinfo{volume}{76}},
  \bibinfo{pages}{409} (\bibinfo{year}{2016}), \eprint{1601.03217}.

\bibitem[{\citenamefont{Chakrabarti et~al.}(2015)\citenamefont{Chakrabarti,
  Mondal, and Mukherjee}}]{Chakrabarti:2015lba}
\bibinfo{author}{\bibfnamefont{D.}~\bibnamefont{Chakrabarti}},
  \bibinfo{author}{\bibfnamefont{C.}~\bibnamefont{Mondal}}, \bibnamefont{and}
  \bibinfo{author}{\bibfnamefont{A.}~\bibnamefont{Mukherjee}},
  \bibinfo{journal}{Phys. Rev. D} \textbf{\bibinfo{volume}{91}},
  \bibinfo{pages}{114026} (\bibinfo{year}{2015}), \eprint{1505.02013}.

\bibitem[{\citenamefont{Gutsche et~al.}(2017)\citenamefont{Gutsche,
  Lyubovitskij, and Schmidt}}]{Gutsche:2016gcd}
\bibinfo{author}{\bibfnamefont{T.}~\bibnamefont{Gutsche}},
  \bibinfo{author}{\bibfnamefont{V.~E.} \bibnamefont{Lyubovitskij}},
  \bibnamefont{and} \bibinfo{author}{\bibfnamefont{I.}~\bibnamefont{Schmidt}},
  \bibinfo{journal}{Eur. Phys. J. C} \textbf{\bibinfo{volume}{77}},
  \bibinfo{pages}{86} (\bibinfo{year}{2017}), \eprint{1610.03526}.

\bibitem[{\citenamefont{Mondal et~al.}(2016)\citenamefont{Mondal, Kumar,
  Dahiya, and Chakrabarti}}]{Mondal:2016xsm}
\bibinfo{author}{\bibfnamefont{C.}~\bibnamefont{Mondal}},
  \bibinfo{author}{\bibfnamefont{N.}~\bibnamefont{Kumar}},
  \bibinfo{author}{\bibfnamefont{H.}~\bibnamefont{Dahiya}}, \bibnamefont{and}
  \bibinfo{author}{\bibfnamefont{D.}~\bibnamefont{Chakrabarti}},
  \bibinfo{journal}{Phys. Rev. D} \textbf{\bibinfo{volume}{94}},
  \bibinfo{pages}{074028} (\bibinfo{year}{2016}), \eprint{1608.01095}.

\bibitem[{\citenamefont{Maji et~al.}(2016)\citenamefont{Maji, Mondal,
  Chakrabarti, and Teryaev}}]{Maji:2015vsa}
\bibinfo{author}{\bibfnamefont{T.}~\bibnamefont{Maji}},
  \bibinfo{author}{\bibfnamefont{C.}~\bibnamefont{Mondal}},
  \bibinfo{author}{\bibfnamefont{D.}~\bibnamefont{Chakrabarti}},
  \bibnamefont{and} \bibinfo{author}{\bibfnamefont{O.~V.}
  \bibnamefont{Teryaev}}, \bibinfo{journal}{JHEP}
  \textbf{\bibinfo{volume}{01}}, \bibinfo{pages}{165} (\bibinfo{year}{2016}),
  \eprint{1506.04560}.

\bibitem[{\citenamefont{Choudhary et~al.}(2022)\citenamefont{Choudhary, Gurjar,
  Chakrabarti, and Mukherjee}}]{Choudhary:2022den}
\bibinfo{author}{\bibfnamefont{P.}~\bibnamefont{Choudhary}},
  \bibinfo{author}{\bibfnamefont{B.}~\bibnamefont{Gurjar}},
  \bibinfo{author}{\bibfnamefont{D.}~\bibnamefont{Chakrabarti}},
  \bibnamefont{and}
  \bibinfo{author}{\bibfnamefont{A.}~\bibnamefont{Mukherjee}},
  \bibinfo{journal}{2206.12206}  (\bibinfo{year}{2022}), \eprint{2206.12206}.

\bibitem[{\citenamefont{Goeke et~al.}(2005)\citenamefont{Goeke, Metz, and
  Schlegel}}]{Goeke:2005hb}
\bibinfo{author}{\bibfnamefont{K.}~\bibnamefont{Goeke}},
  \bibinfo{author}{\bibfnamefont{A.}~\bibnamefont{Metz}}, \bibnamefont{and}
  \bibinfo{author}{\bibfnamefont{M.}~\bibnamefont{Schlegel}},
  \bibinfo{journal}{Phys. Lett. B} \textbf{\bibinfo{volume}{618}},
  \bibinfo{pages}{90} (\bibinfo{year}{2005}), \eprint{hep-ph/0504130}.

\bibitem[{\citenamefont{Bacchetta
  et~al.}(2008{\natexlab{b}})\citenamefont{Bacchetta, Conti, and
  Radici}}]{Bacchetta:2008af}
\bibinfo{author}{\bibfnamefont{A.}~\bibnamefont{Bacchetta}},
  \bibinfo{author}{\bibfnamefont{F.}~\bibnamefont{Conti}}, \bibnamefont{and}
  \bibinfo{author}{\bibfnamefont{M.}~\bibnamefont{Radici}},
  \bibinfo{journal}{Phys. Rev. D} \textbf{\bibinfo{volume}{78}},
  \bibinfo{pages}{074010} (\bibinfo{year}{2008}{\natexlab{b}}),
  \eprint{0807.0323}.

\bibitem[{\citenamefont{Brodsky and Gardner}(2006)}]{Brodsky:2006ha}
\bibinfo{author}{\bibfnamefont{S.~J.} \bibnamefont{Brodsky}} \bibnamefont{and}
  \bibinfo{author}{\bibfnamefont{S.}~\bibnamefont{Gardner}},
  \bibinfo{journal}{Phys. Lett. B} \textbf{\bibinfo{volume}{643}},
  \bibinfo{pages}{22} (\bibinfo{year}{2006}), \eprint{hep-ph/0608219}.

\bibitem[{\citenamefont{Ji et~al.}(2003)\citenamefont{Ji, Ma, and
  Yuan}}]{Ji:2002xn}
\bibinfo{author}{\bibfnamefont{X.-d.} \bibnamefont{Ji}},
  \bibinfo{author}{\bibfnamefont{J.-P.} \bibnamefont{Ma}}, \bibnamefont{and}
  \bibinfo{author}{\bibfnamefont{F.}~\bibnamefont{Yuan}},
  \bibinfo{journal}{Nucl. Phys. B} \textbf{\bibinfo{volume}{652}},
  \bibinfo{pages}{383} (\bibinfo{year}{2003}), \eprint{hep-ph/0210430}.

\bibitem[{\citenamefont{Ahmady et~al.}(2019)\citenamefont{Ahmady, Mondal, and
  Sandapen}}]{Ahmady:2019yvo}
\bibinfo{author}{\bibfnamefont{M.}~\bibnamefont{Ahmady}},
  \bibinfo{author}{\bibfnamefont{C.}~\bibnamefont{Mondal}}, \bibnamefont{and}
  \bibinfo{author}{\bibfnamefont{R.}~\bibnamefont{Sandapen}},
  \bibinfo{journal}{Phys. Rev. D} \textbf{\bibinfo{volume}{100}},
  \bibinfo{pages}{054005} (\bibinfo{year}{2019}), \eprint{1907.06561}.

\bibitem[{\citenamefont{Brodsky
  et~al.}(2002{\natexlab{b}})\citenamefont{Brodsky, Hwang, and
  Schmidt}}]{Brodsky:2002rv}
\bibinfo{author}{\bibfnamefont{S.~J.} \bibnamefont{Brodsky}},
  \bibinfo{author}{\bibfnamefont{D.~S.} \bibnamefont{Hwang}}, \bibnamefont{and}
  \bibinfo{author}{\bibfnamefont{I.}~\bibnamefont{Schmidt}},
  \bibinfo{journal}{Nucl. Phys. B} \textbf{\bibinfo{volume}{642}},
  \bibinfo{pages}{344} (\bibinfo{year}{2002}{\natexlab{b}}),
  \eprint{hep-ph/0206259}.

\bibitem[{\citenamefont{Lu and Ma}(2004)}]{Lu:2004hu}
\bibinfo{author}{\bibfnamefont{Z.}~\bibnamefont{Lu}} \bibnamefont{and}
  \bibinfo{author}{\bibfnamefont{B.-Q.} \bibnamefont{Ma}},
  \bibinfo{journal}{Phys. Rev. D} \textbf{\bibinfo{volume}{70}},
  \bibinfo{pages}{094044} (\bibinfo{year}{2004}), \eprint{hep-ph/0411043}.

\bibitem[{\citenamefont{Wang et~al.}(2017)\citenamefont{Wang, Wang, and
  Lu}}]{Wang:2017onm}
\bibinfo{author}{\bibfnamefont{Z.}~\bibnamefont{Wang}},
  \bibinfo{author}{\bibfnamefont{X.}~\bibnamefont{Wang}}, \bibnamefont{and}
  \bibinfo{author}{\bibfnamefont{Z.}~\bibnamefont{Lu}}, \bibinfo{journal}{Phys.
  Rev. D} \textbf{\bibinfo{volume}{95}}, \bibinfo{pages}{094004}
  (\bibinfo{year}{2017}), \eprint{1702.03637}.

\bibitem[{\citenamefont{Pasquini and Schweitzer}(2014)}]{Pasquini:2014ppa}
\bibinfo{author}{\bibfnamefont{B.}~\bibnamefont{Pasquini}} \bibnamefont{and}
  \bibinfo{author}{\bibfnamefont{P.}~\bibnamefont{Schweitzer}},
  \bibinfo{journal}{Phys. Rev. D} \textbf{\bibinfo{volume}{90}},
  \bibinfo{pages}{014050} (\bibinfo{year}{2014}), \eprint{1406.2056}.

\bibitem[{\citenamefont{Gamberg and Schlegel}(2010)}]{Gamberg:2009uk}
\bibinfo{author}{\bibfnamefont{L.}~\bibnamefont{Gamberg}} \bibnamefont{and}
  \bibinfo{author}{\bibfnamefont{M.}~\bibnamefont{Schlegel}},
  \bibinfo{journal}{Phys. Lett. B} \textbf{\bibinfo{volume}{685}},
  \bibinfo{pages}{95} (\bibinfo{year}{2010}), \eprint{0911.1964}.

\bibitem[{\citenamefont{Fischer and Alkofer}(2003)}]{Fischer:2003rp}
\bibinfo{author}{\bibfnamefont{C.~S.} \bibnamefont{Fischer}} \bibnamefont{and}
  \bibinfo{author}{\bibfnamefont{R.}~\bibnamefont{Alkofer}},
  \bibinfo{journal}{Phys. Rev. D} \textbf{\bibinfo{volume}{67}},
  \bibinfo{pages}{094020} (\bibinfo{year}{2003}), \eprint{hep-ph/0301094}.

\bibitem[{\citenamefont{Boer et~al.}(2003{\natexlab{b}})\citenamefont{Boer,
  Brodsky, and Hwang}}]{Boer:2002ju}
\bibinfo{author}{\bibfnamefont{D.}~\bibnamefont{Boer}},
  \bibinfo{author}{\bibfnamefont{S.~J.} \bibnamefont{Brodsky}},
  \bibnamefont{and} \bibinfo{author}{\bibfnamefont{D.~S.} \bibnamefont{Hwang}},
  \bibinfo{journal}{Phys. Rev. D} \textbf{\bibinfo{volume}{67}},
  \bibinfo{pages}{054003} (\bibinfo{year}{2003}{\natexlab{b}}),
  \eprint{hep-ph/0211110}.

\bibitem[{\citenamefont{Ellis et~al.}(2009)\citenamefont{Ellis, Hwang, and
  Kotzinian}}]{Ellis:2008in}
\bibinfo{author}{\bibfnamefont{J.~R.} \bibnamefont{Ellis}},
  \bibinfo{author}{\bibfnamefont{D.~S.} \bibnamefont{Hwang}}, \bibnamefont{and}
  \bibinfo{author}{\bibfnamefont{A.}~\bibnamefont{Kotzinian}},
  \bibinfo{journal}{Phys. Rev. D} \textbf{\bibinfo{volume}{80}},
  \bibinfo{pages}{074033} (\bibinfo{year}{2009}), \eprint{0808.1567}.

\bibitem[{\citenamefont{Aybat and Rogers}(2011)}]{Aybat:2011zv}
\bibinfo{author}{\bibfnamefont{S.~M.} \bibnamefont{Aybat}} \bibnamefont{and}
  \bibinfo{author}{\bibfnamefont{T.~C.} \bibnamefont{Rogers}},
  \bibinfo{journal}{Phys. Rev. D} \textbf{\bibinfo{volume}{83}},
  \bibinfo{pages}{114042} (\bibinfo{year}{2011}), \eprint{1101.5057}.

\bibitem[{\citenamefont{Echevarria et~al.}(2014)\citenamefont{Echevarria,
  Idilbi, Kang, and Vitev}}]{Echevarria:2014xaa}
\bibinfo{author}{\bibfnamefont{M.~G.} \bibnamefont{Echevarria}},
  \bibinfo{author}{\bibfnamefont{A.}~\bibnamefont{Idilbi}},
  \bibinfo{author}{\bibfnamefont{Z.-B.} \bibnamefont{Kang}}, \bibnamefont{and}
  \bibinfo{author}{\bibfnamefont{I.}~\bibnamefont{Vitev}},
  \bibinfo{journal}{Phys. Rev. D} \textbf{\bibinfo{volume}{89}},
  \bibinfo{pages}{074013} (\bibinfo{year}{2014}), \eprint{1401.5078}.

\bibitem[{\citenamefont{Echevarria et~al.}(2013)\citenamefont{Echevarria,
  Idilbi, Sch\"afer, and Scimemi}}]{Echevarria:2012pw}
\bibinfo{author}{\bibfnamefont{M.~G.} \bibnamefont{Echevarria}},
  \bibinfo{author}{\bibfnamefont{A.}~\bibnamefont{Idilbi}},
  \bibinfo{author}{\bibfnamefont{A.}~\bibnamefont{Sch\"afer}},
  \bibnamefont{and} \bibinfo{author}{\bibfnamefont{I.}~\bibnamefont{Scimemi}},
  \bibinfo{journal}{Eur. Phys. J. C} \textbf{\bibinfo{volume}{73}},
  \bibinfo{pages}{2636} (\bibinfo{year}{2013}), \eprint{1208.1281}.

\bibitem[{\citenamefont{Kishore et~al.}(2020)\citenamefont{Kishore, Mukherjee,
  and Rajesh}}]{Kishore:2019fzb}
\bibinfo{author}{\bibfnamefont{R.}~\bibnamefont{Kishore}},
  \bibinfo{author}{\bibfnamefont{A.}~\bibnamefont{Mukherjee}},
  \bibnamefont{and} \bibinfo{author}{\bibfnamefont{S.}~\bibnamefont{Rajesh}},
  \bibinfo{journal}{Phys. Rev. D} \textbf{\bibinfo{volume}{101}},
  \bibinfo{pages}{054003} (\bibinfo{year}{2020}), \eprint{1908.03698}.

\bibitem[{\citenamefont{Anselmino et~al.}(2006)\citenamefont{Anselmino,
  Boglione, Collins, D'Alesio, Efremov, Goeke, Kotzinian, Menzel, Metz, Murgia
  et~al.}}]{anselmino2006comparing}
\bibinfo{author}{\bibfnamefont{M.}~\bibnamefont{Anselmino}},
  \bibinfo{author}{\bibfnamefont{M.}~\bibnamefont{Boglione}},
  \bibinfo{author}{\bibfnamefont{J.~C.} \bibnamefont{Collins}},
  \bibinfo{author}{\bibfnamefont{U.}~\bibnamefont{D'Alesio}},
  \bibinfo{author}{\bibfnamefont{A.}~\bibnamefont{Efremov}},
  \bibinfo{author}{\bibfnamefont{K.}~\bibnamefont{Goeke}},
  \bibinfo{author}{\bibfnamefont{A.}~\bibnamefont{Kotzinian}},
  \bibinfo{author}{\bibfnamefont{S.}~\bibnamefont{Menzel}},
  \bibinfo{author}{\bibfnamefont{A.}~\bibnamefont{Metz}},
  \bibinfo{author}{\bibfnamefont{F.}~\bibnamefont{Murgia}},
  \bibnamefont{et~al.}, \bibinfo{journal}{hep-ph/0511017} pp.
  \bibinfo{pages}{236--243} (\bibinfo{year}{2006}), \eprint{hep-ph/0511017}.

\bibitem[{\citenamefont{Alexeev et~al.}(2019)}]{COMPASS:2018ofp}
\bibinfo{author}{\bibfnamefont{M.~G.} \bibnamefont{Alexeev}}
  \bibnamefont{et~al.} (\bibinfo{collaboration}{COMPASS}),
  \bibinfo{journal}{Nucl. Phys. B} \textbf{\bibinfo{volume}{940}},
  \bibinfo{pages}{34} (\bibinfo{year}{2019}), \eprint{1809.02936}.

\bibitem[{\citenamefont{Bacchetta et~al.}(2000)\citenamefont{Bacchetta,
  Boglione, Henneman, and Mulders}}]{Bacchetta:1999kz}
\bibinfo{author}{\bibfnamefont{A.}~\bibnamefont{Bacchetta}},
  \bibinfo{author}{\bibfnamefont{M.}~\bibnamefont{Boglione}},
  \bibinfo{author}{\bibfnamefont{A.}~\bibnamefont{Henneman}}, \bibnamefont{and}
  \bibinfo{author}{\bibfnamefont{P.~J.} \bibnamefont{Mulders}},
  \bibinfo{journal}{Phys. Rev. Lett.} \textbf{\bibinfo{volume}{85}},
  \bibinfo{pages}{712} (\bibinfo{year}{2000}), \eprint{hep-ph/9912490}.

\bibitem[{\citenamefont{Boffi et~al.}(2009)\citenamefont{Boffi, Efremov,
  Pasquini, and Schweitzer}}]{Boffi:2009sh}
\bibinfo{author}{\bibfnamefont{S.}~\bibnamefont{Boffi}},
  \bibinfo{author}{\bibfnamefont{A.~V.} \bibnamefont{Efremov}},
  \bibinfo{author}{\bibfnamefont{B.}~\bibnamefont{Pasquini}}, \bibnamefont{and}
  \bibinfo{author}{\bibfnamefont{P.}~\bibnamefont{Schweitzer}},
  \bibinfo{journal}{Phys. Rev. D} \textbf{\bibinfo{volume}{79}},
  \bibinfo{pages}{094012} (\bibinfo{year}{2009}), \eprint{0903.1271}.

\bibitem[{\citenamefont{Anselmino
  et~al.}(2005{\natexlab{b}})\citenamefont{Anselmino, Boglione, D'Alesio,
  Kotzinian, Murgia, and Prokudin}}]{Anselmino:2005nn}
\bibinfo{author}{\bibfnamefont{M.}~\bibnamefont{Anselmino}},
  \bibinfo{author}{\bibfnamefont{M.}~\bibnamefont{Boglione}},
  \bibinfo{author}{\bibfnamefont{U.}~\bibnamefont{D'Alesio}},
  \bibinfo{author}{\bibfnamefont{A.}~\bibnamefont{Kotzinian}},
  \bibinfo{author}{\bibfnamefont{F.}~\bibnamefont{Murgia}}, \bibnamefont{and}
  \bibinfo{author}{\bibfnamefont{A.}~\bibnamefont{Prokudin}},
  \bibinfo{journal}{Phys. Rev. D} \textbf{\bibinfo{volume}{71}},
  \bibinfo{pages}{074006} (\bibinfo{year}{2005}{\natexlab{b}}),
  \eprint{hep-ph/0501196}.

\bibitem[{\citenamefont{Ji et~al.}(2006)\citenamefont{Ji, Qiu, Vogelsang, and
  Yuan}}]{Ji:2006br}
\bibinfo{author}{\bibfnamefont{X.}~\bibnamefont{Ji}},
  \bibinfo{author}{\bibfnamefont{J.-W.} \bibnamefont{Qiu}},
  \bibinfo{author}{\bibfnamefont{W.}~\bibnamefont{Vogelsang}},
  \bibnamefont{and} \bibinfo{author}{\bibfnamefont{F.}~\bibnamefont{Yuan}},
  \bibinfo{journal}{Phys. Lett. B} \textbf{\bibinfo{volume}{638}},
  \bibinfo{pages}{178} (\bibinfo{year}{2006}), \eprint{hep-ph/0604128}.

\bibitem[{\citenamefont{Anselmino
  et~al.}(2007{\natexlab{a}})\citenamefont{Anselmino, Boglione, Prokudin, and
  Turk}}]{Anselmino:2006rv}
\bibinfo{author}{\bibfnamefont{M.}~\bibnamefont{Anselmino}},
  \bibinfo{author}{\bibfnamefont{M.}~\bibnamefont{Boglione}},
  \bibinfo{author}{\bibfnamefont{A.}~\bibnamefont{Prokudin}}, \bibnamefont{and}
  \bibinfo{author}{\bibfnamefont{C.}~\bibnamefont{Turk}},
  \bibinfo{journal}{Eur. Phys. J. A} \textbf{\bibinfo{volume}{31}},
  \bibinfo{pages}{373} (\bibinfo{year}{2007}{\natexlab{a}}),
  \eprint{hep-ph/0606286}.

\bibitem[{\citenamefont{Ji et~al.}(2005)\citenamefont{Ji, Ma, and
  Yuan}}]{Ji:2004wu}
\bibinfo{author}{\bibfnamefont{X.-d.} \bibnamefont{Ji}},
  \bibinfo{author}{\bibfnamefont{J.-p.} \bibnamefont{Ma}}, \bibnamefont{and}
  \bibinfo{author}{\bibfnamefont{F.}~\bibnamefont{Yuan}},
  \bibinfo{journal}{Phys. Rev. D} \textbf{\bibinfo{volume}{71}},
  \bibinfo{pages}{034005} (\bibinfo{year}{2005}), \eprint{hep-ph/0404183}.

\bibitem[{\citenamefont{Ji et~al.}(2004)\citenamefont{Ji, Ma, and
  Yuan}}]{Ji:2004xq}
\bibinfo{author}{\bibfnamefont{X.-d.} \bibnamefont{Ji}},
  \bibinfo{author}{\bibfnamefont{J.-P.} \bibnamefont{Ma}}, \bibnamefont{and}
  \bibinfo{author}{\bibfnamefont{F.}~\bibnamefont{Yuan}},
  \bibinfo{journal}{Phys. Lett. B} \textbf{\bibinfo{volume}{597}},
  \bibinfo{pages}{299} (\bibinfo{year}{2004}), \eprint{hep-ph/0405085}.

\bibitem[{\citenamefont{Echevarr\'\i{}a
  et~al.}(2013)\citenamefont{Echevarr\'\i{}a, Idilbi, and
  Scimemi}}]{Echevarria:2012js}
\bibinfo{author}{\bibfnamefont{M.~G.} \bibnamefont{Echevarr\'\i{}a}},
  \bibinfo{author}{\bibfnamefont{A.}~\bibnamefont{Idilbi}}, \bibnamefont{and}
  \bibinfo{author}{\bibfnamefont{I.}~\bibnamefont{Scimemi}},
  \bibinfo{journal}{Phys. Lett. B} \textbf{\bibinfo{volume}{726}},
  \bibinfo{pages}{795} (\bibinfo{year}{2013}), \eprint{1211.1947}.

\bibitem[{\citenamefont{Anselmino
  et~al.}(2011{\natexlab{a}})\citenamefont{Anselmino, Boglione, D'Alesio,
  Melis, Murgia, Nocera, and Prokudin}}]{PhysRevD.83.114019}
\bibinfo{author}{\bibfnamefont{M.}~\bibnamefont{Anselmino}},
  \bibinfo{author}{\bibfnamefont{M.}~\bibnamefont{Boglione}},
  \bibinfo{author}{\bibfnamefont{U.}~\bibnamefont{D'Alesio}},
  \bibinfo{author}{\bibfnamefont{S.}~\bibnamefont{Melis}},
  \bibinfo{author}{\bibfnamefont{F.}~\bibnamefont{Murgia}},
  \bibinfo{author}{\bibfnamefont{E.~R.} \bibnamefont{Nocera}},
  \bibnamefont{and} \bibinfo{author}{\bibfnamefont{A.}~\bibnamefont{Prokudin}},
  \bibinfo{journal}{Phys. Rev. D} \textbf{\bibinfo{volume}{83}},
  \bibinfo{pages}{114019} (\bibinfo{year}{2011}{\natexlab{a}}).

\bibitem[{\citenamefont{Kretzer et~al.}(2001)\citenamefont{Kretzer, Leader, and
  Christova}}]{Kretzer:2001pz}
\bibinfo{author}{\bibfnamefont{S.}~\bibnamefont{Kretzer}},
  \bibinfo{author}{\bibfnamefont{E.}~\bibnamefont{Leader}}, \bibnamefont{and}
  \bibinfo{author}{\bibfnamefont{E.}~\bibnamefont{Christova}},
  \bibinfo{journal}{Eur. Phys. J. C} \textbf{\bibinfo{volume}{22}},
  \bibinfo{pages}{269} (\bibinfo{year}{2001}), \eprint{hep-ph/0108055}.

\bibitem[{\citenamefont{Anselmino
  et~al.}(2013{\natexlab{b}})\citenamefont{Anselmino, Boglione, D'Alesio,
  Melis, Murgia, and Prokudin}}]{Anselmino:2013vqa}
\bibinfo{author}{\bibfnamefont{M.}~\bibnamefont{Anselmino}},
  \bibinfo{author}{\bibfnamefont{M.}~\bibnamefont{Boglione}},
  \bibinfo{author}{\bibfnamefont{U.}~\bibnamefont{D'Alesio}},
  \bibinfo{author}{\bibfnamefont{S.}~\bibnamefont{Melis}},
  \bibinfo{author}{\bibfnamefont{F.}~\bibnamefont{Murgia}}, \bibnamefont{and}
  \bibinfo{author}{\bibfnamefont{A.}~\bibnamefont{Prokudin}},
  \bibinfo{journal}{Phys. Rev. D} \textbf{\bibinfo{volume}{87}},
  \bibinfo{pages}{094019} (\bibinfo{year}{2013}{\natexlab{b}}),
  \eprint{1303.3822}.

\bibitem[{\citenamefont{Ji et~al.}(2021)\citenamefont{Ji, Liu, Sch\"afer, and
  Yuan}}]{Ji:2020jeb}
\bibinfo{author}{\bibfnamefont{X.}~\bibnamefont{Ji}},
  \bibinfo{author}{\bibfnamefont{Y.}~\bibnamefont{Liu}},
  \bibinfo{author}{\bibfnamefont{A.}~\bibnamefont{Sch\"afer}},
  \bibnamefont{and} \bibinfo{author}{\bibfnamefont{F.}~\bibnamefont{Yuan}},
  \bibinfo{journal}{Phys. Rev. D} \textbf{\bibinfo{volume}{103}},
  \bibinfo{pages}{074005} (\bibinfo{year}{2021}), \eprint{2011.13397}.

\bibitem[{\citenamefont{Anselmino
  et~al.}(2011{\natexlab{b}})\citenamefont{Anselmino, Boglione, D'Alesio,
  Melis, Murgia, Nocera, and Prokudin}}]{Anselmino:2011ch}
\bibinfo{author}{\bibfnamefont{M.}~\bibnamefont{Anselmino}},
  \bibinfo{author}{\bibfnamefont{M.}~\bibnamefont{Boglione}},
  \bibinfo{author}{\bibfnamefont{U.}~\bibnamefont{D'Alesio}},
  \bibinfo{author}{\bibfnamefont{S.}~\bibnamefont{Melis}},
  \bibinfo{author}{\bibfnamefont{F.}~\bibnamefont{Murgia}},
  \bibinfo{author}{\bibfnamefont{E.~R.} \bibnamefont{Nocera}},
  \bibnamefont{and} \bibinfo{author}{\bibfnamefont{A.}~\bibnamefont{Prokudin}},
  \bibinfo{journal}{Phys. Rev. D} \textbf{\bibinfo{volume}{83}},
  \bibinfo{pages}{114019} (\bibinfo{year}{2011}{\natexlab{b}}),
  \eprint{1101.1011}.

\bibitem[{\citenamefont{Wang et~al.}(2018)\citenamefont{Wang, Mao, and
  Lu}}]{Wang:2018naw}
\bibinfo{author}{\bibfnamefont{X.}~\bibnamefont{Wang}},
  \bibinfo{author}{\bibfnamefont{W.}~\bibnamefont{Mao}}, \bibnamefont{and}
  \bibinfo{author}{\bibfnamefont{Z.}~\bibnamefont{Lu}}, \bibinfo{journal}{Eur.
  Phys. J. C} \textbf{\bibinfo{volume}{78}}, \bibinfo{pages}{643}
  (\bibinfo{year}{2018}), \eprint{1805.03017}.

\bibitem[{\citenamefont{Anselmino
  et~al.}(2007{\natexlab{b}})\citenamefont{Anselmino, Boglione, D'Alesio,
  Kotzinian, Murgia, Prokudin, and Turk}}]{Anselmino:2007fs}
\bibinfo{author}{\bibfnamefont{M.}~\bibnamefont{Anselmino}},
  \bibinfo{author}{\bibfnamefont{M.}~\bibnamefont{Boglione}},
  \bibinfo{author}{\bibfnamefont{U.}~\bibnamefont{D'Alesio}},
  \bibinfo{author}{\bibfnamefont{A.}~\bibnamefont{Kotzinian}},
  \bibinfo{author}{\bibfnamefont{F.}~\bibnamefont{Murgia}},
  \bibinfo{author}{\bibfnamefont{A.}~\bibnamefont{Prokudin}}, \bibnamefont{and}
  \bibinfo{author}{\bibfnamefont{C.}~\bibnamefont{Turk}},
  \bibinfo{journal}{Phys. Rev. D} \textbf{\bibinfo{volume}{75}},
  \bibinfo{pages}{054032} (\bibinfo{year}{2007}{\natexlab{b}}),
  \eprint{hep-ph/0701006}.

\end{thebibliography}
\end{document}